\documentclass{article} % For LaTeX2e
\PassOptionsToPackage{numbers, compress}{natbib}
\usepackage[utf8]{inputenc} % allow utf-8 input
\usepackage{graphicx}
\usepackage[T1]{fontenc}    % use 8-bit T1 fonts
\usepackage{hyperref}       % hyperlinks
\hypersetup{hidelinks, linktoc=all, colorlinks=false} % hyperlinks w/o colored boxes
\usepackage{xurl}            % simple URL typesetting
\usepackage{booktabs}       % professional-quality tables
\usepackage{longtable}      % for latex longtables in the IMP rules
\usepackage{amsfonts}       % blackboard math symbols
\usepackage{amsthm}         % for theorems and definition
\usepackage{nicefrac}       % compact symbols for 1/2, etc.
\usepackage{microtype}      % microtypography
\usepackage[table]{xcolor}  % for coloring table cells
\usepackage{multirow}
\usepackage[most]{tcolorbox}  % to draw customized boxes
\tcbuselibrary{listings}      % to use listing inside tcolorbox
\tcbuselibrary{skins}         % to set title position in tcolorbox
\usepackage{footnote}
\usepackage[numbers]{natbib}% citing with numbers and et al
\usepackage{xspace}         % for adding space after commands
\usepackage{wrapfig}        % for adding text wrapped figures
\usepackage{mathpartir}     % for inferrule
\usepackage{amsmath}        % for math symbols
\usepackage{amssymb}        % for math symbols
\usepackage[font=footnotesize,labelfont=bf]{caption} % for changing figure and table caption font size
\usepackage[font=footnotesize]{subcaption}     % for subcaption environment
\usepackage{listings}       % for listings environment
\usepackage{outlines}       % for bulleted list with subitems
\usepackage{tikz}           % for drawing figures with Tikz
\usetikzlibrary{calc,positioning,fit,tikzmark,shapes.multipart,shapes,arrows.meta} %% for using polar coords in Tikz pictures
\usepackage{pifont}
\usepackage{titletoc}       % for partial TOCs in the Appendix
\usepackage{authblk}        % for author affliation grouping
\usepackage[only,llbracket,rrbracket]{stmaryrd}
\usepackage{mathrsfs}
\usepackage{mathtools}
\usepackage{pgfplots}
\pgfplotsset{compat=1.18}
\usepackage{enumitem}       % for enumerating with custom commands
\usepackage{placeins}       % for preventing floats crossing section boundaries
\usepackage{threeparttable} % for table footer
\usepackage{xparse}         % for multiple default args for Latex commands
\usepackage{bm}             % for bold math symbols
\usepackage{siunitx}        % for advanced table column formatting
\usepackage{colortbl}       % for row coloring
\usepackage{adjustbox}      % for resizing table to fit specific width
\usepackage{xfp}      

\definecolor{gray}{RGB}{211,211,211}
\newcommand{\jbasicstyle}{\small\ttfamily} % Style of code

\newcommand{\jnumberstyle}{\scriptsize}

\lstdefinelanguage{pseudo}
{
  morekeywords={},
  keywordstyle=\bfseries,
  lineskip=-0.1em,
  numbers=left, % none for no numbers
  numberstyle=\jnumberstyle,
  numbersep=4pt,
  basicstyle=\jbasicstyle,
  breaklines=true,
  breakautoindent=true,
  tabsize=2,
  columns=fullflexible,
  morecomment=*[l][\textsl]{//},
  mathescape=true,
  xleftmargin=10pt,
}

\lstdefinelanguage{todo-comment}
{
  morekeywords={},
  keywordstyle=\bfseries,
  lineskip=-0.1em,
  numbers=none,
  basicstyle=\scriptsize\ttfamily,
  breaklines=true,
  breakautoindent=true,
  tabsize=2,
  columns=fullflexible,
  morecomment=*[l][\textsl]{//},
  mathescape=true,
  xleftmargin=0pt,
}

\definecolor{keywordcolor}{rgb}{0,0,1}      % Blue for primary keywords
\definecolor{modifiercolor}{rgb}{0.5,0,0.5} % Purple for modifiers
\definecolor{datatypecolor}{rgb}{0.82,0.16,0.46} % Reddish for data types
\definecolor{methodcolor}{rgb}{0.25,0.5,0.35} % Greenish for method-related keywords
\definecolor{byzantine}{rgb}{0.74, 0.2, 0.64}  % if, else
\definecolor{cadetblue}{rgb}{0.37, 0.62, 0.63}  % data type
\definecolor{cadet}{rgb}{0.0, 0.42, 0.24}
\definecolor{brown(web)}{rgb}{0.65, 0.16, 0.16}  % string
\definecolor{bluegray}{rgb}{0.2, 0.2, 0.6}

\lstdefinelanguage{java-pretty}
{
  language=java,
  numbers=left,
  basicstyle=\scriptsize\ttfamily,
  numberstyle=\scriptsize,
  breaklines=true,
  columns=fullflexible,
  xleftmargin=18pt,
  tabsize=2,
  showstringspaces=false,
  deletekeywords={public, private, protected, static, final, class, interface, abstract, implements, extends, if, else, while, do, for, switch, case, default, break, continue, return, int, long, double, float, boolean, char, void, String,this},
  morekeywords=[1]{if, else, while, do, for, switch, case, default, break, continue, return},
  keywordstyle=[1]\color{byzantine}\bfseries,
  morekeywords=[2]{public, private, protected, def, object, static, final, class, interface, abstract, implements, extends},
  keywordstyle=[2]\color{bluegray}\bfseries,
  morekeywords=[3]{int, Int, long, double, float, boolean, char, void, String, @Override, @Test},
  keywordstyle=[3]\color{cadet}\bfseries,
  morekeywords=[4]{class, interface, extends, implements, new, super, throw, throws, try, catch, finally},
  keywordstyle=[4]\color{methodcolor},
  morecomment=[l]{//},
  commentstyle=\color{cadet},
  stringstyle=\color{brown(web)},
}

\lstdefinelanguage{imp-pretty-no-lines}
{
  language=imp-srp,
  numbers=none,
  basicstyle=\scriptsize\ttfamily,
  numberstyle=\scriptsize,
  breaklines=true,
  columns=fullflexible,
  aboveskip=0pt,
  belowskip=0pt,
  xleftmargin=0pt,
  tabsize=2,
  showstringspaces=false,
  deletekeywords={public, private, protected, static, final, class, interface, abstract, implements, extends, if, else, while, do, for, switch, case, default, break, continue, return, int, long, double, float, boolean, char, void, String,this},
  morekeywords=[1]{if, else, while, do, for, switch, case, default, break, continue, return},
  keywordstyle=[1]\color{byzantine}\bfseries,
  morekeywords=[2]{public, private, protected, static, final, class, interface, abstract, implements, extends},
  keywordstyle=[2]\color{bluegray}\bfseries,
  morekeywords=[3]{int, long, double, float, boolean, char, void, String, @Override, @Test},
  keywordstyle=[3]\color{cadet}\bfseries,
  morekeywords=[4]{class, interface, extends, implements, new, super, throw, throws, try, catch, finally},
  keywordstyle=[4]\color{methodcolor},
  morecomment=[l]{//},
  moredelim=**[is][\spanDel]{<<DEL>>}{<<END>>},           
  moredelim=**[is][\spanAdd]{<<ADD>>}{<<END>>},   
  commentstyle=\color{cadet},
  stringstyle=\color{brown(web)},
}

\lstdefinelanguage{java-pretty-small-framed}
{
  language=java,
  basicstyle=\tiny\ttfamily,
  breaklines=true,
  columns=fullflexible,
  tabsize=2,
  showstringspaces=false,
  deletekeywords={public, private, protected, static, final, class, interface, abstract, implements, extends, if, else, while, do, for, switch, case, default, break, continue, return, int, long, double, float, boolean, char, void, String,this},
  morekeywords=[1]{if, else, while, do, for, switch, case, default, break, continue, return},
  keywordstyle=[1]\color{byzantine}\bfseries,
  morekeywords=[2]{public, private, protected, static, final, class, interface, abstract, implements, extends},
  keywordstyle=[2]\color{bluegray}\bfseries,
  morekeywords=[3]{int, long, double, float, boolean, char, void, String, @Override, @Test},
  keywordstyle=[3]\color{cadet}\bfseries,
  morekeywords=[4]{class, interface, extends, implements, new, super, throw, throws, try, catch, finally},
  keywordstyle=[4]\color{methodcolor},
  morecomment=[l]{//},
  commentstyle=\color{cadet},
  stringstyle=\color{brown(web)},
}

\lstdefinelanguage{imp-grammar}{
	language=java,
	keywords=[2]{SYNTACTIC_ASSIGN, SYNTACTIC_IF, SYNTACTIC_ELSE, SYNTACTIC_WHILE, SYNTACTIC_ADD, SYNTACTIC_SUB, SYNTACTIC_MUL, SYNTACTIC_DIV, SYNTACTIC_MOD, SYNTACTIC_LT, SYNTACTIC_LTEQ, SYNTACTIC_GT, SYNTACTIC_GTEQ, SYNTACTIC_EQ, SYNTACTIC_NEQ, SYNTACTIC_NOT, SYNTACTIC_AND, SYNTACTIC_OR, Rule},
  keywordstyle=[2]\bfseries,
	keywords=[3]{Order, Asc},
	keywordstyle=[3]\color{purple},
	basicstyle=\scriptsize\ttfamily,
}

\lstdefinelanguage{operational-semantics}
{
  language=java,
  basicstyle=\tiny\ttfamily,
  breaklines=true,
  columns=fullflexible,
  tabsize=2,
  showstringspaces=false,
  deletekeywords={public, private, protected, static, final, class, interface, abstract, implements, extends, if, else, while, do, for, switch, case, default, break, continue, return, int, long, double, float, boolean, char, void, String,this},
  morekeywords=[1]{if, else, while, do, for, switch, case, default, break, continue, return},
  keywordstyle=[1]\color{byzantine}\bfseries,
  morekeywords=[2]{public, private, protected, static, final, class, interface, abstract, implements, extends},
  keywordstyle=[2]\color{bluegray}\bfseries,
  morekeywords=[3]{int, long, double, float, boolean, char, void, String, @Override, @Test},
  keywordstyle=[3]\color{cadet}\bfseries,
  morekeywords=[4]{class, interface, extends, implements, new, super, throw, throws, try, catch, finally},
  keywordstyle=[4]\color{methodcolor},
  morecomment=[l]{//},
  commentstyle=\color{cadet},
  stringstyle=\color{brown(web)},
}

\lstdefinelanguage{operational-semantics-small-framed}
{
  language=java,
  basicstyle=\tiny\ttfamily,
  breaklines=false,
  columns=fullflexible,
  tabsize=2,
  showstringspaces=false,
  deletekeywords={public, private, protected, static, final, class, interface, abstract, implements, extends, if, else, while, do, for, switch, case, default, break, continue, return, int, long, double, float, boolean, char, void, String,this},
  morekeywords=[1]{if, else, while, do, for, switch, case, default, break, continue, return},
  keywordstyle=[1]\color{byzantine}\bfseries,
  morekeywords=[2]{public, private, protected, static, final, class, interface, abstract, implements, extends},
  keywordstyle=[2]\color{bluegray}\bfseries,
  morekeywords=[3]{int, long, double, float, boolean, char, void, String, @Override, @Test},
  keywordstyle=[3]\color{cadet}\bfseries,
  morekeywords=[4]{class, interface, extends, implements, new, super, throw, throws, try, catch, finally},
  keywordstyle=[4]\color{methodcolor},
  morecomment=[l]{//},
  commentstyle=\color{cadet},
  stringstyle=\color{brown(web)},
}

\lstdefinelanguage{bnf-grammar}{
  language=java,
  basicstyle=\scriptsize\ttfamily,
  numbers=left,
  numberstyle=\scriptsize,
  xleftmargin=18pt,
  morekeywords=[1]{::=,|,program,stmt,stmt_list,decl_stmt,assign_stmt,assgn_stmt,id,type,exp,literal,letter,digit,type,decl,BOOL, MATHOP,LOGICALOP,LOGNOT,SP,ID,block,aexp,bexp,RELOP,id_list,ids,program, class_decls, class_decl, field_decls, field_decl, fields, field_list, method_decls, method_decl, params, param, param_list, var_decls, var_decl, vars, var_list, var_defs, var_def, stmt, assign_stmt, if_stmt, while_stmt, for_stmt, return_stmt, break_stmt, block, stmts, method_call_stmt, field_access_stmt, method_call_list, field_access_list, arg_list, args, expr, literal, unary_expr, binary_expr, method_call_expr, field_access_expr, array_index_expr, BOOL, ARRAY, CHAR, STRING, INTEGER, MATHOP, RELOP, NOT, LOGICALOP, ID, letter, digit, SP, alphanumeric, alphanumeric_list, mathop, relop, bool, lognot, logicalop
},
  keywordstyle=[1]\color{blue},
  commentstyle=\color{green},
  stringstyle=\color{red},
  breaklines=true,
  backgroundcolor=\color{white},
  showstringspaces=false,
  rulecolor=\color{black},
  morecomment=[l]{\#},
}

\lstdefinelanguage{bnf-grammar-tiny}{
  language=java,
  basicstyle=\tiny\ttfamily,
  numbers=left,
  numberstyle=\tiny,
  xleftmargin=18pt,
  morekeywords=[1]{::=,|,program,stmt,stmt_list,decl_stmt,assign_stmt,id,type,exp,literal,letter,digit,type,decl,BOOL, MATHOP,LOGICALOP,LOGNOT,SP,ID,block,aexp,bexp,RELOP,id_list,ids,program, class_decls, class_decl, field_decls, field_decl, fields, field_list, method_decls, method_decl, params, param, param_list, var_decls, var_decl, vars, var_list, var_defs, var_def, stmt, assign_stmt, if_stmt, while_stmt, for_stmt, return_stmt, break_stmt, block, stmts, method_call_stmt, field_access_stmt, method_call_list, field_access_list, arg_list, args, expr, literal, unary_expr, binary_expr, method_call_expr, field_access_expr, array_index_expr, BOOL, ARRAY, CHAR, STRING, INTEGER, MATHOP, RELOP, NOT, LOGICALOP, ID, letter, digit, SP, alphanumeric, alphanumeric_list
},
  keywordstyle=[1]\color{blue},
  commentstyle=\color{green},
  stringstyle=\color{red},
  breaklines=true,
  backgroundcolor=\color{white},
  showstringspaces=false,
  rulecolor=\color{black},
  morecomment=[l]{\#},
}

\lstdefinelanguage{k-framework}{
  language=java,
  basicstyle=\scriptsize\ttfamily,
  numbers=left,
  numberstyle=\scriptsize,
  xleftmargin=18pt,
  morekeywords=[1]{syntax, configuration, rule, imports, import},
  keywordstyle=[1]\textbf,
  morekeywords=[2]{module, endmodule},
  keywordstyle=[2]\textbf,
  breaklines=true,
  backgroundcolor=\color{white},
  showstringspaces=false,
  rulecolor=\color{black},
  morecomment=[l]{\#},
}

\lstdefinelanguage{imp}{
	language=java,
	keywords=[2]{int, while, if, else, return, break, halt},
        keywordstyle=[2]\bfseries,
	keywords=[3]{Order, Asc},
	keywordstyle=[3]\color{purple},
	basicstyle=\scriptsize\ttfamily,
}

\lstdefinestyle{impstyle}{
  language=imp,
  basicstyle=\ttfamily\footnotesize,
  keywordstyle=\bfseries,
  keywordstyle=[2]\itshape,
  commentstyle=\color{gray}\ttfamily,
  stringstyle=\color{brown},
  showstringspaces=false,
  columns=fullflexible,
  keepspaces=true,
  breaklines=true,
  frame=none,
  tabsize=2,
  aboveskip=0.5em,
  belowskip=0.5em,
  literate={*}{{\textbf{*}}}1
}

\newcommand{\spanDel}[1]{
\begingroup\setlength{\fboxsep}{1ex}\colorbox{red!10}{#1}\endgroup
}

\newcommand{\spanAdd}[1]{
\begingroup\setlength{\fboxsep}{1ex}\colorbox{green!10}{#1}\endgroup
}

\lstdefinelanguage{imp-srp}{
	language=java,
	keywords=[2]{int, while, if, else, return, break, halt},
        keywordstyle=[2]\bfseries,
	keywords=[3]{Order, Asc},
	keywordstyle=[3]\color{purple},
	basicstyle=\scriptsize\ttfamily,
        moredelim=**[is][\spanDel]{<<DEL>>}{<<END>>},
        moredelim=**[is][\spanAdd]{<<ADD>>}{<<END>>},
}

\lstdefinelanguage{imp-unseen}{
	language=java,
	keywords=[2]{int, while, if, else, return, break, halt},
        keywordstyle=[2]\bfseries,
        inputencoding=utf8,
        extendedchars=true,
        literate=
             {𐕂}{{\char"10542}}1
             {𐕕}{{\char"10555}}1
             {𐕊}{{\char"1054A}}1
             {𐕃}{{\char"10543}}1
             {𐕐}{{\char"10550}}1
             {𐕏}{{\char"1054F}}1
             {𐔸}{{\char"10538}}1
             {𐕟}{{\char"1055F}}1,
	basicstyle=\scriptsize\ttfamily,
}

\definecolor{light-purple}{RGB}{151,156,171}
\definecolor{cherryblossompink}{rgb}{1.0, 0.72, 0.77}
\definecolor{blue-color}{RGB}{40,166,189}
\definecolor{pink-color}{RGB}{237,46,104}
\definecolor{dark-grey-color}{RGB}{79,91,102}
\definecolor{darkbyzantium}{rgb}{0.36, 0.22, 0.33}
\definecolor{bluebell}{rgb}{0.64, 0.64, 0.82}
\definecolor{airforceblue}{rgb}{0.36, 0.54, 0.66}
\definecolor{response}{RGB}{245,198,165}
\newcommand{\promptsubsection}[1]{
\setlength{\parskip}{6pt} \noindent\textbf{{#1}:}
}

\newtcolorbox[list inside=prompt,auto counter,number within=subsection]{prompt}[1][]{
    colbacktitle=cherryblossompink,
    colframe=cherryblossompink,
    coltitle=darkbyzantium,
    fontupper=\scriptsize\ttfamily,
    boxsep=5pt,
    left=0pt,
    right=0pt,
    top=0pt,
    bottom=0pt,
    boxrule=1pt,
    enhanced,
    breakable,
    skin first=enhanced,
    skin middle=enhanced,
    skin last=enhanced,
    #1,
}

\newtheoremstyle{myDef}% name of the style to be used
  {2.5mm}% measure of space to leave above the theorem. E.g.: 3pt
  {2.5mm}% measure of space to leave below the theorem. E.g.: 3pt
  {}% name of font to use in the body of the theorem
  {0pt}% measure of space to indent
  {}% name of head font
  {.  }% punctuation between head and body
  { }% space after theorem head; " " = normal interword space
  {\textbf{\thmname{#1}\thmnumber{ #2}}{\thmnote{ (#3)}}}
\theoremstyle{myDef}
\newtheorem{myDef}{Definition}[section]

\definecolor{keyblue}{RGB}{235,242,250}   % very light
\definecolor{keyborder}{RGB}{120,160,210}

\newtcolorbox{keyfinding}[1][Key Findings]{
  enhanced,
  title=#1,
  boxrule=0.5pt,
  colframe=black!90,
  colback=keyblue,
  arc=1pt,
  left=0pt,
  right=0pt,
  top=1pt,
  before skip=11pt,
  after skip=11pt,
  bottom=1pt,
  fonttitle=\bfseries\small,
  fontupper=\keyFindingFont
}

\newtcolorbox{accuracyBox}{
  sharp corners,
  shadow=false,
  boxrule=0.8pt,
  colframe=black!100,
  colback=white,
  arc=0pt,
  left=1pt,
  right=1pt,
  top=0pt,
  bottom=-2pt,
  halign=center,
  valign=center,
  before skip=4pt,
  after skip=4pt,
  fontupper=\footnotesize,
  before upper=\[,
  after upper=\],
}
\makesavenoteenv{accuracyBox}   % for footnote handling inside tcolorbox

\newcommand{\DefMacro}[2]{\expandafter\newcommand\csname rmk-#1\endcsname{#2}}
\newcommand{\UseMacro}[1]{\csname rmk-#1\endcsname}

\definecolor{SkyBlueAlpha}{HTML}{B0E5ED}
\definecolor{LavenderAlpha}{HTML}{FBC2E4}
\definecolor{annotatecolor}{rgb}{0.59,0,0.09}
\definecolor{predstate}{HTML}{8A4B4B}
\definecolor{predrule}{HTML}{8A4B4B}
\definecolor{predtrace}{HTML}{8A4B4B}
\newcommand{\TableFont}{\scriptsize}

\newcommand{\dataset}{{\small\textsc{PLSemanticsBench}}\xspace}
\newcommand{\datasetLinkAnon}{{\footnotesize\url{https://EngineeringSoftware.github.io/PLSemanticsBench}}}
\newcommand{\PaperTitle}{LLMs Lean on Priors, Not Programming Language Semantics}

\newcommand{\pep}{PredExe\xspace}

\newcommand{\ig}{IG\xspace}
\newcommand{\iga}{IGA\xspace}

\newcommand{\OPTask}{Final-State Prediction\xspace}
\newcommand{\op}{\textbf{\textcolor{predstate}{PredState}}\xspace}
\newcommand{\srpTask}{semantic-rule prediction\xspace}
\newcommand{\SRPTask}{Semantic-Rule Prediction\xspace}
\newcommand{\srp}{\textbf{\textcolor{predrule}{PredRule}}\xspace}
\newcommand{\etpTask}{execution-trace prediction\xspace}

\newcommand{\etp}{\textbf{\textcolor{predtrace}{PredTrace}}\xspace}

\newcommand{\COT}{CoT\xspace}

\newcommand{\ebnf}{EBNF\xspace}
\newcommand{\humanwrit}{Human-Written\xspace}
\newcommand{\llmtrans}{LLM-Translated\xspace}
\newcommand{\fuzzgen}{Fuzzer-Generated\xspace}
\newcommand{\lang}{$\text{C}^{\star}$\xspace}

\newcommand{\nlruleSmall}{{\small$\texttt{NL}\!\to\!\texttt{Rule}$}\xspace}
\newcommand{\rulenlSmall}{{\small$\texttt{Rule}\!\to\!\texttt{NL}$}\xspace}

\newcommand{\hypothesisOne}{\textsc{H}1\xspace}
\newcommand{\hypothesisTwo}{\textsc{H}2\xspace}
\newcommand{\hypothesisThree}{\textsc{H}3\xspace}
\newcommand{\hypothesisFour}{\textsc{H}4\xspace}
\newcommand{\Shift}{Shift\xspace}
\newcommand{\shift}{shift\xspace}
\newcommand{\Shifts}{Shifts\xspace}
\newcommand{\shifts}{shifts\xspace}
\newcommand{\NumModels}{11\xspace}

\newcommand{\NumDatasets}{three\xspace}

\newcommand{\NumExpAverage}{three\xspace}

\newcommand{\LlamaContextLen}{128K}
\newcommand{\QwenFourteenContextLen}{128K}
\newcommand{\QwenThirtyTwoContextLen}{128K}
\newcommand{\GPTFourOContextLen}{128K}
\newcommand{\GPTOThreeContextLen}{200K}
\newcommand{\GPTFiveContextLen}{400K}
\newcommand{\DeepSeekContextLen}{128K}
\newcommand{\QwQContextLen}{130K}
\newcommand{\GeminiContextLen}{1000K}
\newcommand{\LlamaMaxGen}{4K}
\newcommand{\QwenMaxGen}{8K}
\newcommand{\GPTFourOMaxGen}{16K}
\newcommand{\GPTOThreeMaxGen}{8K}
\newcommand{\GPTFiveMaxGen}{16K}
\newcommand{\DeepSeekMaxGen}{8K}
\newcommand{\QwQMaxGen}{8K}
\newcommand{\GeminiMaxGen}{8K}
\newcommand{\CoarseGrainSuccPercent}{90\%\xspace}

\newcommand{\NonStdSemanticDegPercent}{40--60\xspace}

\newcommand{\MaxETPGeminiPerf}{35\%\xspace}
\newcommand{\numNotationCompRuns}{three\xspace}
\newcommand{\numNotationCompSamples}{200\xspace}
\newcommand{\numNotationCompChoices}{five\xspace}
\newcommand{\numNotationCompSosStdMinPerc}{80\xspace}
\newcommand{\numNotationCompSosStdMaxPerc}{90\%\xspace}
\newcommand{\Code}[1]{{\ifmmode{\mathtt{#1}}\else$\mathtt{#1}$\fi}}
\newcommand{\CodeIn}[1]{\texttt{#1}}

\newcommand{\eg}{e.g.\xspace}
\newcommand{\ie}{i.e.\xspace}
\newcommand{\etc}{etc.\xspace}
\newcommand{\MyPara}[1]{\vspace{0.5pt}\textbf{#1}.\hspace{1mm}}
\newcommand{\MyParaTwo}[1]{\vspace{0pt}\textbf{#1}\hspace{1mm}}
\newcommand{\MyHypothesis}[2]{\vspace{-2pt}\hspace{1.5mm}{\large\bm{$\star$}}~\textbf{#1}~(#2):\hspace{1mm}}
\DeclareRobustCommand{\uncircledFootNote}[1]{\tikz[baseline=(char.base)]{\node[shape=circle, draw, minimum size=4pt, inner sep=1pt, text=black] (char) {\footnotesize #1};}}

\DeclareRobustCommand{\uncircledCaption}[1]{\tikz[baseline=(char.base)]{\node[shape=circle, draw, minimum size=4pt, inner sep=1pt, text=black] (char) {\tiny #1};}}

\DeclareRobustCommand{\circledCaption}[1]{\tikz[baseline=(char.base)]{\node[shape=circle, draw, minimum size=4pt, inner sep=1pt, fill=black, text=white] (char) {\tiny #1};}}
\newcommand{\redcircled}[1]{H\xspace}
\newcommand{\keyFindingBullet}{{\small\raisebox{0.1ex}{\bm{$\star$}}}\hspace{2pt}}%{{\scriptsize$\blacktriangleright$}\xspace}
\newcommand{\metricBullet}[1][0.25ex]{{\raisebox{#1}{\scriptsize{$\blacktriangleright$}}}}
\newcommand{\KTool}{{\small$\mathbb{K}$}-framework\xspace}
\newcommand{\Sos}{{\small$\mathbb{S}$}\xspace}
\newcommand{\SosSmall}{{\scriptsize$\mathbb{S}$}\xspace}

\newcommand{\Kos}{{\small$\mathbb{K}$}\xspace}
\newcommand{\KosSmall}{{\scriptsize$\mathbb{K}$}\xspace}

\newcommand{\SosKeyFinding}{{\keyFindingFont$\mathbb{S}$}\xspace}

\newcommand{\IMP}{\lang}
\newcommand{\imp}{\lang}

\newcommand{\LLM}{LLM\xspace}
\newcommand{\LLMs}{LLMs\xspace}

\newcommand{\Antlr}{ANTLR4\xspace}
\newcommand{\lLang}{\mathcal{L}}
\newcommand{\rAssign}{\textsc{S-Assign}}
\newcommand{\rAssignStep}{\textsc{S-AssignStep}}

\newcommand{\fConfig}[2]{\langle #1,#2\rangle}
\newcommand{\fState}[1]{\sigma_{#1}}
\newcommand{\fVar}[1]{\texttt{x}_{#1}}
\newcommand{\fLit}[1]{\texttt{v}_{#1}}
\newcommand{\fExp}[1]{\texttt{e}_{#1}}
\newcommand{\fStmt}[1]{\texttt{s}_{#1}}
\newcommand{\fTrans}[6]{\fConfig{#1}{#2}\!\xrightarrow{#6}_{#5}\!\fConfig{#3}{#4}}
\newcommand{\fTransE}[4]{\fTrans{#1}{#2}{#3}{#2}{E}{#4}}
\newcommand{\fTransS}[5]{\fTrans{#1}{#2}{#3}{#4}{S}{#5}}

\newcommand{\fStepS}[1]{\!\xrightarrow{#1}_S\!}

\newcommand{\fRule}{r}
\newcommand{\fRuleList}[1]{\text{R}_{#1}}
\newcommand{\fRules}{\mathscr{R}}
\newcommand{\fDomain}[1]{\text{\textbf{dom}}(#1)}
\newcommand{\wrapFootNoteSize}[1]{\begin{footnotesize}#1\end{footnotesize}}
\newcommand{\wrapScriptSize}[1]{\begin{scriptsize}#1\end{scriptsize}}
\newcommand{\keyFindingFont}{\small}

\newcommand{\fSize}[1]{\lvert#1\rvert}
\newcommand{\fConcat}{\oplus}
\newcommand{\fConcatBig}{\bigoplus}
\newcommand{\fTrace}[1]{\tau_{\scalebox{0.6}{$#1$}}}
\newcommand{\fGrammar}{\mathit{G}}

\newcommand{\fProject}[2]{\fProjectNone_{#1}(#2)}
\newcommand{\fProjectNone}{\pi}

\newcommand{\fProgram}{\mathcal{P}}

\newcommand{\fSemantics}{\Psi}
\newcommand{\rotSquare}{\raisebox{-0.25ex}{\makebox[0pt][c]{\rotatebox{45}{$\bm{\square}$}}}}
\newcommand{\fVariantPlaceHold}[1][0.6]{\scalebox{#1}{$\rotSquare$}\hspace{1pt}}
\newcommand{\fFormalPlaceHold}[1][0.7]{\scalebox{#1}{$\bm{\square}$}}
\newcommand{\fVarStd}[1][1]{\scalebox{#1}{\text{\textbf{std}}}}
\newcommand{\fVarSwap}[1][1]{\scalebox{#1}{\text{\textbf{swap}}}}
\newcommand{\fVarObf}[1][1]{\scalebox{#1}{\text{\textbf{obf}}}}
\NewDocumentCommand{\fSemanticsUnSpec}{O{0.6} O{0.5} O{1}}{%
  {\scalebox{#3}{$\Psi$}}^{%
    \fFormalPlaceHold[#1],\;%
    \fVariantPlaceHold[#2]%
  }%
}
\NewDocumentCommand{\fSemanticsSos}{O{0.5} O{1}}{%
  {\scalebox{#2}{$\Psi$}}^{%
    \text{\SosSmall},\;%
    \fVariantPlaceHold[#1]%
  }%
}
\newcommand{\fSplit}{\mathcal{D}}
\newcommand{\fSet}[1]{\{#1\}}
\newcommand{\fFormalSet}{\fSet{$\Sos,\Kos{}$}}
\newcommand{\fVariantSet}{\fSet{\fVarStd[0.9],\fVarSwap[0.9],\fVarObf[0.9]}}
\newcommand{\fSplitSet}{\fSet{\text{\humanwrit, \llmtrans, \fuzzgen}}}
\newcommand{\fSemanticsUnSpecSub}{\fSemanticsUnSpec[0.4][0.4][0.7]}
\newcommand{\fSeqOp}{\otimes}
\newcommand{\fExec}[2]{{\llbracket #1 \rrbracket}_{#2}}
\newcommand{\fExecAcc}[4]{{\llbracket #1 \rrbracket}^{#2}_{#3}(#4)}
\newcommand{\fExecAccTwo}[3]{{\llbracket #1 \rrbracket}^{#2}_{#3}}
\NewDocumentCommand{\metricProgramComplexity}{O{\square} O{0.9} O{0.6}}{%
  \ensuremath{%
    \scalebox{#2}{$\bm{\Omega}$}_{\scalebox{#3}{$\bm{#1}$}}%
  }%
}
\newcommand{\metricMaxnestif}{$\Omega_\text{If}$\xspace}
\newcommand{\metricMaxnestwhile}{$\Omega_\text{Loop}$\xspace}
\newcommand{\metricMaxtakenif}{$\hat\Omega_\text{If}$\xspace}
\newcommand{\metricMaxtakenwhile}{$\hat\Omega_\text{Loop}$\xspace}
\newcommand{\metricNumAssign}{$\hat\Omega_\text{Assign}$\xspace}
\newcommand{\metricLoc}{$\Omega_\text{Loc}$\xspace}
\newcommand{\metricTracelen}{$\hat\Omega_\text{Trace}$\xspace}
\newcommand{\metricCC}{$\Omega_\text{CC}$\xspace}
\newcommand{\metricVol}{$\Omega_\text{Vol}$\xspace}
\newcommand{\metricVocab}{$\Omega_\text{Voc}$\xspace}
\newcommand{\metricDepdeg}{$\Omega_\text{DD}$\xspace}
\newcommand{\orIqr}{$\Theta(\Delta)$\xspace}
\newcommand{\metricMaxnestifBold}{$\bm{\Omega_\textbf{\text{If}}}$\xspace}
\newcommand{\metricMaxnestwhileBold}{$\bm{\Omega_\textbf{\text{Loop}}}$\xspace}
\newcommand{\metricMaxtakenifBold}{$\bm{\hat\Omega_\textbf{\text{If}}}$\xspace}
\newcommand{\metricMaxtakenwhileBold}{$\bm{\hat\Omega_\textbf{\text{Loop}}}$\xspace}
\newcommand{\metricNumAssignBold}{$\bm{\hat\Omega_\textbf{\text{Assign}}}$\xspace}
\newcommand{\metricLocBold}{$\bm{\Omega_\textbf{\text{Loc}}}$\xspace}
\newcommand{\metricTracelenBold}{$\bm{\hat\Omega_\textbf{\text{Trace}}}$\xspace}
\newcommand{\metricCCBold}{$\bm{\Omega_\textbf{\text{CC}}}$\xspace}
\newcommand{\metricVolBold}{$\bm{\Omega_\textbf{\text{Vol}}}$\xspace}
\newcommand{\metricVocabBold}{$\bm{\Omega_\textbf{\text{Voc}}}$\xspace}
\newcommand{\metricDepdegBold}{$\bm{\Omega_\textbf{\text{DD}}}$\xspace}
\newcommand{\metricDeltaCond}{\Delta_\text{cnd}}
\newcommand{\metricDeltaRobust}{\Delta_\text{is}}
\newcommand{\metricDeltaCondFootNote}{{\footnotesize\Delta_\text{cnd}}}
\newcommand{\metricDeltaRobustFootNote}{{\footnotesize\Delta_\text{is}}}

\newcommand{\metricAccuracyNK}{\text{Acc}_\text{na}}
\newcommand{\metricAccuracyStd}{\text{Acc}_\text{\fVarStd}}
\newcommand{\metricAccuracySwap}{\text{Acc}_\text{\fVarSwap}}
\newcommand{\metricAccuracyObf}{\text{Acc}_\text{\fVarObf}}
\newcommand{\keywordMut}{KeywordSwap\xspace}

\newcommand{\keywordObf}{KeywordObf\xspace}

\newcommand{\standardSem}{standard\xspace}
\newcommand{\nstandardSem}{nonstandard\xspace}

\newcommand{\standardSemCap}{Standard\xspace}

\newcommand{\Program}{$p$}

\newcommand{\MutatedProgramKS}{$p'_{ks}$}
\newcommand{\MutatedProgramKO}{$p'_{ko}$}
\newcommand{\OperationalSemantics}{$\fSemantics$}

\newcommand{\SOSOperationalSemantics}[1][0.7]{$\fSemantics^{\scalebox{#1}{\Sos},\;\fVariantPlaceHold[0.4]}$}
\newcommand{\KOperationalSemantics}[1][0.7]{$\fSemantics^{\scalebox{#1}{\Kos},\;\fVariantPlaceHold[0.4]}$}
\newcommand{\StdSOSOperationalSemantics}[1][0.7]{$\fSemantics^{\scalebox{#1}{\Sos},\textbf{std}}$}
\newcommand{\StdOperationalSemantics}[1][0.7]{$\fSemantics^{\scalebox{#1}{$\square$},\textbf{std}}$}
\newcommand{\KeywordMutOperationalSemantics}[1][0.7]{$\fSemantics^{\scalebox{#1}{$\square$},\textbf{swap}}$}
\newcommand{\KeywordMutSOSOperationalSemantics}[1][0.7]{$\fSemantics^{\scalebox{#1}{\Sos},\textbf{swap}}$}
\newcommand{\KeywordObfOperationalSemantics}[1][0.7]{$\fSemantics^{\scalebox{#1}{$\square$},\textbf{obf}}$}
\newcommand{\MutatedOperationalSemanticsA}{${s'_{A}}$}
\newcommand{\MutatedOperationalSemanticsU}{${s'_{U}}$}
\newcommand{\uk}{\boldmath(\OperationalSemantics,\Program)}

\newcommand{\ksMk}{\boldmath(\KeywordMutOperationalSemantics, \MutatedProgramKS)\xspace}
\newcommand{\koMk}{\boldmath(\KeywordObfOperationalSemantics, \MutatedProgramKO)\xspace}
\newcommand{\nk}{\boldmath\Program}
\newcommand{\nkAcc}{\textbf{Acc$_{\mathrm{na}}$}}
\newcommand{\CAADD}{\raisebox{-0.5ex}{\includegraphics[height=1em, trim=30 10 10 40, clip]{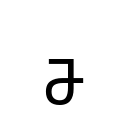}}}

\newcommand{\CASUB}{\raisebox{-0.5ex}{\includegraphics[height=1.1em, trim=30 10 10 40, clip]{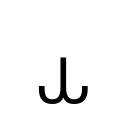}}}
\newcommand{\CAMUL}{\raisebox{-0.5ex}{\includegraphics[height=1.1em, trim=30 10 10 40, clip]{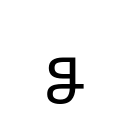}}}
\newcommand{\CADIV}{\raisebox{-0.5ex}{\includegraphics[height=1.1em, trim=30 10 10 40, clip]{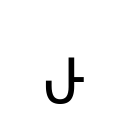}}}
\newcommand{\CAMOD}{\raisebox{-0.5ex}{\includegraphics[height=1.1em, trim=30 10 10 40, clip]{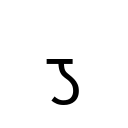}}}
\newcommand{\CAASSIGN}{\raisebox{-0.5ex}{\includegraphics[height=1.1em, trim=30 10 10 40, clip]{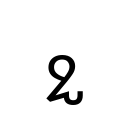}}}
\newcommand{\CALT}{\raisebox{-0.5ex}{\includegraphics[height=1.1em, trim=30 10 10 40, clip]{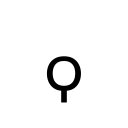}}}
\newcommand{\CAGT}{\raisebox{-0.5ex}{\includegraphics[height=1.1em, trim=30 10 10 40, clip]{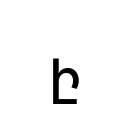}}}
\newcommand{\CALTEQ}{\raisebox{-0.5ex}{\includegraphics[height=1.1em, trim=30 10 10 40, clip]{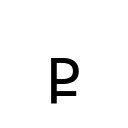}}}
\newcommand{\CAGTEQ}{\raisebox{-0.5ex}{\includegraphics[height=1.1em, trim=30 10 10 40, clip]{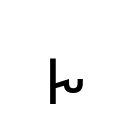}}}
\newcommand{\CAEQ}{\raisebox{-0.5ex}{\includegraphics[height=1.1em, trim=30 10 10 40, clip]{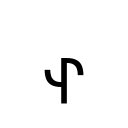}}}
\newcommand{\CANEQ}{\raisebox{-0.5ex}{\includegraphics[height=1.1em, trim=30 10 10 40, clip]{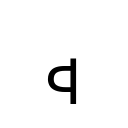}}}
\newcommand{\CAAND}{\raisebox{-0.5ex}{\includegraphics[height=1.1em, trim=30 10 10 40, clip]{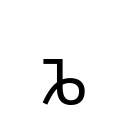}}}
\newcommand{\CAOR}{\raisebox{-0.5ex}{\includegraphics[height=1.1em, trim=30 10 10 40, clip]{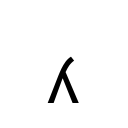}}}
\newcommand{\CANOT}{\raisebox{-0.5ex}{\includegraphics[height=1.1em, trim=30 10 10 40, clip]{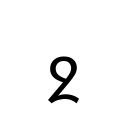}}}
\newcommand{\CABREAK}{\raisebox{-0.5ex}{\includegraphics[height=1.1em, trim=30 10 10 40, clip]{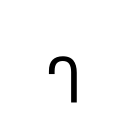}}}
\newcommand{\CAIF}{\raisebox{-0.5ex}{\includegraphics[height=1.1em, trim=30 10 10 40, clip]{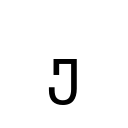}}}
\newcommand{\CAELSE}{\raisebox{-0.5ex}{\includegraphics[height=1.1em, trim=30 10 10 40, clip]{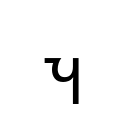}}}
\newcommand{\CAWHILE}{\raisebox{-0.5ex}{\includegraphics[height=1.1em, trim=30 10 10 40, clip]{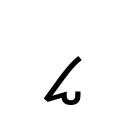}}}
\newcommand{\CAHALT}{\raisebox{-0.5ex}{\includegraphics[height=1.1em, trim=30 10 10 40, clip]{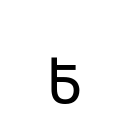}}}
\newcommand{\CACONTINUE}{\raisebox{-0.5ex}{\includegraphics[height=1.1em, trim=30 10 10 40, clip]{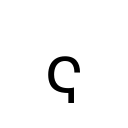}}}
\newcommand{\delt}[1]{%
  \pgfmathtruncatemacro{\DeltaInt}{round(#1)}%
  \pgfmathtruncatemacro{\DeltaAbsInt}{abs(\DeltaInt)}%
  \ifnum\DeltaAbsInt<10
    \edef\DeltaFmt{0\DeltaAbsInt}%
  \else
    \edef\DeltaFmt{\DeltaAbsInt}%
  \fi
  \ifnum\DeltaInt>0
    \textcolor{green!45!black}{\scriptsize(+\DeltaFmt)}%
  \else\ifnum\DeltaInt<0
    \textcolor{red!60!black}{\scriptsize(\texttt{-}\DeltaFmt)}%
  \else
    \textcolor{black!55}{\scriptsize(000)}%
  \fi\fi
}

\newcommand{\accdfrom}[2]{%
  \pgfmathtruncatemacro{\DeltaTmp}{round((#1) - (#2))}%
  #1\,\delt{\DeltaTmp}%
}
\newcommand{\accdrobust}[2]{%
  \pgfmathtruncatemacro{\DeltaTmp}{round((#1) - (#2))}%
  #1\,\delt{\DeltaTmp}%
}
\newcommand{\llamaBig}{\textsc{Llama}\hspace{0pt}-3.3 70B\xspace}
\newcommand{\llamaBigSmall}{{\footnotesize\textsc{Llama}\hspace{0pt}-3.3 70B}\xspace}
\newcommand{\gemini}{\textsc{Gemini}\hspace{0pt}-2.5-pro\xspace}
\newcommand{\geminiSmall}{{\footnotesize\textsc{Gemini}\hspace{0pt}-2.5-pro}\xspace}

\newcommand{\qwenCoder}[1]{\textsc{Qwen}\hspace{0pt}2.5-\textsc{Inst}\hspace{0pt} #1B\xspace}
\newcommand{\qwenCoderSmall}[1]{{\footnotesize\textsc{Qwen}\hspace{0pt}2.5-\textsc{Inst}\hspace{0pt} #1B}\xspace}
\newcommand{\dpskQwenSmall}[1]{{\footnotesize\textsc{DS}\hspace{0pt}-\textsc{Qwen}\hspace{0pt} #1B}\xspace}
\newcommand{\dpskQwen}[1]{\textsc{DS-Qwen}\hspace{0pt} #1B\xspace}

\newcommand{\qwq}{\textsc{QwQ}\hspace{0pt} 32B\xspace}
\newcommand{\qwqSmall}{{\footnotesize\textsc{QwQ}\hspace{0pt} 32B}\xspace}

\newcommand{\dpskLlama}[1]{\textsc{DS}\hspace{0pt}-\textsc{Llama}\hspace{0pt} #1B\xspace}
\newcommand{\dpskLlamaSmall}[1][70]{{\footnotesize\textsc{DS}\hspace{0pt}-\textsc{Llama}\hspace{0pt} #1B}\xspace}
\newcommand{\gptfo}{\textsc{GPT}\hspace{0pt}-4o-mini\xspace}
\newcommand{\gptfom}{\textsc{GPT}\hspace{0pt}-4o-mini\xspace}
\newcommand{\gptfomSmall}{{\footnotesize\textsc{GPT}\hspace{0pt}-4o-mini}\xspace}

\newcommand{\gptfivemini}{\textsc{GPT}\hspace{0pt}-5-mini\xspace}
\newcommand{\gptfiveminiSmall}{{\footnotesize\textsc{GPT}\hspace{0pt}-5-mini}\xspace}
\newcommand{\othree}{\textsc{o3}\hspace{0pt}-mini\xspace}
\newcommand{\othreeSmall}{{\footnotesize\textsc{o3}\hspace{0pt}-mini}\xspace}
\input{tables/numbers-imp-dataset-stats.tex}
\input{tables/numbers-op-IMP-SOS.tex}
\input{tables/numbers-mk-IMP-SOS-stats.tex}
\input{tables/numbers-srp-IMP-SOS.tex}
\input{tables/numbers-etp-IMP-SOS.tex}
\input{tables/numbers-benchmark-stats.tex}
\input{tables/numbers-op-IMP-K}
\input{tables/numbers-etp-IMP-K}
\input{tables/numbers-srp-IMP-K}
\input{tables/numbers-op-IMP-SOS-synthetic-cpp}
\input{tables/numbers-op-IMP-K-synthetic-cpp}
\input{tables/numbers-op-IMP-K-fuzzer-generated}
\input{tables/numbers-op-IMP-SOS-fuzzer-generated}
\input{tables/macros-table-imp-dataset.tex}
\input{tables/macros-table-op-IMP-SOS.tex}
\input{tables/macros-table-mk-IMP-SOS.tex}
\input{tables/macros-table-srp-IMP-SOS.tex}
\input{tables/macros-table-etp-IMP-SOS.tex}
\input{tables/macros-table-mutation-rules}
\input{tables/macros-table-mutation-caucasian-rules}
\input{tables/macros-table-srp-etp-IMP-SOS.tex}
\input{tables/macros-table-benchmark.tex}
\input{tables/macros-table-etp-uk-IMP-SOS-percentage-trace-match}
\input{tables/macros-table-etp-mk-IMP-SOS-unseen-percentage-trace-match}
\input{tables/macros-table-etp-mk-IMP-SOS-addSub_mulDiv_negateRelation-percentage-trace-match}
\input{tables/macros-table-etp-uk-IMP-K-percentage-trace-match}
\input{tables/macros-table-etp-mk-IMP-K-unseen-percentage-trace-match}
\input{tables/macros-table-etp-mk-IMP-K-addSub_mulDiv_negateRelation-percentage-trace-match}
\input{tables/macros-table-srp-IMP-K-IMP-SOS-xmatch-accuracy}
\input{tables/macros-table-etp-IMP-K-IMP-SOS-xmatch-accuracy}
\input{tables/macros-table-etp-IMP-K-IMP-SOS-approx-match-accuracy}
\input{tables/macros-table-etp-IMP-K-IMP-SOS-final-state-match}
\input{tables/macros-table-op-odds-per-iqr}
\input{tables/macros-table-op-nk-odds-per-iqr}
\input{tables/macros-table-datasets-stats}
\input{tables/macros-table-op-IMP-K-IMP-SOS-human-written-synthetic-cpp-fuzzer-generated-acc}
\input{tables/macros-table-op-IMP-K-IMP-SOS-human-written-synthetic-cpp-fuzzer-generated-var-acc}
\input{tables/macros-table-odds-per-iqr-nk-human-written-synthetic-cpp-fuzzer-generated}
\input{tables/macros-table-odds-per-iqr-uk-human-written-synthetic-cpp-fuzzer-generated}
\newcommand{\TableNotationPrimerCaption}{Notation primer.\label{tab:notation-primer}}
\newcommand{\TablePredRuleProcessCaption}{
Processing of statements sampled from \IMP programs for the \srp task.
The pair <Statement, State> is transformed into the pair <PredRule
Program, PredRule State>. The transformed pair is used in constructing
the prompt for the \srp task.
\label{tab:appendix-srp-data}
}
\newcommand{\TableEvalModelsCaption}{
Evaluated \LLMs grouped by non-reasoning and reasoning
variants. Max Gen. is the maximum output generation tokens.
\label{tab:evaluated-models}
}
\newcommand{\TableFuzzerSettingsCaption}{
Settings for the fuzzer knobs used to generate \IMP programs for
the \fuzzgen dataset.
\label{tab:appendix-fuzzer-settings}
}
\newcommand{\TableNonStandardRulesCaption}{
Mutations and obfuscations applied to the \standardSem
semantics to derive the \nstandardSem semantics \keywordMut
and \keywordObf.
\label{tab:mutation-rules-combined}
}
\DefMacro{TCap-models-op-IMP-K-IMP-SOS-human-written-synthetic-cpp-fuzzer-generated-acc}{
Accuracies of the models on the \op task, using \Sos and
\Kos-semantics, for both the \standardSem and
\nstandardSem variants across the \humanwrit, the \llmtrans,
and the \fuzzgen datasets. The best performing models in every column
within a dataset are shown in boldface font. The cases where models
under \standardSem semantics perform better/worse than with
no-semantics are shaded green/red.
\label{tab:op-k-sos-all-datasets-acc}
}
\DefMacro{TCap-models-op-IMP-K-IMP-SOS-human-written-synthetic-cpp-fuzzer-generated-var-acc}{
Average percentage of variables predicted correctly per program on the
\op task.
\label{tab:op-k-sos-all-datasets-var-acc}
}
\DefMacro{TCap-models-etp-uk-IMP-SOS-percentage-trace-match}{Percentage of execution-trace match for \etp for \uk, \ksMk, and \koMk with \Sos.}
\DefMacro{TCap-models-etp-mk-IMP-SOS-addSub_mulDiv_negateRelation-percentage-trace-match}{Percentage of trace match for \etp for \ksMk with \Sos.}
\DefMacro{TCap-models-etp-mk-IMP-SOS-unseen-percentage-trace-match}{Percentage of trace match for \etp for \koMk with \Sos.}
\DefMacro{TCap-models-etp-uk-IMP-K-percentage-trace-match}{Percentage of execution-trace match for \etp for \uk, \ksMk, and \koMk with \Kos-semantics.}
\DefMacro{TCap-models-etp-mk-IMP-K-addSub_mulDiv_negateRelation-percentage-trace-match}{Percentage of trace match for \etp for \ksMk with K-semantics.}
\DefMacro{TCap-models-etp-mk-IMP-K-unseen-percentage-trace-match}{Percentage of trace match for \etp for \koMk with K-semantics.}
\DefMacro{TCap-benchmark-dataset-stats}{Median code-complexity statistics summarizing the \humanwrit, \llmtrans, and \fuzzgen datasets.}
\DefMacro{TCap-models-srp-etp-IMP-SOS}{Models' exact-match accuracies on \srp and \etp tasks. \label{tab:srp-etp}}
\DefMacro{TCap-models-pcp-op-IMP-SOS}{Models' accuracies on \pep and \op tasks. \label{tab:pep-pop}}
\DefMacro{TCap-models-pcp-op-IMP-K}{Models' accuracies on \pep and \op tasks with K-semantics. \label{tab:pep-pop-k}}
\DefMacro{TCap-models-pcp-IMP-K-IMP-SOS-accuracy}{Models' accuracies on the \pep task under \Kos-semantics and \Sos for \standardSem, \keywordMut and \keywordObf cases. \label{tab:pep-k-sos}}
\DefMacro{TCap-models-op-IMP-K-IMP-SOS-acc}{
Accuracies of the models on the \op task (\op),
using the \KTool and \Sos semantics, for both the \standardSem and
\nstandardSem variants.
\label{tab:op-k-sos}
}
\DefMacro{TCap-models-op-IMP-K-IMP-SOS-var-acc}{The average percentage of declared variables in \IMP programs whose final states are predicted correctly by models evaluated with K-semantics and \Sos.}
\DefMacro{TCap-models-op-IMP-K-IMP-SOS-acc-synthetic-cpp}{Model's accuracies on \op tasks with K-semantics and \Sos on the LLM Translated dataset. \label{tab:pep-k-sos-llm-translated}}
\DefMacro{TCap-models-op-IMP-K-IMP-SOS-var-acc-synthetic-cpp}{The average percentage of declared variables in \IMP programs whose final states are predicted correctly by models evaluated with K-semantics and \Sos on the LLM Translated dataset.}
\DefMacro{TCap-models-op-IMP-K-IMP-SOS-acc-fuzzer-generated}{Model's accuracies on \op tasks with K-semantics and \Sos on the Fuzzer Generated dataset. \label{tab:pep-k-sos-fuzzer-generated}}
\DefMacro{TCap-models-op-IMP-K-IMP-SOS-var-acc-fuzzer-generated}{The average percentage of declared variables in \IMP programs whose final states are predicted correctly by models evaluated with K-semantics and \Sos on the Fuzzer Generated dataset.}
\DefMacro{TCap-models-srp-IMP-K-IMP-SOS-xmatch-accuracy}{
The exact-match accuracies of the models on the \srpTask task under
\Sos and \Kos-semantics on the \humanwrit dataset. The cases where
the models under one of the \nstandardSem semantics perform
better/worse relative to their counterpart are shaded green/red.
\label{tab:srp-k-sos}
}
\DefMacro{TCap-models-etp-IMP-K-IMP-SOS-xmatch-accuracy}{
The exact-match accuracies of the models on the \etpTask task under
\Sos and \Kos-semantics on the \humanwrit dataset.
\label{tab:etp-k-sos-xmatch}}
\DefMacro{TCap-models-etp-IMP-K-IMP-SOS-approx-match-accuracy}{Models' approximate-match accuracies on the \etp task with \Sos and \Kos-semantics. \label{tab:etp-k-sos-approx-match}}
\DefMacro{TCap-models-etp-IMP-K-IMP-SOS-final-state-match}{Models' final-state only match accuracies on the \etp task under the \standardSem and \nstandardSem semantics with \Sos and \Kos-semantics formalization. \label{tab:etp-k-sos-final-state-match}}
\DefMacro{TCap-models-op-odds-per-iqr}{Models' performance odds per IQR of code-complexity metrics.}
\DefMacro{TCap-models-op-nk-odds-per-iqr}{Models' performance odds per IQR of code-complexity metrics for \op nk.}
\DefMacro{TCap-models-datasets-stats}{
Median code-complexity statistics of our  dataset splits. Control-flow complexity is characterized using extended
cyclomatic complexity (\wrapFootNoteSize{\metricCC{}}), maximum nested
if–else (\wrapFootNoteSize{\metricMaxnestif{}}) and nested loop
(\wrapFootNoteSize{\metricMaxnestwhile{}}) depths, maximum taken nested
if–else (\wrapFootNoteSize{\metricMaxtakenif{}}), and taken nested loop
(\wrapFootNoteSize{\metricMaxtakenwhile{}}) depths. Program size complexity is
measured using lines of code (\wrapFootNoteSize{\metricLoc{}}), Halstead
metrics Volume (\wrapFootNoteSize{\metricVol{}}) and
Vocabulary (\wrapFootNoteSize{\metricVocab{}}), and execution trace
length (\wrapFootNoteSize{\metricTracelen{}}). Data-flow complexity
is analyzed using DepDegree (\wrapFootNoteSize{\metricDepdeg{}}) and the
total number of assignments to variables in execution traces
(\wrapFootNoteSize{\metricNumAssign{}}).  Metrics computed under
dynamic-analysis are shown with a hat.
\label{tab:dataset-median-stats}}
\newcommand{\TableMutationRuleCaption}{The 3 unit mutation patterns in \keywordMut.\label{tab:mutation-rule}}
\newcommand{\TableMutationCaucasianRuleCaption}{
Some of the obfuscations of operators and keywords in \standardSem
semantics used in \keywordObf
semantics.\label{tab:mutation-caucasian-rule}}
\newcommand{\TableMutationCaucasianRuleFullCaption}{Complete list of obfuscations of operators and keywords in \standardSem semantics to \keywordObf semantics.\label{tab:mutation-caucasian-rule-full}}
\newcommand{\TableSRPCategoryMappingCaption}{Rule categorization for \srp analysis.\label{tab:appendix-srp-mapping}}
\newcommand{\TableIMPSOSRulesCaption}{Summary of small-step SOS rules used to formalize \IMP.\label{tab:appendix-imp-sos-rules}}
\DefMacro{TCap-models-ig-uk-IMP-SOS-EBNF}{
Results on \ig task with unmutated EBNF syntax.  Comp indicates
whether the program can be compiled (not applicable to Python).  $P$
denotes the number of \imp programs that are correctly handled by the
generated interpreter.  $F_{syn}$ counts the number of cases where the
interpreter incorrectly reports a syntax error when it should not.
$F_{sem_w}$ refers to cases where the interpreter executes the program
without error but produces incorrect results.  $F_{sem_r}$ (for valid
programs) counts cases where the interpreter reports a semantic error
when it should not.  $F_{sem}$ (for invalid programs) counts cases
where the interpreter fails to report a semantic error as it should.
$F_{o}$ includes any other types of failures.  $TO$ indicates a
timeout, which is set to 5 seconds in the experiments.
\label{tab:ig-ebnf}}

\DefMacro{TCap-models-ig-uk-IMP-SOS-ANTLR}{
Results on \ig task with unmutated ANTLR syntax.  Comp indicates
whether the program can be compiled (not applicable to Python).  $P$
denotes the number of \imp programs that are correctly handled by the
generated interpreter.  $F_{syn}$ counts the number of cases where the
interpreter incorrectly reports a syntax error when it should not.
$F_{sem_w}$ refers to cases where the interpreter executes the program
without error but produces incorrect results.  $F_{sem_r}$ (for valid
programs) counts cases where the interpreter reports a semantic error
when it should not.  $F_{sem}$ (for invalid programs) counts cases
where the interpreter fails to report a semantic error as it should.
$F_{o}$ includes any other types of failures.  $TO$ indicates a
timeout, which is set to 5 seconds in the experiments.
\label{tab:ig-antlr}}

\DefMacro{TCap-models-iga-uk-IMP-SOS}{
Results on \iga task with unmutated ANTLR syntax.  Comp indicates
whether the program can be compiled (not applicable to Python).  $P$
denotes the number of \imp programs that are correctly handled by the
generated interpreter.  $F_{syn}$ counts the number of cases where the
interpreter incorrectly reports a syntax error when it should not.
$F_{sem_w}$ refers to cases where the interpreter executes the program
without error but produces incorrect results.  $F_{sem_r}$ (for valid
programs) counts cases where the interpreter reports a semantic error
when it should not.  $F_{sem}$ (for invalid programs) counts cases
where the interpreter fails to report a semantic error as it should.
$F_{o}$ includes any other types of failures.  $TO$ indicates a
timeout, which is set to 5 seconds in the experiments.
\label{tab:iga-antlr}}

\DefMacro{TCap-models-iga-mk-ks-IMP-SOS}{
    IGA task with mutated \keywordMut semantics.
\label{tab:iga-mk-ks-IMP-SOS}}

\DefMacro{TCap-models-iga-mk-ko-IMP-SOS}{
    IGA task with mutated \keywordObf semantics.
\label{tab:iga-mk-ko-IMP-SOS}}

\DefMacro{TCap-models-igaf-uk-IMP-SOS}{
    IGA uk with feedback.
\label{tab:igaf-uk-IMP-SOS}}

\DefMacro{TCap-models-igaf-mk-ks-IMP-SOS}{
    IGA ks with feedback.
\label{tab:igaf-mk-ks-IMP-SOS}}

\DefMacro{TCap-models-igaf-mk-ko-IMP-SOS}{
    IGA ko with feedback.
\label{tab:igaf-mk-ko-IMP-SOS}}

\DefMacro{TCap-models-op-odds-per-iqr-uk-human-written}{
Odds ratio per interquartile range (\orIqr) for each code-complexity
metric for the \op task \textbf{with the
standard \IMP semantics (\Kos-semantics and \Sos)}.
\orIqr for a metric is the odds ratio for a correct
final-state prediction when that metric increases from its
$\text{25}^\text{th}$ to its $\text{75}^\text{th}$ percentile, holding
other metrics fixed. Reported only for models with <90\% accuracy on
the \op task
(Table~\ref{tab:analysis-op-srp-combined}, left panel) to mitigate class
imbalance. The largest absolute values in each row are shown in
boldface font.
\label{tab:or-per-iqr-uk}
}

\DefMacro{TCap-models-op-odds-per-iqr-nk-human-written}{
Odds ratio per interquartile range (\orIqr) for each code-complexity
metric for the \op task \textbf{without semantics}.
\orIqr for a metric is the odds ratio for a correct
final-state prediction when that metric increases from its
$\text{25}^\text{th}$ to its $\text{75}^\text{th}$ percentile, holding
other metrics fixed. Reported only for models with <90\% accuracy on
the \op task
(Table~\ref{tab:analysis-op-srp-combined}, left panel) to mitigate class
imbalance. The largest absolute values in each row are shown in
boldface font.
\label{tab:or-per-iqr-nk}
}
\newcommand{\TableNotationComprehensionCaption}{
Hierarchical organization of \Sos semantics rules into families
and constructs, with associated categories and semantic roles, used to
sample near-miss distractors for the formal-semantics notation
comprehension tasks.
\label{tab:rule-families-full}
}
\newcommand{\TableNotationComprehensionKCaption}{
Hierarchical organization of \Kos semantics rules into families
and constructs, with associated categories and semantic roles, used to
sample near-miss distractors for the formal-semantics notation
comprehension tasks.
\label{tab:rule-families-full-k}
}
\newcommand{\TablePredStatePredRuleCombinedCaption}{
\textbf{Global and state mutation-free rule-conditioned reasoning.}
\emph{Left:} predict final program state accuracy via rule composition.
\emph{Right:} predict semantic-rules for program execution in absence of state mutation.
Best per column is bold.
\label{tab:analysis-op-srp-combined}
}
\newcommand{\TablePredStateComplexCaption}{
\op results on the \llmtrans and \fuzzgen datasets. Best per column within each dataset is bold.
\label{tab:analysis-op-complex}
}

\newcommand{\TablePredTraceNonZeroCaption}{
\etp results on the \humanwrit (least complex) dataset split.
Best per column is bold.
\label{tab:analysis-etp-non-zero}
}

\newcommand{\FigBackgroundGrammarCaption}{Syntax in Backus-Naur Form~\cite{mcckracken_bnf}.}
\newcommand{\FigBackgroundSOSSemanticsCaption}{Semantics in Small-step operational semantics~\citep{plotkin_small_step_sos}.}

\newcommand{\FigBackgroundSOSCaption}{
Formal syntax and semantics of an example language $\lLang$.
}

\newcommand{\FigSRPUKFirstMisMatchIMPKRadarCaption}{\standardSemCap semantics. \label{figure:srp-mismatch-k-uk}}
\newcommand{\FigSRPMKSwapFirstMisMatchIMPKRadarCaption}{\keywordMut semantics. \label{figure:srp-mismatch-k-swap}}
\newcommand{\FigSRPMKUnseenFirstMisMatchIMPKRadarCaption}{\keywordObf semantics. \label{figure:srp-mismatch-k-unseen}}
\newcommand{\FigSRPUKFirstMisMatchIMPSOSRadarCaption}{\standardSemCap semantics. \label{figure:srp-mismatch-sos-uk}}
\newcommand{\FigSRPMKSwapFirstMisMatchIMPSOSRadarCaption}{\keywordMut semantics. \label{figure:srp-mismatch-sos-swap}}

\newcommand{\FigSRPFirstMisMatchRadarCaption}{
First-point-of-mismatch rate by category for the \srp task with
the \Kos-semantics (top) and \Sos (bottom) on the \humanwrit
dataset.
\label{figure:srp-first-mismatch-radar}}

\newcommand{\FigOverviewCaption}{
Overview of the \dataset construction workflow and evaluation tasks
designed to probe rule-conditioned reasoning under formal semantics.
Each program is written in \lang with EBNF syntax and paired with two
semantic systems—small-step operational semantics (\Sos) and K
semantics (\Kos) (\circledCaption{1}). The standard semantics are systematically
transformed into two nonstandard variants, \keywordMut and
\keywordObf, which preserve rule structure while perturbing the
symbol--meaning mapping (\circledCaption{2}). These semantics are used to
derive ground-truth execution traces for each transformed program
(\circledCaption{3}--\circledCaption{4}). The transformed program and its
semantics (\circledCaption{5}) are then provided to models as prompts.
Tasks (\circledCaption{6}--\circledCaption{8}) range from predicting final
states (\op; \textbf{\hypothesisOne}), to selecting applicable rules when state does not
mutate (\srp; \textbf{\hypothesisTwo}), to generating full execution traces to test
long-horizon rule conditioning (\etp; \textbf{\hypothesisThree}). An analogous pipeline is
used for \Sos by replacing the execution engine, enabling controlled
comparisons across semantic formalisms and shifts (\textbf{\hypothesisFour}).
}
\newcommand{\FigIMPEBNFCaption}{Complete syntax of \IMP used in our experiments in EBNF.\label{figure:appendix-imp-syntax}}
\DefMacro{DataExample}{An example of re-writing a \CodeIn{C++} program into an \IMP program in the \humanwrit dataset.}
\DefMacro{cProgramExample}{The \texttt{C++} solution to the problem ``MBCPP/962'' in BabelCode MBPP and one public test case. The public test we use is \CodeIn{sumEven(3, 8)==18}.}
\DefMacro{impProgramExample}{The \IMP program re-written from the C++ solution.}

\newcommand{\FigDatasetMetrics}{
Distributions of the code-complexity metrics extended cyclomatic
complexity (\metricCC), maximum nested if–else (\metricMaxnestif) and
nested loop (\metricMaxnestwhile) depths , maximum taken nested
if–else (\metricMaxtakenif), and taken nested loop
(\metricMaxtakenwhile) depths, the program data-flow complexity
metrics DepDegree (\metricDepdeg) and the total number of assignments
to variables in execution traces (\metricNumAssign), and finally the
program size complexity metrics, lines of code (\metricLoc), Halstead
metrics Volume (\metricVol) and Vocabulary(\metricVocab), and
execution trace length (\metricTracelen).
\label{figure:appendix-dataset-stats}}
\newcommand{\FigPOPDendrogram}{
Dendrogram of models for the \op task on the \humanwrit dataset
under no-semantics and \standardSem semantics (\Kos-semantics
and \Sos). We show the top two most distinguishable metrics per
cluster, identified using the Cohen's d one-vs-rest test. The
silhouette score is 0.58 thus indicating a good clustering structure.
\label{figure:appendix-op-dendrogram}}

\newcommand{\FigPOPExtendedAnalysisCaption}{
Analyzing the impact of different code-complexity metrics on \LLM
performance in the \op task.
\label{figure:appendix-pop-ext-analysis}}
\newcommand{\FigPOPExtendedAnalysisDescCaption}{
Workflow of the \op task. \IMP programs, along with
optional semantics (K-semantics or \Sos) and syntax, are: (1)~executed in
the \KTool to obtain the gold final states of all declared variables,
and (2)~used to construct a prompt for the \LLMs to predict those
final states. The gold and predicted states are then compared, scored
as \CodeIn{1} for a match and \CodeIn{0} otherwise, and accumulated
into a result vector.
\label{figure:appendix-pop-ext-analysis-workflow}}
\newcommand{\FigPOPExtendedAnalysisModelingCaption}{
Modeling \LLM performance on \IMP programs. We treat each \LLM as a
black box and apply \textbf{Elastic Net regression} using
code-complexity metrics as predictors. \textbf{Partial Least Squares}
(PLS) is employed for dimensionality reduction and to address
multicollinearity. The magnitude and sign of the regression weights
provide insight into the potential impact of each metric on
the classifier's performance and hence to an extent the \LLM's
performance.
\label{figure:appendix-pop-ext-analysis-modeling}}

\newcommand{\FigAppendixIMPFuzzGenExampleCaption}{
An example \IMP program (\CodeIn{fuzz\_100.imp}) from the
\fuzzgen dataset. Its code-complexity
metric profile is: control-flow complexity (\metricCC= 62,
\metricMaxnestif= 5, \metricMaxnestwhile= 6, \metricMaxtakenif= 3,
\metricMaxtakenwhile= 5), data-flow complexity (\metricDepdeg= 2603,
 \metricNumAssign= 86), and program-size complexity (\metricLoc=
492, \metricVol= 37140, \metricVocab= 91, \metricTracelen= 249).
The \gemini model successfully predicted the final program-state of
this program in the \op task.
\label{fig:appendix:imp-fuzzgen-example}}

\newcommand{\FigNotationCompImpSos}{Formal notation comprehension of \Sos.
\label{figure:notation-comp-sos}}
\newcommand{\FigNotationCompImpK}{Formal notation comprehension of \Kos.
\label{figure:notation-comp-k}}
\newcommand{\FigNotationCompCombined}{%
Formal-semantics notation comprehension results. Models solve two multiple-choice
(\numNotationCompChoices choices) tasks: mapping natural-language descriptions to
semantic rules (\nlruleSmall) and vice-versa (\rulenlSmall).
Panels (a,b) show results on \nlruleSmall/\rulenlSmall tasks under $\fSemanticsUnSpec$\xspace
($\;\fVariantPlaceHold\in\!\fVariantSet$) for $\fFormalPlaceHold\!=\!$ \SosSmall and \KosSmall
respectively,
averaged over \numNotationCompRuns runs of \numNotationCompSamples samples. Panels (c,d)
and (e,f) plot \nlruleSmall confusion matrices under \StdSOSOperationalSemantics and
\KeywordMutSOSOperationalSemantics respectively for the three most frequently
mispredicted rules for {\scriptsize\qwenCoder{14}} and {\scriptsize\gptfom}. For example in (c),
{\scriptsize\qwenCoder{14}} under \StdSOSOperationalSemantics selects rule~\texttt{22} for rule~\texttt{21}'s description in 100\% of
cases.
\label{figure:notation-comp-combined}}

\newcommand{\FigNotationCompImpSosConfMatQwenStd}{{\qwenCoder{14} std}.
\label{figure:notation-comp-conf-mat-qwen}}
\newcommand{\FigNotationCompImpSosConfMatGptStd}{{\gptfom std}.
\label{figure:notation-comp-conf-mat-gpt}}
\newcommand{\FigNotationCompImpSosConfMatQwenSwap}{{\qwenCoder{14} swap}.
\label{figure:notation-comp-conf-mat-qwen}}
\newcommand{\FigNotationCompImpSosConfMatGptSwap}{{\gptfom swap}.
\label{figure:notation-comp-conf-mat-gpt}}

\usepackage[accepted]{icml2026}

\usepackage{amsmath,amsfonts,bm}

\def\eqref#1{equation~\ref{#1}}
\def\1{\bm{1}}

\DeclareMathAlphabet{\mathsfit}{\encodingdefault}{\sfdefault}{m}{sl}
\SetMathAlphabet{\mathsfit}{bold}{\encodingdefault}{\sfdefault}{bx}{n}

\icmltitlerunning{\PaperTitle}

\begin{document}

\twocolumn[
  \icmltitle{\PaperTitle}

  \begin{icmlauthorlist}
    \icmlauthor{Aditya Thimmaiah}{ut}
    \icmlauthor{Jiyang Zhang}{ut}
    \icmlauthor{Jayanth Srinivasa}{cisco}
    \icmlauthor{Junyi Jessy Li}{ut}
    \icmlauthor{Milos Gligoric}{ut}
  \end{icmlauthorlist}
  
  \icmlaffiliation{ut}{The University of Texas at Austin}
  \icmlaffiliation{cisco}{Cisco Research}
  \icmlcorrespondingauthor{Aditya Thimmaiah}{\texttt{auditt@utexas.edu}}
  \icmlkeywords{Machine Learning, ICML}

  \vskip 0.3in
]
% Footnote indentation
\makeatletter
\renewcommand\@makefntext[1]{%
  \noindent
  \hb@xt@1em{\hss\@makefnmark}#1}
\makeatother

% 2mm gap between table and captions
\captionsetup[table]{skip=0.05cm}
% 2mm gap between figure and captions
\captionsetup[figure]{skip=0.2cm}
% 2mm gap between figure and captions
\captionsetup[subfigure]{skip=0.2cm}
\setlength{\columnsep}{2mm}%
\setlength{\textfloatsep}{8pt plus 1pt minus 1pt}
\setlength{\intextsep}{6pt plus 1pt minus 1pt}
\setlength{\parskip}{4pt}

% math mode paddings
\setlength{\abovedisplayskip}{-2pt}
\setlength{\belowdisplayskip}{-6pt}
\setlength{\abovedisplayshortskip}{0pt}
\setlength{\belowdisplayshortskip}{-6pt}

\printAffiliationsAndNotice{} % leave empty if you don’t want a notice

\begin{abstract}
  Recent work asks whether large language models (\LLMs) condition their
  reasoning on explicit rules rather than statistical regularities from
  pretraining. Program execution provides a canonical instance: formal
  semantics define behavior through symbolic transition rules that can be
  systematically altered under distribution shift.
  We investigate whether \LLMs can condition their reasoning on formal
  semantics through program execution and introduce \dataset, pairing
  featherweight C programs with two semantic systems—small-step
  operational semantics and K semantics—and probing four
  capabilities: composing rules for final states, selecting rules when
  state is unmutated, sustaining such conditioning over long traces, and
  following supplied rules under novel semantics. To decouple semantic
  reasoning from syntactic familiarity, we redefine familiar operators to
  induce symbol-meaning conflict and introduce novel symbols defined only
  through the supplied rules, and stress-test models on \humanwrit{},
  \llmtrans{}, and \fuzzgen{} splits with increasing structural complexity.
  Across \NumModels frontier \LLMs, strong final-state accuracy under
  standard semantics (up to \CoarseGrainSuccPercent) drops sharply—by as
  much as \NonStdSemanticDegPercent{}\%\ points—under semantic mutations
  and increasing structural complexity. Only a handful of models achieve
  non-zero long-horizon conditioning accuracy, and even the best systems
  reach just \MaxETPGeminiPerf. Together, these results suggest that
  contemporary \LLMs often rely on pretrained lexical associations rather
  than systematically conditioning on supplied formal rules. \dataset 
  is publicly available at \datasetLinkAnon.
  \end{abstract}

%\newpage
\section{Introduction}

Modern large language models (\LLMs) increasingly solve programming
tasks that appear to require reasoning about program behavior, from
predicting outputs~\cite{Cruxeval,NiETAL24Next} to repairing and
generating~\cite{wang2025efficoder} code. This raises a natural
question: do such models rely primarily on statistical regularities
acquired during pretraining, or can they flexibly condition their
reasoning on explicitly provided behavioral rules?

Consider an integer arithmetic operation with two alternative
semantics for the standard `\texttt{+}' symbol, a scenario 
frequently encountered in operator overloading~\cite{ravasi2020pylops,tritonPython26}. Under the first,
\texttt{2+2} behaves conventionally as \textit{addition}; under the second, the
same symbol is defined to perform \textit{subtraction}. A system that reasons
from syntax alone (learned priors) would produce the same answer in both
cases. A system that conditions on the supplied semantic definitions
would change its behavior immediately. This contrast captures a broader
scientific issue:

\noindent
\emph{Can \LLMs adapt their reasoning to externally specified formal
systems, even when those systems conflict with entrenched priors
learned from data?}

Formal semantics~\cite{pierce2002tapl} offers a uniquely controlled
setting for studying this question. The semantics of a programming
language consist of symbolic transition rules governing program-state
evolution.
Correct execution requires repeatedly selecting and composing such
rules over many steps. Crucially, these rules can be modified without
altering surface syntax, allowing one to separate reliance on lexical
cues from genuine conditioning on semantics.
Furthermore, formal semantic rules: (1)~are atomic with uniform
granularity, enabling systematic comparison across programs and model
predictions, and (2)~specify behavior mathematically rather than in
natural language, reducing ambiguity between intended execution and
model instructions.

\MyPara{Tested hypotheses}%
We use program execution as a lens for analyzing \emph{formal semantic
rule-conditioned reasoning in \LLMs}. Rather than asking whether models
can execute programs in familiar languages, we ask whether they
can---\emph{alter} their reasoning under novel formal semantics, apply
individual rules at fine granularity, and sustain such conditioning
across long execution horizons (\eg, loops and nested control flow).
This yields four concrete hypotheses about model capabilities:

\MyHypothesis{\hypothesisOne}{Global Rule Conditioning}%
Models can combine many rule applications to correctly predict final
states.

\MyHypothesis{\hypothesisTwo}{State-Free Rule Conditioning}%
Models can follow rules correctly under state-mutation free execution.

\MyHypothesis{\hypothesisThree}{Long-Horizon Rule Conditioning}%
Models can follow formal rules consistently across long execution
traces.

\MyHypothesis{\hypothesisFour}{Rule Conditioning Under Semantic Shift}%
Models continue to follow supplied rules under novel semantics.

To test these hypotheses, we introduce \dataset, which pairs a
featherweight~\cite{harper2016pfpl} C programming language \lang with
two formal semantic systems—the fine-grained small-step structural
operational semantics (\Sos) and the coarser rewriting-based \Kos
semantics~\cite{rosu_kframework}. The benchmark probes the hypotheses
via three complementary tasks---predicting final program states (\op),
selecting semantic rules governing execution in absence of state
mutation (\srp), and generating full execution traces to probe
long-horizon rule application (\etp)---while using semantic mutations
and program-complexity splits as stressors for robustness.

\MyPara{Reliance on learned priors vs supplied rules}%
We disentangle reliance on supplied rules versus learned priors along
two orthogonal axes: semantic mutation and program-complexity shifts.
A key feature enabled by formal semantics is \emph{\nstandardSem}
variants that systematically perturb symbol meanings. In \keywordMut,
common operators exchange their behavior, creating direct conflicts
with pretrained priors. In \keywordObf, familiar syntax is replaced
with novel symbols whose meanings are defined only through the
supplied rules.
Models are additionally evaluated on human-written,
LLM-translated, and fuzzer-generated programs with varied structural
complexity, stressing deep control flow and unusual data-flow
patterns.

\MyPara{Choice of programming language}%
We use \lang rather than indentation-sensitive languages such as
Python, whose concrete syntax requires recovering block
structure~\cite{adams2013indentation} from layout before abstract
syntax can be constructed, thereby entangling syntactic recovery with
semantic reasoning.
Explicit block delimiters `\{\}' in \lang
%(as in C)
avoid this confound, allowing us to isolate the model's ability to
condition on formal semantics defined over the abstract syntax.

Our experiments across a broad set of frontier and open-weight models
show that while several benefit from access to formal rules under
standard semantics, performance deteriorates sharply under semantic
mutations, increased rule granularity, and long execution horizons,
exposing systematic limits in current models' ability to sustain
reasoning conditioned on externally specified formal systems.

By framing program execution as a controlled probe of rule-conditioned
reasoning, we provide a semantics-driven benchmark for assessing when
\LLMs adapt their behavior to externally specified formal systems.

\section{Background}
\begin{figure}[t]
  \subcaptionbox{%
    \FigBackgroundGrammarCaption%
    \label{figure:simple-grammar}}[\linewidth]{%
\begin{lstlisting}[language=bnf-grammar]
<program>   ::= <stmt>
<stmt>      ::= <assgn_stmt>(*@\label{line:assign-stmt}@*)
<assgn_stmt>::= <id>'='<exp>';'
<exp>       ::= <literal> | <exp>'+'<literal>
<literal>   ::= <digit> | <literal> <digit>
<id>        ::= <letter> | <id> <letter>
<digit>     ::= '0' | ... | '9'
<letter>    ::= 'a' | ... | 'z'
\end{lstlisting}
  }%
  \vspace{3mm}%
  \\
  \subcaptionbox{%
    \FigBackgroundSOSSemanticsCaption%
    \label{figure:simple-semantics}}[\linewidth]{%
    \resizebox{0.95\linewidth}{!}{%
      \input{figures/background/simple_sos_semantics}
    }%
  }%
  % \vspace{3mm}%
  % \\
  \caption{\FigBackgroundSOSCaption}
  \label{figure:background-sos}
\end{figure}

\begin{figure*}[t]%
  \centering%
  \resizebox{0.9\textwidth}{!}{%
    \includegraphics{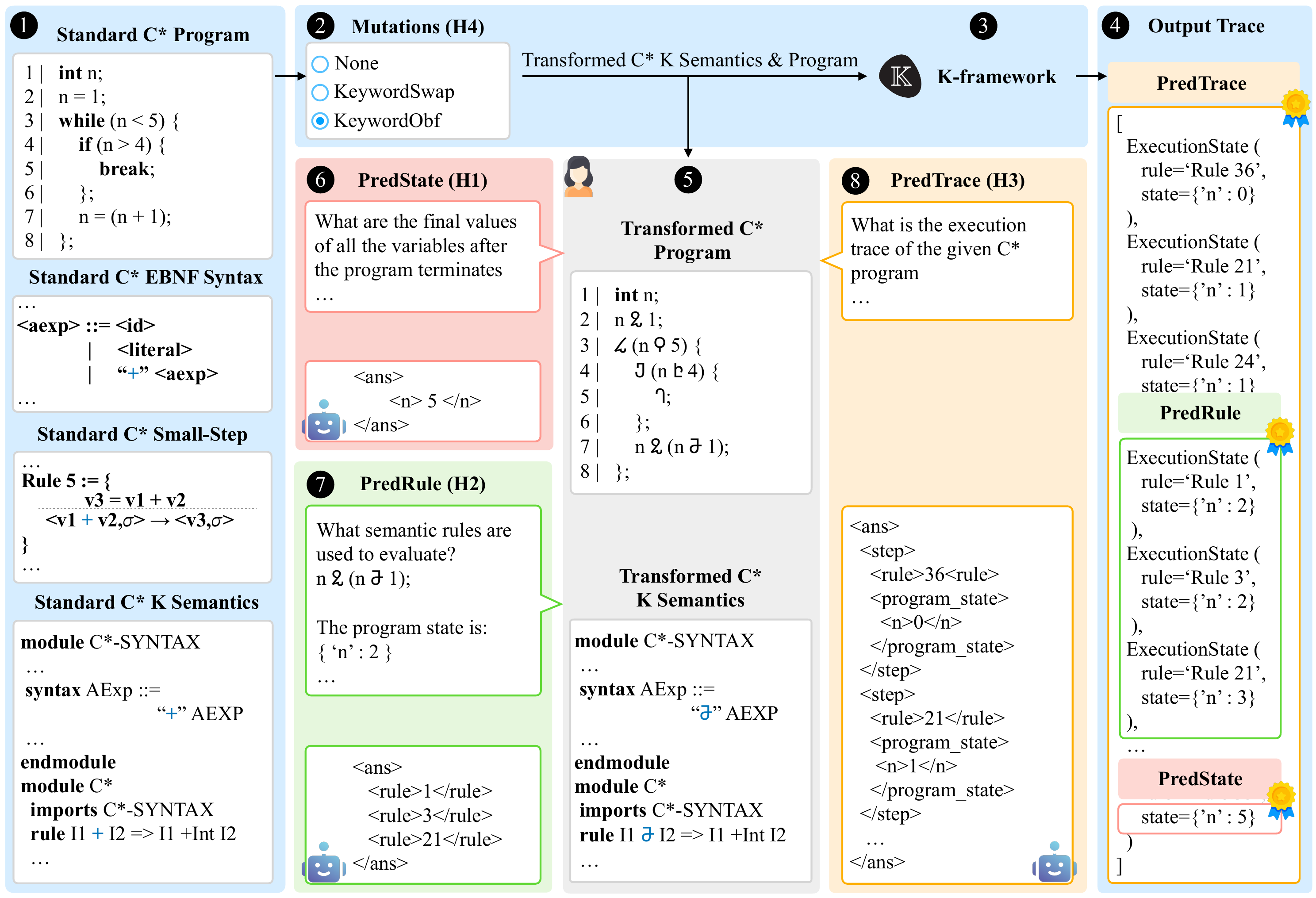}%
  }%
  \caption{\FigOverviewCaption\label{figure:overview}}%
  %\vspace{-2mm}%
\end{figure*}%

\noindent The semantics of a programming language defines program
behavior. Structural operational semantics specifies semantics via
inference rules that govern transitions between \emph{configurations},
each pairing a program fragment with its execution state. We use
small-step semantics (\Sos), where each rule represents one atomic
computation and execution arises from repeated rule applications.
Rules are written in Gentzen-style inference notation
\citep{gentzen_notation}, with premises and side conditions above the
fraction bar and conclusions below.

\begin{wraptable}{r}{0.22\textwidth}%
  \vspace{-0.25cm}%
  \centering%
  \TableFont%%
  \caption{\TableNotationPrimerCaption}%
%  \vspace{-0.3cm}%
  %  \hspace*{-0.15cm}%
  \setlength{\tabcolsep}{1pt}
  \begin{tabular}{@{}cl@{}}%
    \toprule%
    \textbf{Notation} & \textbf{Definition} \\
    \midrule%
    $\fState{}$ & Program state \\
    $\fStmt{}$ & Statement \\
    $\fVar{}$ & Int variable \\
    $\fExp{}$ & Int expression \\
    $\fLit{}$ & Int literal \\
    $\fConfig{operation}{\fState{}}$ & Configuration \\
    $\sigma[\texttt{x} \mapsto \texttt{v}]$ & Store $\texttt{v}$ in $\texttt{x}$\\
    $\fTransE{\fExp{}}{\fState{}}{\fExp{}'}{}$ & Expression-step\\
    $\fTransS{\fStmt{}}{\fState{}}{\fStmt{}'}{\fState{}'}{}$ & Statement-step\\
    $\fConfig{\epsilon}{\fState{}}$ & \texttt{NOP} \\
    \bottomrule%
  \end{tabular}%
  \vspace{-0.3cm}%
\end{wraptable}%
We illustrate semantics formalization in \Sos using a simple
imperative language $\lLang$ whose syntax
(Figure~\ref{figure:simple-grammar}) includes assignments and integer
expressions with addition. Table~\ref{tab:notation-primer} summarizes
the notation used in its formalization
(Figure~\ref{figure:simple-semantics}). Configurations take the form
$\fConfig{c}{\fState{}}$, where $c$ ranges over statements
($\fStmt{}$) and expressions ($\fExp{}$), and the state maps variables
($\fVar{}$) to integer values ($\fLit{}$).

Expression transitions
\wrapFootNoteSize{$\fTransE{\fExp{}}{\fState{}}{\fExp{}'}{\fRule}$}
apply a single rule to reduce an expression and record the ordered list
of rules $\fRule$ used so far; they do not mutate state and terminate in
a literal under transitive--reflexive closure. Statement transitions
\wrapFootNoteSize{$\fTransS{\fStmt{}}{\fState{}}{\fStmt{}'}{\fState{}'}{r}$}
may update the state and iterate until reaching \texttt{NOP}
($\fConfig{\epsilon}{\fState{}}$).
For example, \wrapFootNoteSize{$\rAssignStep$} propagates evaluation
through an assignment by stepping the right-hand-side expression,
while \wrapFootNoteSize{$\rAssign$} applies once that expression
reduces to a literal and commits the value to the state; together,
such rules illustrate how programs execute by repeated configuration
transitions.

\begin{myDef}[Statement Execution]
  \label{def:stmt-execution}
  Let $\fStmt{}$ be a statement derived from a given grammar
  $\fGrammar$ that is semantically valid under an \Sos formalization
  $\fSemantics$, and let $\fRules$ be the set of all rule names in
  $\fSemantics$. Suppose that under $\fSemantics$ and an initial
  program state $\fState{0}$, the statement $\fStmt{}$ reduces to a
  \texttt{NOP} configuration in $n$ statement-steps:
  \wrapFootNoteSize{$\fConfig{\fStmt{}}{\fState{0}}\fStepS{\fRule_1}\fConfig{\fStmt{}'}{\fState{1}}\fStepS{\fRule_2}\dots\fStepS{\fRule_n}\fConfig{\epsilon}{\fState{n}}$}.
  For each $i \in [1,n]$, $\fRule_i\!\in\!\fRules^*$ denotes the
  ordered list of rules required for the $i^{th}$ statement-step, with
  elements indexed as $\fRule_{i,j}$ where $j\!\in\![0, \fSize{\fRule_i})$,
  and $\fState{i}$ denotes the program state after the $i^{th}$-step.
  Let `$\fConcat$' denote the standard list concatentation operator.
  We then define statement execution of $\fStmt{}$
  under $\fSemantics$ and an initial program state $\fState{0}$ as a
  pair:

  \wrapFootNoteSize{
    \(
    \fExec{\fStmt{}}{\fSemantics}(\fState{0}) \triangleq
    (\underbrace{\textstyle \fState{n} \vphantom{\fConcatBig_{j=0}^{\fSize{\fRule_i}}}}_{\mathclap{\textstyle \wrapFootNoteSize{\text{resulting state}}}},\;\;\;
    \underbrace{\textstyle \fConcatBig_{i=1}^{n}\fConcatBig_{j=0}^{\fSize{\fRule_i}-1}[(\fState{i},\fRule_{i,j})]}_{\mathclap{\textstyle \wrapFootNoteSize{\text{execution trace}}}}),\;\;
    \fRule_{i,j} \in \fRules
    \)
  }
\end{myDef}

%% Automatically generated by: table.py make_dataset_stats_table
\begin{table*}[t]
%\vspace{cm}%
\begin{center}
\TableFont
\caption{\UseMacro{TCap-models-datasets-stats}\label{tab:datasets-stats}}
%\vspace{-0.1cm}%
\begin{threeparttable}
\begin{adjustbox}{max width=\linewidth}
\renewcommand{\arraystretch}{1.12}
\begin{tabular}{l c c c  c  c  c  c  c  c  c  c  c  c  c  c }
\toprule
 &
 & \multicolumn{3}{c}{\UseMacro{THead-datasets-stats-NumTokens}\tnote{*}}
 & \multicolumn{5}{c}{\UseMacro{THead-datasets-stats-Control-flow}}
 & \multicolumn{2}{c}{\UseMacro{THead-datasets-stats-Data-flow}}
 & \multicolumn{4}{c}{\UseMacro{THead-datasets-stats-Size}}
\\
 \cmidrule(lr){3-5}
 \cmidrule(lr){6-10}
 \cmidrule(lr){11-12}
 \cmidrule(lr){13-16}
 \UseMacro{THead-datasets-stats-datasets}
 & \UseMacro{THead-datasets-stats-NumPoints}
 & \UseMacro{THead-datasets-stats-NumTokens-std}
 & \UseMacro{THead-datasets-stats-NumTokens-swap}
 & \UseMacro{THead-datasets-stats-NumTokens-obf}
 & \textbf{\UseMacro{THead-datasets-stats-CC}}
 & \textbf{\UseMacro{THead-datasets-stats-MaxNestedIf}}
 & \textbf{\UseMacro{THead-datasets-stats-MaxNestedLoop}}
 & \textbf{\UseMacro{THead-datasets-stats-MaxTakenIf}}
 & \textbf{\UseMacro{THead-datasets-stats-MaxTakenLoop}}
 & \textbf{\UseMacro{THead-datasets-stats-DepDegree}}
 & \textbf{\UseMacro{THead-datasets-stats-NumAssignments}}
 & \textbf{\UseMacro{THead-datasets-stats-LOC}}
 & \textbf{\UseMacro{THead-datasets-stats-Volume}}
 & \textbf{\UseMacro{THead-datasets-stats-Vocabulary}}
 & \textbf{\UseMacro{THead-datasets-stats-TraceLength}}
\\
\bottomrule
\end{tabular}
\end{adjustbox}
\begin{tablenotes}
\item[] * Median token counts using \gptfom; not used for complexity metrics. ** We study tokenization impact on rule-conditioning via a controlled ablation (Appendix~\ref{app:keyword-obf-tokenization}).
\end{tablenotes}
\end{threeparttable}
\end{center}
\vspace{-0.4cm}%
\end{table*}

\setcounter{footnote}{1}
\renewcommand{\thefootnote}{\fnsymbol{footnote}}

\begin{myDef}[Program Execution]
  \label{def:program-execution}
  Let $\fProgram$ be a program derived from a given grammar
  $\fGrammar$ that can be parsed into an ordered
  list of statements $[\fStmt{0},\dots,\fStmt{n}]$.
  We define program execution of $\fProgram$ under a given formalization
  $\fSemantics$ and an initial program state $\fState{0}$ compositionally
  using Definition~\ref{def:stmt-execution} as:
  
  \wrapFootNoteSize{
    \(
    \fExec{\fProgram}{\fSemantics}(\fState{0})\!=\!\fExec{[\fStmt{0},\dots,\fStmt{n}]}{\fSemantics}(\fState{0})\!\triangleq\!
    (
    \fExec{\fStmt{0}}{\fSemantics}\!\fSeqOp\!\dots\!\fSeqOp\!\fExec{\fStmt{n}}{\fSemantics}) (\fState{0})
    \footnote{
      Statements $\fStmt{0},\dots,\fStmt{n}$ are sequenced at the program
      level as single units, regardless of whether their execution expands
      into nested statements (\eg, loops).
    }
    \footnote{
    \wrapFootNoteSize{
      The empty program denotes the base case of the compositional definition:
      \(\forall \fState{}.\fExec{[\phantom{w}]}{\fSemantics}(\fState{}) \triangleq (\fState{}, [\phantom{w}])\)}.
    }
    \)
  }%
  
  \noindent Here, `$\fSeqOp$' is a left-associative sequencing operator defined by
  direct style \emph{Kleisli composition}~\cite{wadler1995} as follows:
  
  \wrapFootNoteSize{\((\fExec{\fStmt{A}}{\fSemantics} \fSeqOp \fExec{\fStmt{B}}{\fSemantics})(\fState{}) \triangleq
    \begin{aligned}[t]
      &\text{let } (\fState{A}, \fTrace{A}) = \fExec{\fStmt{A}}{\fSemantics}(\fState{}) \text{ in}\\
      &\text{let } (\fState{B}, \fTrace{B}) = \fExec{\fStmt{B}}{\fSemantics}(\fState{A}) \text{ in}\\
      &\qquad (\fState{B}, \fTrace{A} \fConcat \fTrace{B})
    \end{aligned}
    \)}
  
  where $\fState{A}$ and $\fState{B}$ are the resulting states
  of \wrapFootNoteSize{$\fExec{\fStmt{A}}{\fSemantics}(\fState{})$} and
  \wrapFootNoteSize{$\fExec{\fStmt{B}}{\fSemantics}(\fState{A})$} respectively,
  and $\fTrace{A}$ and $\fTrace{B}$ are the corresponding execution traces.
  Consequently,
  \wrapFootNoteSize{\(\fExec{\fProgram}{\fSemantics}(\fState{0})\!=\!
    (\fState{n}, \fConcat_{i=0}^{n}\fTrace{i})\)}, where $\fState{n}$
  is the final state---obtained by executing the terminal
  statement $\fStmt{n}$. This holds true for both \Kos and \Sos formalizations.
  We use the notations $\fExecAccTwo{\cdot}{\fState{}}{\fSemantics}$
  and $\fExecAccTwo{\cdot}{\fTrace{}}{\fSemantics}$ to denote
  accessing the final state and execution trace respectively.
  We also introduce the projection
  (\wrapFootNoteSize{$\fProjectNone$}) based notation for
  tuple element access: For a tuple
  \wrapFootNoteSize{$t\!=\!(x_1,\dots,x_k)$, $\fProject{i}{t}\!=\!x_i$} and its
  compositional extension to an ordered list of tuples
  \wrapFootNoteSize{$T\!=\![t_1,\dots,t_n]$} as
  \wrapFootNoteSize{$\fProject{i}{T}\!=\![\,\fProject{i}{t_1},\dots,\fProject{i}{t_n}\,]$}.
\end{myDef}

\section{Benchmark Construction}
\label{sec:dataset}

\begin{table*}
  \centering%
  \TableFont%
  \caption{\TableNonStandardRulesCaption}%
  \begin{threeparttable}
   \vspace{-0.3cm}%
   \setlength{\tabcolsep}{4pt}%
   \renewcommand{\arraystretch}{1.1}
   \begin{tabular}{l c ccccc cccccc ccc ccccc}%
     \toprule%
     \UseMacro{Caucasiantitle1} &
     \multicolumn{1}{c}{\UseMacro{CaucasianAssignment}} &
     \multicolumn{5}{c}{\UseMacro{CaucasianArithmetic}} &
     \multicolumn{6}{c}{\UseMacro{CaucasianRelational}} &
     \multicolumn{3}{c}{\UseMacro{CaucasianLogical}} &
     \multicolumn{5}{c}{\UseMacro{CaucasianKeyword}} \\
     \cmidrule(lr){1-1}
     \cmidrule(lr){2-2}
     \cmidrule(lr){3-7}
     \cmidrule(lr){8-13}
     \cmidrule(lr){14-16}
     \cmidrule(lr){17-21}
     %\midrule%
     \UseMacro{Caucasiantitle2} &
     \texttt{=} &
     \texttt{+} & 
     \texttt{-} &
     \texttt{*} &
     \texttt{/} &
     \texttt{\%} &

     \texttt{<} &
     \texttt{<=} &
     \texttt{>} &
     \texttt{>=} &
     \texttt{==} &
     \texttt{!=} &

     \texttt{!} &
     \texttt{\&\&} &
     \texttt{||} &

     \texttt{if-else} &
     \texttt{while} &
     \texttt{halt} &
     \texttt{break} &
     \texttt{continue}\\
     \UseMacro{Caucasiantitle4}\tnote{*} &
     \texttt{=} &
     \texttt{-} & 
     \texttt{+} &
     \texttt{/} &
     \texttt{*} &
     \texttt{\%} &

     \texttt{>} &
     \texttt{>=} &
     \texttt{<} &
     \texttt{<=} &
     \texttt{!=} &
     \texttt{==} &

     \texttt{!} &
     \texttt{||} &
     \texttt{\&\&} &

     \texttt{if-else} &
     \texttt{while} &
     \texttt{halt} &
     \texttt{break} &
     \texttt{continue}\\
     \UseMacro{Caucasiantitle3}\tnote{**} &
     \CAASSIGN &
     \CAADD &
     \CASUB &
     \CAMUL &
     \CADIV &
     \CAMOD &

     \CALT &
     \CALTEQ &
     \CAGT &
     \CAGTEQ &
     \CAEQ &
     \CANEQ &

     \CANOT &
     \CAAND &
     \CAOR &

     \CAIF-\CAELSE &
     \CAWHILE &
     \CAHALT &
     \CABREAK &
     \CACONTINUE\\
     \bottomrule%
   \end{tabular}
   \begin{tablenotes}
     \scriptsize
     \item * Swaps the semantics of standard operator/keyword symbols; ** Assigns semantics of standard operators/keywords to novel symbols (from the Caucasian-Albanian script).
   \end{tablenotes}
  \end{threeparttable}
  \vspace{-0.3cm}%
\end{table*}

Figure~\ref{figure:overview} shows the benchmark construction process.
We formalize \lang in both \Sos and \Kos semantics (\circledCaption{1}). We
use the \KTool (\circledCaption{3}) to obtain ground-truths (\circledCaption{4}) for
\Kos experiments and a custom \Antlr-based interpreter for those with
\Sos.
The \lang program along with the \Kos or \Sos derived formalization is
used to prompt the \LLMs (\circledCaption{5}).

\subsection{Dataset Curation}
\label{sec:dataset:collection}

\dataset contains \NumDatasets splits, the \humanwrit, the
\llmtrans, and the \fuzzgen.

\MyPara{\humanwrit}%
This set of \lang
programs we manually adapted from C++ solutions to coding problems
sourced from LeetCode~\citep{leetcode},
HumanEval~\citep{humanEval,codegeex},
CodeContests~\citep{codecontest}, and MBPP~\citep{mbpp,
  OrlanskiETAL23Measuring}. We use public test cases as input and
their corresponding oracles as expected outputs.
Variable names are obfuscated by replacing semantically meaningful
identifiers (\eg, \CodeIn{maxIter}) with random strings
(Appendix~\ref{sec:appendix-imp-humanwrit-example}).
We validate correctness by executing the programs with \KTool and verifying
outputs against the test oracles.

\MyPara{\llmtrans}%
This set of \lang programs are translated from C++
programs using \LLMs.  Specifically, we collect the C++ programs from
the CodeForces solutions published on Hugging
Face~\citep{penedo2025codeforces}. We prompt \qwenCoderSmall{32} with the
\lang syntax, semantics constraints, the C++ solution and one
corresponding public test case to generate a valid \lang
program which we subsequently filter via the \KTool
based on successful test execution.

\MyPara{\fuzzgen}%
We construct this with a depth-controlled,
semantics-aware, grammar-based
fuzzer~\citep{YangCsmith,Han2019CodeAlchemist}; a fuzzer is a tool
that automatically generates programs and it is commonly used for
testing compilers and interpreters.
The fuzzer samples statements from---assign,
if–else, while, break, continue, halt---using depth-tapered probabilities—a cosine decay
reduces the chance of generating new \texttt{if/while} as nesting
grows—and legality masks that enforce syntactic and semantic validity
(Appendix~\ref{sec:appendix-imp-fuzzgen-example}).

\MyPara{Program complexity and data statistics}%
We characterize program complexity along three axes—control-flow,
data-flow, and size. For control-flow, we use extended cyclomatic
complexity (\metricCC)~\citep{mccabe-cc}; the \emph{static} maximum
nesting depths of if–else and while (\metricMaxnestif,
\metricMaxnestwhile); and their \emph{dynamic} counterparts measured
along executed paths (\metricMaxtakenif, \metricMaxtakenwhile). For
data-flow, we use DepDegree (\metricDepdeg), which quantifies uses and
redefinitions of declared variables~\citep{Beyer-depdeg}, and the
total number of executed assignments (\metricNumAssign). For size, we
use Halstead Vocabulary and Volume (\metricVocab,
\metricVol)~\citep{Halstead1977} which captures the symbol variety and
program information in bits respectively, lines of code (\metricLoc),
and execution-trace length (\metricTracelen).% under \Sos.

Table~\ref{tab:dataset-median-stats} reports median values of the
complexity metrics per dataset. Across $\approx$165 programs per split, the
median complexity increases progressively from \humanwrit to \llmtrans
to \fuzzgen along all three axes (complexity metrics distributions given in
Appendix~\ref{sec:appendix-metric-dist}).

\subsection{Semantic \Shifts}
\label{sec:dataset:mutation}

To test whether models condition their reasoning on explicitly supplied
semantics—rather than rely on pretraining-induced associations between
surface syntax and behavior—we introduce two semantic \shifts,
\keywordMut and \keywordObf. These are transformations of the
\standardSem \lang semantics (\StdOperationalSemantics;
\wrapFootNoteSize{$\fFormalPlaceHold\!\in\!\fFormalSet$}) that preserve
rule structure while perturbing the mapping between syntactic symbols
and their conventional meanings, enabling controlled tests of whether
models follow the supplied inferential rules when syntactic familiarity
is disrupted.

\MyParaTwo{\keywordMut (\KeywordMutOperationalSemantics).}
\keywordMut swaps the semantic interpretations of selected syntactic
operators in the \standardSem semantics with their \keywordMut
counterparts (Table~\ref{tab:mutation-rules-combined}); for example, it
exchanges addition (\texttt{+}) and subtraction (\texttt{-}), so that an
expression written as \texttt{x}+\texttt{y} is evaluated according to the
subtraction rule. Because \keywordMut preserves surface syntax while
altering operational meaning, correct reasoning requires conditioning on
the explicit transition rules rather than defaulting to
pretraining-derived interpretations of common symbols.

\MyParaTwo{\keywordObf (\KeywordObfOperationalSemantics).}
\keywordObf probes the complementary case in which syntactic familiarity
is removed altogether by systematically replacing standard keywords and
operators in the \standardSem semantics with symbols drawn from the
rarely encountered Caucasian-Albanian script~\citep{GippertSchulze_albanian}
(Table~\ref{tab:mutation-rules-combined}). Under \keywordObf, expressions
such as \texttt{x}\CAADD\texttt{y} execute identically to
\texttt{x}+\texttt{y} under \standardSem semantics, but without relying
on familiar symbolic cues. By eliminating conventional symbol
associations while preserving rule structure, \keywordObf isolates a
model's ability to follow explicit operational definitions in the
absence of syntactic priors.

\begin{figure*}[!t]
  \centering
  \begin{subfigure}[b]{0.4\linewidth}
    \centering
    \resizebox{\linewidth}{!}{\input{figures/tasks/notation_comprehension_sos_tikz}}%
    \caption{\FigNotationCompImpSos}
    \label{fig:notation-comp-sos}
  \end{subfigure}
  \hfill
  \raisebox{2.2cm}{
  \begin{subfigure}[b]{0.15\linewidth}
    \centering
    \includegraphics[width=\linewidth]{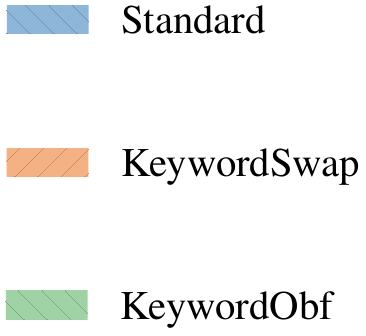}%
  \end{subfigure}}
  \hfill
  \begin{subfigure}[b]{0.4\linewidth}
    \centering
    \includegraphics[width=\linewidth]{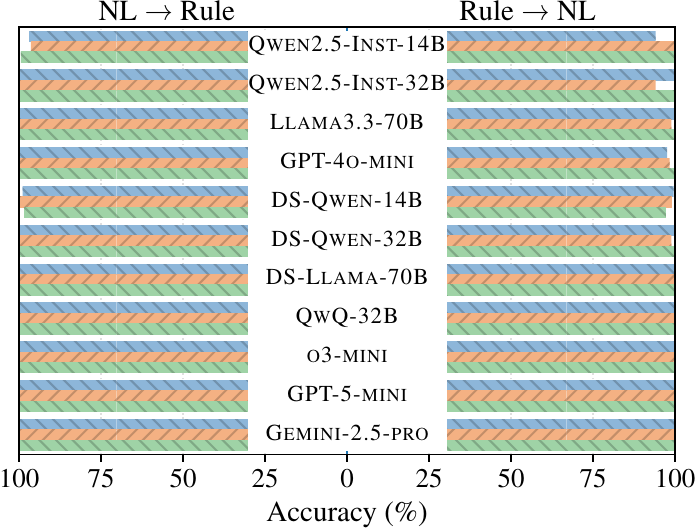}%
    \caption{\FigNotationCompImpK}
    \label{fig:notation-comp-k}
  \end{subfigure}
  \vspace{0.2cm}%
  \\
  \includegraphics[width=0.48\linewidth]{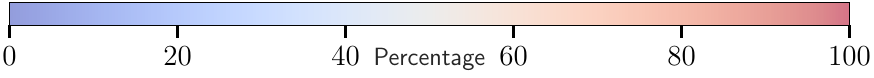}%
  %
  %\vspace{0.2cm}%
  %
  \\
  \begin{subfigure}[b]{0.22\linewidth}
    \centering
    \includegraphics[width=0.9\linewidth]{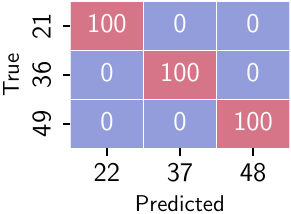}%
    \caption{\FigNotationCompImpSosConfMatQwenStd}
    \label{fig:cm-qwen-std}
  \end{subfigure}
  \hfill
  \begin{subfigure}[b]{0.22\linewidth}
    \centering
    \includegraphics[width=0.9\linewidth]{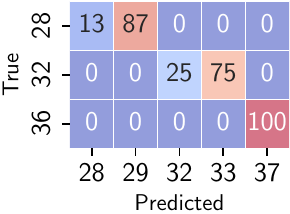}%
    \caption{\FigNotationCompImpSosConfMatGptStd}
    \label{fig:cm-gpt-std}
  \end{subfigure}
  \hfill
  \begin{subfigure}[b]{0.23\linewidth}
    \centering
    \includegraphics[width=0.86\linewidth]{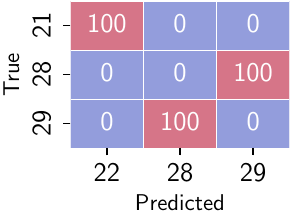}%
    \caption{\FigNotationCompImpSosConfMatQwenSwap}
    \label{fig:cm-qwen-swap}
  \end{subfigure}
  \hfill
  \begin{subfigure}[b]{0.22\linewidth}
    \centering
    \includegraphics[width=0.9\linewidth]{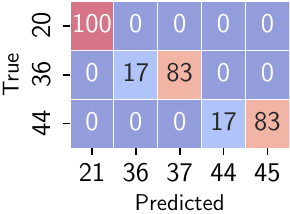}%
    \caption{\FigNotationCompImpSosConfMatGptSwap}
    \label{fig:cm-gpt-swap}
  \end{subfigure}
  \caption{\FigNotationCompCombined}
\end{figure*}

\section{Evaluation Setup}
\label{sec:eval-framework}

\MyPara{Models and inference settings}%
We evaluate eleven frontier \LLMs divided into two classes.  The
\emph{non-reasoning} group consists of \llamaBigSmall
\cite{grattafiori2024llama}, \qwenCoderSmall{14} and
\qwenCoderSmall{32} \cite{hui2024qwen2}, and \gptfomSmall
\cite{gptfo}. The \emph{reasoning} group includes DeepSeek variants
\dpskLlamaSmall, \dpskQwenSmall{14}, and \dpskQwenSmall{32}
\cite{guo2025deepseek}, as well as \othreeSmall and \gptfiveminiSmall
\cite{OpenAI_GPT5mini_2025}, \geminiSmall \cite{gemini25}, and
\qwqSmall \cite{qwq32b}.
We average
all reasoning-model runs (and \gptfomSmall) over \NumExpAverage
trials.  The temperature for all non-reasoning models (except
\gptfomSmall) is set to \texttt{0} (prompts and additional details in
Appendix~\ref{sec:appendix-prompts}).

\MyPara{Preliminary validation: formal-notation understanding}%
Before testing whether models can condition their reasoning on
explicit formal semantic rules, we verify that they can interpret the
notation used to express those rules; otherwise downstream failures
could reflect superficial misunderstanding of the formalism rather
than limitations in rule-conditioned reasoning. We perform this
validation using two auxiliary classification tasks: \nlruleSmall and
\rulenlSmall. In \nlruleSmall, models select the correct formal rule
(out of \numNotationCompChoices choices) given its natural-language
description (human-written); conversely, in \rulenlSmall they identify
the correct description for a given rule. Together, these tasks
isolate notation-level understanding at the granularity of individual
inference rules.

\emph{Dataset}. Multiple-choice distractors are generated via a hierarchical sampling
strategy to prevent reliance on surface lexical cues (\eg,
random sampling could produce distractors involving unrelated
operators or constructs, enabling pattern matching rather than
semantic discrimination).
Rules are grouped—in descending order of sampling preference—into
families, constructs, and semantic roles
(Appendix~\ref{sec:appendix-notation-comprehension}). We generate
\numNotationCompSamples samples per task and semantic variant
(\standardSemCap, \keywordMut, and \keywordObf).

\emph{Analysis}. Figures~\ref{figure:notation-comp-sos}
and~\ref{figure:notation-comp-k} show the results averaged over
\numNotationCompRuns runs under \emph{zero-shot} prompting for \Sos
and \Kos. Under all semantic variants and
formalizations, most models achieve near-ceiling performance on both
\nlruleSmall and \rulenlSmall tasks. The performance of the
\hypertarget{txt-notation-comp-callout-1}{}%
\qwenCoderSmall{14}~(Figure~\ref{figure:notation-comp-sos},
\hyperlink{fig-notation-comp-callout-1}{\uncircledCaption{1}}) and
\hypertarget{txt-notation-comp-callout-2}{}%
\gptfomSmall~(Figure~\ref{figure:notation-comp-sos},
\hyperlink{fig-notation-comp-callout-2}{\uncircledCaption{2}}) models
on the \nlruleSmall task under $\fSemanticsSos$ are the exceptions,
maxing out at
$\approx$\numNotationCompSosStdMinPerc-\numNotationCompSosStdMaxPerc.
Figures~\ref{fig:cm-qwen-std} and~\ref{fig:cm-qwen-swap}, and
Figures~\ref{fig:cm-gpt-std} and~\ref{fig:cm-gpt-swap}, show the
confusion matrices for the top three most mispredicted rules for
\xspace$\;\fVariantPlaceHold\!=$\fVarStd\xspace and \fVarSwap\xspace
respectively, for \qwenCoderSmall{14} and \gptfomSmall on
\nlruleSmall.  Both models primarily confuse structurally adjacent
rules that govern small-step reduction of expressions and computations
for \StdSOSOperationalSemantics and \KeywordMutSOSOperationalSemantics
formalizations. Most rule mispredictions fall within the
\emph{Arithmetic Expression} (7-23) and the \emph{Relational
Comparison} (28-51) families.

In summary, most frontier models exhibit stable notation-level competence across
semantic formalizations and \shifts, indicating that subsequent failures
primarily reflect limitations in rule-conditioned reasoning rather than
inability to parse the formalism itself. When errors occur, the dominant
failure mode is \emph{imprecise discrimination among fine-grained semantic
roles within a construct} (\eg, step vs.\ compute cases), rather than
global breakdown or random guessing. \KOperationalSemantics[0.7]\xspace exhibits fewer such confusions,
consistent with its coarser rule inventory per construct, which reduces
the density of near-miss distractors relative to \SOSOperationalSemantics[0.7].
\section{Experiments and Results}
\label{sec:quant-analysis}

% \newcommand{\delt}[1]{%
%   % Round to integer first
%   \pgfmathtruncatemacro{\DeltaInt}{round(#1)}%
%   \pgfmathtruncatemacro{\DeltaAbsInt}{abs(\DeltaInt)}%
%   % zero-pad to 2 digits
%   \ifnum\DeltaAbsInt<10
%     \edef\DeltaFmt{0\DeltaAbsInt}%
%   \else
%     \edef\DeltaFmt{\DeltaAbsInt}%
%   \fi
%   % color + sign
%   \ifnum\DeltaInt>0
%     \textcolor{green!45!black}{\scriptsize(+\DeltaFmt)}%
%   \else\ifnum\DeltaInt<0
%     \textcolor{red!60!black}{\scriptsize(\texttt{-}\DeltaFmt)}%
%   \else
%     \textcolor{black!55}{\scriptsize(000)}%
%   \fi\fi
% }
% \newcommand{\accd}[2]{#1\,\delt{#2}}
% % needs: \usepackage{pgf}

% % acc + delta where delta is computed from (a - b), rounded to nearest int
% \newcommand{\accdfrom}[2]{%
%   \pgfmathtruncatemacro{\DeltaTmp}{round((#1) - (#2))}%
%   #1\,\delt{\DeltaTmp}%
% }
% % robust delta: Var - Std (still prints Var as the main number)
% \newcommand{\accdrobust}[2]{%
%   \pgfmathtruncatemacro{\DeltaTmp}{round((#1) - (#2))}%
%   #1\,\delt{\DeltaTmp}%
% }

\colorlet{zebra}{blue!5}

\begin{table*}[t]
\begin{center}
\TableFont
\caption{\TablePredStatePredRuleCombinedCaption}

\setlength{\tabcolsep}{3.1pt}
\renewcommand{\arraystretch}{1}

\begin{tabular}{l S[table-format=3, table-column-width=4mm, detect-all=false]  r  r  r  c  r  r  r  S[table-format=1, table-column-width=4mm, detect-all=false]  S[table-format=2, table-column-width=6mm, detect-all=false]  r  r  c  S[table-format=2, table-column-width=6mm, detect-all=false]  r  r}
\toprule
& \multicolumn{8}{c}{\textbf{\op ($\fSplit$=\humanwrit)}}
& 
& \multicolumn{7}{c}{\textbf{\srp}} 
\\
\cmidrule(lr){2-9}\cmidrule(lr){11-17}
%\\
&
& \multicolumn{3}{c}{\textbf{\KosSmall-Formalization}}
& 
& \multicolumn{3}{c}{\textbf{\SosSmall-Formalization}}
& 
& \multicolumn{3}{c}{\textbf{\KosSmall-Formalization}} 
&
& \multicolumn{3}{c}{\textbf{\SosSmall-Formalization}} 
\\
\cmidrule(lr){3-5}\cmidrule(lr){7-9}\cmidrule(lr){11-13}\cmidrule(lr){15-17}
%\\
& \textbf{$\metricAccuracyNK$}
& \textbf{$\metricAccuracyStd$} & \textbf{$\metricAccuracySwap$} & \textbf{$\metricAccuracyObf$}
& 
& \textbf{$\metricAccuracyStd$} & \textbf{$\metricAccuracySwap$} & \textbf{$\metricAccuracyObf$}
&
& \textbf{$\metricAccuracyStd$} & \textbf{$\metricAccuracySwap$} & \textbf{$\metricAccuracyObf$}
&
& \textbf{$\metricAccuracyStd$} & \textbf{$\metricAccuracySwap$} & \textbf{$\metricAccuracyObf$} 
\\
%\\
\textbf{\UseMacro{THead-op-IMP-K-IMP-SOS-acc-models}}
& 
& ($\metricDeltaCond$) & ($\metricDeltaRobust$) & ($\metricDeltaRobust$) 
& 
& ($\metricDeltaCond$) & ($\metricDeltaRobust$) & ($\metricDeltaRobust$)
& 
&
& ($\metricDeltaRobust$) & ($\metricDeltaRobust$)
& 
&
& ($\metricDeltaRobust$) & ($\metricDeltaRobust$) 
\\
\midrule

% =========================
% Non-reasoning
% =========================

%\addlinespace[1pt]
\rowcolor{black!12}
\multicolumn{17}{l}{\textit{Non-reasoning}}\\
\bottomrule
\end{tabular}
\end{center}
%\vspace{-0.6cm}%
\end{table*}

We now evaluate our hypotheses concerning whether \LLMs can condition
their reasoning on explicitly specified formal semantic rules.
Specifically, we test whether models can (\hypothesisOne) compose
rules to obtain correct final states (\S~\ref{sec:analysis-global-rule-composition}), (\hypothesisTwo) select
appropriate rules when execution does not mutate state (\S~\ref{sec:analysis-local-rule-app}),
(\hypothesisThree) sustain such conditioning across long execution
traces (\S~\ref{sec:analysis-long-horizon-reasoning}), and (\hypothesisFour) remain faithful to supplied rules under
semantic \shifts that conflict with learned priors concerning symbol--meaning
associations.

We first introduce two task scoped metrics to test our hypotheses.
Without loss of generality, consider a model's task-scoped accuracy score
\wrapFootNoteSize{\bm{$\mathrm{Acc}^{\fFormalPlaceHold[0.5]}_{\;\fVariantPlaceHold[0.4]}$}}, potentially\footnotemark[7]\footnotetext[7]{Not all of our tasks support the entire \emph{combination} of pairings of formalizations and semantic \shifts.} realizable under the set of formalizations:%
\wrapFootNoteSize{
  \(\fSet{\fSemanticsUnSpec\!\mid\fFormalPlaceHold,\;\fVariantPlaceHold\in (\text{$\fFormalSet$}\times\fVariantSet) \cup \fSet{\bot\footnotemark[8]}}\)
}\footnotetext[8]{We denote `$\bot$' as the absence of formalization and assume that only the
  final state of program execution is computable under $\bot$ while its execution-trace is not.}.
We then define:

\metricBullet\hspace{1pt}\MyParaTwo{Semantic Conditioning~($\metricDeltaCondFootNote$)}%
\wrapFootNoteSize{\bm{$\triangleq\mathrm{Acc}^{\fFormalPlaceHold[0.5]}_{\fVarStd[0.7]}\!-\!\mathrm{Acc}_{\textbf{na}}$}} quantifies the effect of supplying semantics formalization explicitly on a model's accuracy score.
\wrapFootNoteSize{\bm{$\mathrm{Acc}_{\textbf{na}}$}} is the model's accuracy when \wrapFootNoteSize{$\fFormalPlaceHold,\;\fVariantPlaceHold\!=\!\bot$} \ie, when provided with no formalization while
\wrapFootNoteSize{\bm{$\mathrm{Acc}^{\fFormalPlaceHold[0.5]}_{\fVarStd[0.7]}$}} is the
accuracy under \StdOperationalSemantics[0.6].
A positive score is indicative of a model's ability to condition its
reasoning on explicitly supplied formal semantics.

\metricBullet\hspace{1pt}\MyParaTwo{Semantic \Shift Sensitivity~($\metricDeltaRobustFootNote$)}%
\wrapFootNoteSize{\bm{$\triangleq\mathrm{Acc}^{\fFormalPlaceHold[0.5]'}_{\;\fVariantPlaceHold[0.5]}\!-\!\mathrm{Acc}^{\fFormalPlaceHold[0.5]}_{\fVarStd[0.7]}$}} quantifies the effect of semantic \shifts on a model's accuracy score.
\wrapFootNoteSize{\bm{$\mathrm{Acc}^{\fFormalPlaceHold[0.5]'}_{\;\fVariantPlaceHold[0.5]}$}} is the model's accuracy when \wrapFootNoteSize{$\;\fVariantPlaceHold\in\!\fSet{\fVarSwap,\fVarObf}$} \ie, under semantic \shifts while
\wrapFootNoteSize{\bm{$\mathrm{Acc}^{\fFormalPlaceHold[0.5]}_{\fVarStd[0.7]}$}} is the
accuracy under standard semantics as before, with both being realized under the same semantic
framework \ie, \wrapFootNoteSize{$\forall\fFormalPlaceHold,\fFormalPlaceHold'\!\in\!\text{\fSet{\Sos,\Kos{}}};\fFormalPlaceHold\!=\!\fFormalPlaceHold'$}.
A large negative score is indicative of a model's inability to override
pretrained symbol priors when operator meanings are perturbed.

\subsection{Global Rule-Conditioned Reasoning (\hypothesisOne)}
\label{sec:analysis-global-rule-composition}

\MyPara{Motivation}%
\hypothesisOne posits that models can condition their reasoning using
explicitly supplied formal semantic rules to determine program final
states (program execution reasoning at \emph{coarser granularity}).
We introduce the \op task requiring predicting final states by
composing rule applications across control and data flow, thereby
testing \hypothesisOne by probing whether semantics guide multi-step
execution reasoning rather than acting as inert context.

\MyPara{Dataset}%
From Definition~\ref{def:program-execution} the final program state
of a program \wrapFootNoteSize{$\fProgram$},
with an initial program state $\fState{0}$, and
under a semantic formalization \wrapFootNoteSize{$\fSemanticsUnSpec$}
is
\wrapFootNoteSize{$\fExecAcc{\fProgram}{\fState{}}{\fSemanticsUnSpecSub}{\fState{0}}$}
which we use as the ground-truth for \op.
Supposing $\fSplit'$ be a subset of a \op dataset $\fSplit$ for which a model's
results are well formed, then we define accuracy over the $\fSplit$ as:
\setcounter{footnote}{4}
\renewcommand{\thefootnote}{\fnsymbol{footnote}}
\begin{accuracyBox}
\mathrm{Acc}^{\fFormalPlaceHold[0.4]\;}_{\;\fVariantPlaceHold[0.4]} \triangleq
\frac{1}{|\fSplit|}
\sum_{(\fProgram,\fState{0})\in\fSplit'}
\mathbf{1}\!\left[\fExecAcc{\fProgram}{\fState{}}{\fSemanticsUnSpecSub}{\fState{0}}=
\fState{\text{pred}}\big|_{\fSemanticsUnSpecSub, \fProgram, \fState{0}}\right]\footnotemark[2]\footnotemark[3]
\end{accuracyBox}
\footnotetext[2]{We generally drop the superscript `$\fFormalPlaceHold$' for Acc when the type of formalization is readily inferrable.}
\footnotetext[3]{
  \wrapScriptSize{
    $\fState{A}\!=\!\fState{B}\Leftrightarrow \fDomain{\fState{A}}\!=\!\fDomain{\fState{B}}\;\land\;\forall \text{x}\in\fDomain{\fState{A}}.\fState{A}(\text{x})\!=\!\fState{B}(\text{x})$
  }
}
Where $\fFormalPlaceHold,\;\fVariantPlaceHold\in\!(\fFormalSet\!\times\!\fVariantSet)\!\cup\!\fSet{\bot}$,
$\fDomain{\fState{0}}\text{=}\varnothing$. {\footnotesize$\fSplit\!\in\!\fSplitSet$} and \wrapFootNoteSize{$\fState{\text{pred}}\big|_{\fSemanticsUnSpecSub,\fProgram,\fState{0}}$} is the model's final state prediction for the program \wrapFootNoteSize{$\fProgram$}, with an initial
state $\fState{0}$, and formalization \wrapFootNoteSize{$\fSemanticsUnSpec$}.

\MyPara{Analysis}%
Table~\ref{tab:analysis-op-srp-combined} (left-side) and
Table~\ref{tab:analysis-op-complex} show the accuracy percentages for
the \humanwrit dataset, and the structurally more complex \llmtrans
and \fuzzgen datasets respectively under \emph{one-shot} prompting.
Partial correctness percentage discussed in Appendix~\ref{sec:appendix-op-analysis-metric-avg-var-prediction}.

\textbf{Does providing formal rules change global composition ($\metricDeltaCondFootNote$)?}
On the \humanwrit split (Table~\ref{tab:analysis-op-srp-combined}, left-side),
$\metricDeltaCondFootNote$ sharply separates model classes:
reasoning-oriented models gain \wrapFootNoteSize{9--16}\% points
(\eg, \dpskQwenSmall{32}, \dpskLlamaSmall), pushing \geminiSmall
to \wrapFootNoteSize{$\geq$99\%} accuracy, while non-reasoning models
lose \wrapFootNoteSize{5--25\%} points.
On the \fuzzgen split (Table~\ref{tab:analysis-op-complex}), even
frontier models show generally negative $\metricDeltaCondFootNote$
indicating that structural scale overwhelms global rule-conditioned
composition.

\textbf{Do models override pretrained symbol biases ($\metricDeltaRobust$)?}
Across both tables, \fVarSwap\ causes far larger drops than
\fVarObf—often \wrapFootNoteSize{40--70\%} points—even when standard
accuracy is high. On \humanwrit programs
(Table~\ref{tab:analysis-op-srp-combined}, left-side),
\gptfiveminiSmall and \dpskQwenSmall{32} lose
\wrapFootNoteSize{20--70\%} points under \fVarSwap, while \fVarObf\ causes
modest degradation. Because \fVarSwap\ \emph{preserve surface syntax while
changing operator meaning}, these gaps show that most models fail to
override pretrained symbol associations in favor of supplied rules.
\geminiSmall stands out with \wrapFootNoteSize{$\geq$98\%}
accuracy even under \fVarSwap.

\textbf{Structural factors limiting rule composition ($\metricProgramComplexity$).}
Moving from \humanwrit to \llmtrans and \fuzzgen programs
(Table~\ref{tab:analysis-op-complex}) induces systematic accuracy
collapses—often exceeding \wrapFootNoteSize{40\%} points—highlighting
the fragility of long-horizon rule composition under scale.
Multivariate regression (Appendix~\ref{sec:appendix-op-analysis-metric-impact})
isolates distinct stressors: control-flow depth dominates on human programs,
while data-flow and size-related metrics govern translated and fuzzed inputs,
implicating long execution traces and global state tracking as primary
bottlenecks.

\textbf{Impact of \COT prompting.}
On \humanwrit
programs (Table~\ref{tab:analysis-op-srp-combined}, left-side), chain-of-thought (\COT)
boosts non-reasoning models under standard semantics by nearly
\wrapFootNoteSize{50} points.
However, these gains vanish under \fVarSwap\ and shrink to
\wrapFootNoteSize{$\approx$40} points for \fVarObf\, indicating that \COT
aids long-horizon execution but does not overcome pretrained operator
biases.

\begin{keyfinding}
  \keyFindingBullet Reasoning-oriented models benefit substantially from access to formal
  semantics, while non-reasoning models often degrade.
  
  \keyFindingBullet \keywordMut causes far larger drops
  than \keywordObf, revealing persistent reliance on pretrained
  symbol--semantics associations rather than strict conditioning on
  provided rules.
  
  \keyFindingBullet Performance drops with program complexity—
  deep control flow, heavy data dependencies—highlighting structural limits in global rule-conditioned
  reasoning.

  \keyFindingBullet \COT prompting boosts non-reasoning
  models on standard semantics but fails under semantic swaps,
  indicating reasoning traces alone do not overcome
  pretrained operator biases.
\end{keyfinding}

\colorlet{zebra}{blue!5}

\begin{table}[t]
\begin{center}
\TableFont
\caption{\TablePredStateComplexCaption}
\begin{adjustbox}{max width=\linewidth}
\begin{threeparttable}
\setlength{\tabcolsep}{2.1pt}
\renewcommand{\arraystretch}{1}
\begin{tabular}{l S[table-format=2, table-column-width=4mm, detect-all=false]  r  r  r S[table-format=1, table-column-width=1mm, detect-all=false]  r  r  r}
\toprule
&
& \multicolumn{3}{c}{\textbf{\KosSmall-Formalization}}
&
& \multicolumn{3}{c}{\textbf{\SosSmall-Formalization}}\\
\cmidrule(lr){3-5}\cmidrule(lr){7-9}

& \textbf{$\metricAccuracyNK$}
& \textbf{$\metricAccuracyStd$} & \textbf{$\metricAccuracySwap$} & \textbf{$\metricAccuracyObf$}
&
& \textbf{$\metricAccuracyStd$} & \textbf{$\metricAccuracySwap$} & \textbf{$\metricAccuracyObf$}\\

\textbf{\UseMacro{THead-op-IMP-K-IMP-SOS-acc-models}}\tnote{*}
& 
& ($\metricDeltaCond$) & ($\metricDeltaRobust$) & ($\metricDeltaRobust$) 
&
& ($\metricDeltaCond$) & ($\metricDeltaRobust$) & ($\metricDeltaRobust$) \\
\midrule

% =========================
% LLM transformation
% =========================
\rowcolor{black!12}
\multicolumn{9}{l}{\textit{\UseMacro{THead-op-IMP-K-IMP-SOS-acc-synthetic-cpp-dataset}}}\\
\bottomrule
\end{tabular}
\begin{tablenotes}
\scriptsize
\item[] * Only the best scoring models on \op for the \humanwrit are considered.
\end{tablenotes}
\end{threeparttable}
\end{adjustbox}
\end{center}
\end{table}

\subsection{State-Free Rule-Conditioned Reasoning (\hypothesisTwo)}
\label{sec:analysis-local-rule-app}

\MyPara{Motivation}%
\hypothesisTwo targets a more elementary capability than its
predecessor: selecting the correct operational rules at individual
steps when program state does not mutate. By removing long-horizon
state propagation, this setting isolates whether models ground local
decisions in supplied formal semantics rather than surface syntax or
pretrained operator associations. We test this hypothesis via
the \srp task, asking whether semantic \shifts still disrupt rule
selection when execution does not mutate state.

\MyPara{Dataset}%
From Definition~\ref{def:program-execution} the execution trace
$\fTrace{\fProgram}$ of a program \wrapFootNoteSize{$\fProgram$},
an initial program state $\fState{0}$, under a semantic formalization \wrapFootNoteSize{$\fSemanticsUnSpec$}
is
\wrapFootNoteSize{$\fExecAcc{\fProgram}{\fTrace{}}{\fSemanticsUnSpecSub}{\fState{0}}$}.
The programs (\wrapFootNoteSize{$\fProgram$}) in the \srp dataset are constructed from those in
the \humanwrit split satisfying the invariant
\wrapFootNoteSize{
\(
\forall\fState{}'\!\in\!\fProject{1}{\fTrace{\fProgram}}\!\land\!\fState{}'\!\neq\!\fExecAcc{\fProgram}{\fState{}}{\fSemanticsUnSpecSub}{\fState{0}},\fState{}'\!=\!\fState{0}
\)
}
\ie, the initial program state is unmutated throughout program
execution barring the terminal statement-step. We use the ordered list
of semantic rules $\fConcat\fProject{2}{\fTrace{\fProgram}}$ as the ground-truth.
Supposing \wrapFootNoteSize{$\fSplit'$} is a subset of a \srp dataset \wrapFootNoteSize{$\fSplit$}
for which the model's predictions are well formed, then
we define accuracy over \wrapFootNoteSize{$\fSplit$} as:
\begin{accuracyBox}
\mathrm{Acc}^{\fFormalPlaceHold[0.4]\;}_{\;\fVariantPlaceHold[0.4]} \triangleq
\frac{1}{|\fSplit|}
\hspace{-1mm}\sum_{(\fProgram,\fState{0})\in\fSplit'}
\hspace{-3mm}\mathbf{1}\!\left[\fConcat\fProject{2}{\fExecAcc{\fProgram}{\fTrace{}}{\fSemanticsUnSpecSub}{\fState{0}}}\!=\!\fRuleList{\text{pred}}\big|_{\fSemanticsUnSpecSub, \fProgram, \fState{0}}\right]
\end{accuracyBox}
$\fFormalPlaceHold,\;\fVariantPlaceHold\;\!\in\!\fFormalSet\!\times\!\fVariantSet$
and $\fDomain{\fState{0}}$ may not be
$\varnothing$. \wrapFootNoteSize{$\fRuleList{\text{pred}}\big|_{\fSemanticsUnSpecSub,\fProgram, \fState{0}}$}
is the model predicted ordered list of semantic rules for the program
\wrapFootNoteSize{$\fProgram$}.

\MyPara{Analysis}%
Table~\ref{tab:analysis-op-srp-combined} (right-side) shows the
accuracy percentages for \srp under \emph{one-shot} prompting. Details
about \srp split construction and rule prediction failure rates can be
found in Appendix~\ref{sec:appendix-srp-dataset}
and~\ref{sec:appendix-srp-mis-predicted-rules}

\textbf{Pretrained symbol biases during local rule selection
($\metricDeltaRobust$)}
Across \srp (Table~\ref{tab:analysis-op-srp-combined}, right-side), most systems
exhibit strongly negative $\metricDeltaRobust$, implying that even when
state evolution is removed, local rule selection remains dominated by
symbol priors rather than formal definitions.
Only a narrow subset of frontier models maintain or improve accuracy
under mutation (\eg, only \geminiSmall under \SOSOperationalSemantics\
shows modest improvement), indicating that faithful local rule
conditioning under distribution shift is rare.

\textbf{Is robust local rule conditioning a general capability?}
The pattern of $\metricDeltaRobust$ reveals sharp stratification rather
than smooth scaling: reasoning-oriented models are typically more
stable than non-reasoning ones, but large drops persist even among
strong systems (\eg, \othreeSmall\ under \fVarSwap\xspace drops by \wrapFootNoteSize{$\approx$30\%}).
This heterogeneity suggests that local rule-conditioned reasoning is not
yet a broadly learned behavior across \LLM families.

\begin{keyfinding}
  \keyFindingBullet \srp isolates rule select/app w/o state tracking;
  most models nevertheless fail under \keywordMut, showing persistent
  reliance on pretrained symbol semantics rather than strict
  conditioning on provided rules.

  \keyFindingBullet Semantic \shift robust rule selection
  emerges only in a small subset of frontier systems, indicating that
  this capability is rare and not a generic consequence of scale or
  reasoning prompts (\COT has no impact).
\end{keyfinding}

\subsection{Long-Horizon Rule-Conditioned Reasoning (\hypothesisThree)}
\label{sec:analysis-long-horizon-reasoning}

\MyPara{Motivation}%
\hypothesisOne examined global outcomes, while \hypothesisTwo
targeted state-free rule selection. \hypothesisThree asks whether
\LLMs can \emph{sustain} rule-conditioned reasoning throughout full
program executions. We test this via the \etp task, which—like \op—
targets long-horizon reasoning but at a finer, execution-trace
granularity: models must generate complete sequences of semantic rule
applications and intermediate states, isolating whether they can
repeatedly re-ground their reasoning in explicit operational
definitions while maintaining the long-range dependencies induced by
loops, branching, and mutable stores.

%% Automatically generated by: table.py make_two_results_table_same_task

\colorlet{zebra}{blue!5}

\begin{table}[t]
\begin{center}
\TableFont
\caption{\TablePredTraceNonZeroCaption}
%\begin{adjustbox}{max width=\linewidth}
\begin{threeparttable}
\setlength{\tabcolsep}{4pt}
\renewcommand{\arraystretch}{1}
\begin{tabular}{l S[table-format=2, table-column-width=4mm, detect-all=false]  r  r S[table-format=1, table-column-width=1mm, detect-all=false]  S[table-format=2, table-column-width=4mm, detect-all=false]  r  r}
\toprule
& \multicolumn{3}{c}{\textbf{\KosSmall-Formalization}}
&
& \multicolumn{3}{c}{\textbf{\SosSmall-Formalization}}\\
\cmidrule(lr){2-4}\cmidrule(lr){6-8}

& \textbf{$\metricAccuracyStd$} & \textbf{$\metricAccuracySwap$} & \textbf{$\metricAccuracyObf$}
&
& \textbf{$\metricAccuracyStd$} & \textbf{$\metricAccuracySwap$} & \textbf{$\metricAccuracyObf$}\\

\textbf{\UseMacro{THead-op-IMP-K-IMP-SOS-acc-models}}\tnote{*}
& & ($\metricDeltaRobust$) & ($\metricDeltaRobust$) 
&
& & ($\metricDeltaRobust$) & ($\metricDeltaRobust$) \\
\midrule

\rowcolor{zebra}
\UseMacro{THead-etp-IMP-K-IMP-SOS-xmatch-accuracy-Qwen-QwQ-32B-da}
 & \UseMacro{res-Qwen-QwQ-32B-da-etp-uk-IMP-K-xmatch-accuracy}
 & \accdrobust{\UseMacro{res-Qwen-QwQ-32B-da-etp-mk-IMP-K-addSub_mulDiv_negateRelation-xmatch-accuracy}}
            {\UseMacro{res-Qwen-QwQ-32B-da-etp-uk-IMP-K-xmatch-accuracy}}
 & \accdrobust{\UseMacro{res-Qwen-QwQ-32B-da-etp-mk-IMP-K-unseen-xmatch-accuracy}}
            {\UseMacro{res-Qwen-QwQ-32B-da-etp-uk-IMP-K-xmatch-accuracy}}
 &
 & \UseMacro{res-Qwen-QwQ-32B-da-etp-uk-IMP-SOS-xmatch-accuracy}
 & \accdrobust{\UseMacro{res-Qwen-QwQ-32B-da-etp-mk-IMP-SOS-addSub_mulDiv_negateRelation-xmatch-accuracy}}
            {\UseMacro{res-Qwen-QwQ-32B-da-etp-uk-IMP-SOS-xmatch-accuracy}}
 & \accdrobust{\UseMacro{res-Qwen-QwQ-32B-da-etp-mk-IMP-SOS-unseen-xmatch-accuracy}}
            {\UseMacro{res-Qwen-QwQ-32B-da-etp-uk-IMP-SOS-xmatch-accuracy}}
\\
\bottomrule
\end{tabular}
\begin{tablenotes}
\scriptsize
\item[] * Only models with non-zero scores on \etp are shown.
\end{tablenotes}
\end{threeparttable}
%\end{adjustbox}
\end{center}
\end{table}

\MyPara{Dataset}
The execution trace
\wrapFootNoteSize{$\fExecAcc{\fProgram}{\fTrace{}}{\fSemanticsUnSpecSub}{\fState{0}}$}
of a program \wrapFootNoteSize{$\fProgram$} (Definition~\ref{def:program-execution}), with
an initial program state $\fState{0}$, and under a semantic formalization
\wrapFootNoteSize{$\fSemanticsUnSpec$} is used as the ground-truth in
\etp.
If \wrapFootNoteSize{$\fSplit'$} be the subset of a \etp dataset \wrapFootNoteSize{$\fSplit$}
for which the model's predictions are well formed,
we define the accuracy over a dataset \wrapFootNoteSize{$\fSplit$} for this analysis as:
\begin{accuracyBox}
\mathrm{Acc}^{\fFormalPlaceHold[0.4]\;}_{\;\fVariantPlaceHold[0.4]} \triangleq
\frac{1}{|\fSplit|}
\sum_{(\fProgram,\fState{0})\in\fSplit'}
\mathbf{1}\!\left[\fExecAcc{\fProgram}{\fTrace{}}{\fSemanticsUnSpecSub}{\fState{0}}\!=\!\fTrace{\text{pred}}\big|_{\fSemanticsUnSpecSub, \fProgram, \fState{0}}\right]
\end{accuracyBox}
$\fFormalPlaceHold,\;\fVariantPlaceHold\;\!\in\!\fFormalSet\!\times\!\fVariantSet$ and $\fDomain{\fState{0}}\!=\!\varnothing$.
{\footnotesize$\fSplit$=\humanwrit} and $\fTrace{\text{pred}}$\wrapFootNoteSize{$\big|_{\fSemanticsUnSpecSub,\fProgram,\fState{0}}$}
is the model predicted execution trace for the
program \wrapFootNoteSize{$\fProgram$}.

\MyPara{Analysis}
Table~\ref{tab:analysis-etp-non-zero} shows the accuracy scores ($\%$) for
\etp under \emph{one-shot} prompting.

\textbf{Can models sustain rule conditioning over long horizons?}
\etp sharply exposes the fragility of long-horizon rule conditioning:
only four models achieve non-zero accuracy at all, and even these remain
far from reliable.
Under \Kos, all surviving models exhibit negative
$\metricDeltaRobustFootNote$ (\eg, \qwqSmall\ and \othreeSmall\ lose
\wrapFootNoteSize{2--16} points under \keywordMut), indicating that
symbol–meaning conflicts rapidly derail multi-step rule application.

\textbf{Is long-horizon robustness a rare capability?}
The distribution of $\metricDeltaRobustFootNote$ is highly skewed: most models
collapse to zero accuracy before robustness can even be meaningfully
measured, while the few remaining systems show sharply divergent
behavior.
For instance, \othreeSmall\ and \gptfiveminiSmall\ degrade under semantic
swaps, whereas \geminiSmall\ improves, indicating that the ability to
sustain rule conditioning across dozens of steps is \emph{not} a smooth
function of scale or reasoning prompts, but instead appears only in a
small subset of frontier models.

\begin{keyfinding}
  \keyFindingBullet Sustaining rule-conditioned
  reasoning over long execution horizons is extremely brittle: most
  models fail entirely once state tracking and repeated rule application
  are required.

  \keyFindingBullet Only \gemini\ exhibits consistently positive
  $\metricDeltaRobust$ under \SosKeyFinding, indicating that mutation-robust
  long-horizon rule conditioning is a rare and specialized capability.
\end{keyfinding}

\section{Related Work}

\subsection{Code Reasoning and Execution Benchmarks}

Recent benchmarks evaluate \LLMs' ability to reason about program execution
and behavior (CRUXEval~\citep{Cruxeval}, CRUXEval-X~\citep{xu2025cruxeval},
LiveCodeBench~\citep{Livecodebench},
BigCodeBench~\citep{Bigcodebench}, REval~\citep{chen_reval},
CoCoNUT~\citep{coconut}, CodeMind~\citep{liu_codemind},
SURGE~\citep{lyu2025surge}, and \LLMs as code
executors~\citep{Wang2024largelanguagemodels}), trace-trained models
(CWM~\citep{meta2025cwm}), and code-reasoning generalization
studies~\citep{EvaluatingYangETAL25}.  These works evaluate end-to-end
inputs/outputs or traces under \emph{fixed} language semantics;
\dataset instead supplies formal inference rules and uses execution
as a controlled lens for whether models condition step-level reasoning
on those rules.

\subsection{Execution-Aware Training}

A growing body of work argues that exposing \LLMs to program
executions improves downstream performance, including execution-guided
synthesis~\citep{chen2018execution}, NExT~\citep{NiETAL24Next},
SemCoder~\citep{ding2024semcoder}, TRACED~\citep{ding2024traced}, and
CodeI/O~\citep{li2025codeio}.  
\citet{jin2024emergent} further report that
representations of formal trace semantics emerge in transformer hidden
states under next-token training.  The implicit hypothesis is that
models internalize program semantics from such training.  \dataset
provides the missing diagnostic by directly supplying formal semantic
rules and measuring whether models condition their reasoning on those
rules.
 
\subsection{Perturbing Programs vs.\ Perturbing Semantics}

EquiBench~\citep{EquibenchWeiETAL25},
SeqCoBench~\citep{maveli2025seqcobench}, SPAT~\citep{yu2022dataaug},
CodeARC~\citep{CodeARCWeiETAL25}, and \citet{orvalho2025large} mutate
\emph{programs} under semantics-preserving transformations to test whether 
models track
underlying behavior across syntactic variants.  We invert the setup:
programs remain syntactically identical while the externally supplied
formal semantics are altered, isolating
reliance on pretrained symbol--semantics associations from sensitivity
to surface form.  K-framework formalizations of
C~\citep{ellison_rosu_c_popl, C_star_Hathhorn},
Java~\citep{K_Java_Bogdanas}, and Python~\citep{rosu_chen_python_cpp}
make this methodology directly extensible to richer languages.

\subsection{Rule Following and Conflicts with Priors}

RuleBreakers~\citep{chan2025rulebreakers} and
\citet{sun2024beyondinstruction} probe whether \LLMs follow
natural-language inferential rules; in NLP more broadly,
CheckList~\citep{ribeiro2020checklist}, Contrast
Sets~\citep{gardner2020contrast}, HANS~\citep{mccoy2019hans}, NLI
stress tests~\citep{naik2018stress}, and semantic sensitivity
probes~\citep{arakelyan2024semantic} expose heuristic shortcuts via
input perturbations, paralleling texture-bias diagnostics in
vision~\citep{geirhos2019imagenet}.  \dataset transposes this question
to \emph{formal} rule following: complete operational semantics are
supplied and we test whether
reasoning conditions on those rules when they redefine standard
operator meaning---a conflict that arises in practice with operator
overloading, DSLs, and proof assistants.

\section{Conclusion}

We introduced \dataset, a semantics-driven benchmark for studying
whether large language models (\LLMs) condition their reasoning on explicit
formal rules rather than pretrained syntactic priors. Using program
execution as a controlled probe, a programming language with two
semantic formalisms and \shifts, we isolate four capabilities: global
rule composition, state-free rule selection, long-horizon
conditioning, and robustness to semantic shift.

Across \NumModels frontier models, performance drops sharply under
semantic \shifts and long horizons despite high standard-semantics
accuracy; only a small subset shows robustness to novel rules. These
results position inferential rule conditioning as a largely unsolved
capability axis motivating models to adapt to externally specified
formal systems rather than entrenched lexical associations.

\section*{Acknowledgments}

We thank Cheng Ding, Ivan Grigorik, Michael Y. Levin, Yan Levin,
Tong-Nong Lin, Karl Palmskog, Zijian Yi, Zhiqiang Zang, Linghan
Zhong and the anonymous reviewers for helpful feedback and discussions.

Computational resources were provided by the Texas Advanced
Computing Center at The University of Texas at Austin\footnotemark[2].
This work was supported in part by the U.S. National Science
Foundation (NSF) Nos.~CCF-2217696, CCF-2313027,
CCF-2403036, CCF-2421782; the NSF--Simons AI Institute for Cosmic
Origins\footnotemark[3] funded by NSF
award AST-2421782; the Simons Foundation (MPS-AI-00010515);
and a sponsored research award by
Cisco Research.

The views expressed are those of the authors and do not necessarily
reflect those of sponsors.

\footnotetext[2]{TACC: \url{http://www.tacc.utexas.edu}}
\footnotetext[3]{CosmicAI: \url{https://www.cosmicai.org}}
\section*{Impact Statement}

This paper introduces the first large-scale study of whether large
language models can condition their reasoning on explicitly provided
formal semantics, using program execution as a canonical setting for
investigating this capability.
We present new evaluation tasks and datasets that probe models'
ability to select and compose inference rules across full executions
and under controlled semantic perturbations, establishing a foundation
for systematic study of rule-grounded reasoning in programming
languages.

Overall, this paper positions inferential rule conditioning as a new
capability axis for evaluating learning-based systems for programming
languages, with the long-term goal of building models that reason more
faithfully about formal specifications.

\bibliography{bib}
\bibliographystyle{icml2026}

% math mode paddings
\setlength{\abovedisplayskip}{2pt}
\setlength{\belowdisplayskip}{2pt}
\setlength{\abovedisplayshortskip}{2pt}
\setlength{\belowdisplayshortskip}{2pt}

\newpage
\onecolumn
\appendix

\section*{Appendix}
\addcontentsline{toc}{section}{Appendix}

\begingroup
\setlength{\parindent}{0pt}
\setlength{\parskip}{8pt}

\noindent\hyperref[sec:appendix-operator-overloading]{Operator Overloading Conflicts Learned Priors}\dotfill\pageref{sec:appendix-operator-overloading}\par
\noindent\hyperref[sec:appendix-imp-formalization]{\lang\ Formalization}\dotfill\pageref{sec:appendix-imp-formalization}\par
\noindent\hyperref[sec:appendix-imp-example]{\lang Program Example}\dotfill\pageref{sec:appendix-imp-example}\par
\noindent\hyperref[sec:appendix-metric-dist]{Code-Complexity Distributions}\dotfill\pageref{sec:appendix-metric-dist}\par
\noindent\hyperref[sec:appendix-prompts]{Experiments Details}\dotfill\pageref{sec:appendix-prompts}\par
\noindent\hyperref[app:keyword-obf-tokenization]{Is Keyword Obfuscation a Tokenization Artifact?}\dotfill\pageref{app:keyword-obf-tokenization}\par
\noindent\hyperref[sec:appendix-task-extended]{Task Extended Analysis}\dotfill\pageref{sec:appendix-task-extended}\par
\noindent\hyperref[sec::appendix-external-assets]{Use of External Assets}\dotfill\pageref{sec::appendix-external-assets}\par
\endgroup
\clearpage

\section{Operator Overloading Conflicts Learned Priors}
\label{sec:appendix-operator-overloading}

\begin{figure}[h]
    \centering
    \begin{subfigure}{0.45\textwidth}
        \centering
\begin{lstlisting}[language=java-pretty]
class X:
    def __init__(self, v):
        self.v = v
    def __add__(self, other):
        return X(self.v - other.v)
print((X(10) + X(3)).v)  # 7
\end{lstlisting}
        \caption{Python (redefining existing operator).}
    \end{subfigure}
    \begin{subfigure}{0.45\textwidth}
        \centering
\begin{lstlisting}[language=java-pretty]
object Main:
  extension (x: Int)
    def ~>(y: Int): Int = x - y

  @main def run(): Unit =
    println(10 ~> 3) // 7
\end{lstlisting}
        \caption{Scala (defining new operator).}
    \end{subfigure}
    \caption{Operator Overloading}
\end{figure}

There are real-world situations where operators and other language constructs 
can have very different semantic meaning relative to that assumed during 
training/pre-training. Very popular languages such as C++, Haskell, Julia, Python, Scala, Swift, 
\etc, support operator overloading where new semantics can be assigned 
to existing operators (+, -, *, \etc) or to completely new symbols (Haskell, Julia, Scala), 
previously unencountered during training.

\section{\lang Formalization}
\label{sec:appendix-imp-formalization}

Here we describe the syntax and semantics of \IMP used in all our
experiments.

\subsection{\lang Syntax Description}
\label{sec:appendix-imp-ebnf}

\begin{figure}[h]
\begin{lstlisting}[language=bnf-grammar]
<program>   ::= <stmt_list>
<stmt_list> ::= (<stmt> ';')*
<stmt>      ::= 'int' <id>
            | <id> '=' <aexp> 
            | 'if' '(' <bexp> ')' '{' <stmt_list> '}' 'else' '{' <stmt_list> '}'
            | 'while' '(' <bexp> ')' '{' <stmt_list> '}'
            | 'loop' '(' <bexp> ')' '{' <stmt_list> '}'
            | 'halt'
            | 'continue'
            | 'break'
            | 'LE'
<aexp>      ::= <id>
            | <literal>
            | '(' <aexp>? <mathop> <aexp> ')'
<bexp>      ::= '(' <bool> ')'
            | '(' <aexp> <relop> <aexp> ')'
            | '(' <lognot> <bexp> ')'
            | '(' <bexp> <logicalop> <bexp> ')'
<bool>      ::= 'true' | 'false'
<mathop>    ::= '+' | '-' | '*' | '/' | '%'
<relop>     ::= '<' | '<=' | '>' | '>=' | '==' | '!='
<lognot>    ::= '!'
<logicalop> ::= '&&' | '||'
<id>        ::= <letter> (<letter> | <digit>)*
<literal>   ::= <digit>+
<letter>    ::= 'a' | 'b' | 'c' | 'd' | 'e' | 'f' | 'g' | 'h' | 'i' | 'j'
            | 'k' | 'l' | 'm' | 'n' | 'o' | 'p' | 'q' | 'r' | 's' | 't'
            | 'u' | 'v' | 'w' | 'x' | 'y' | 'z'
            | 'A' | 'B' | 'C' | 'D' | 'E' | 'F' | 'G' | 'H' | 'I' | 'J'
            | 'K' | 'L' | 'M' | 'N' | 'O' | 'P' | 'Q' | 'R' | 'S' | 'T'
            | 'U' | 'V' | 'W' | 'X' | 'Y' | 'Z'
<digit>     ::= '0' | '1' | '2' | '3' | '4' | '5' | '6' | '7' | '8' | '9'
\end{lstlisting}
  \caption{\FigIMPEBNFCaption}
%  \vspace{-0.7cm}%
\end{figure}

The \IMP syntax used in all our experiments is given in \ebnf in
Figure~\ref{figure:appendix-imp-syntax}. The terminals are shown in
red while the non-terminals are shown in blue.

\subsection{Small-step Operational Semantics (\Sos) Rules for \IMP}
\label{sec:appendix:imp-rules}

\begin{wraptable}{r}{0.48\textwidth}%
\centering%
\TableFont%
\vspace{-0.25cm}%
\caption{Metavariables used in the \Sos formalization of \IMP.\label{tab:table-metavariable}}%
\begin{tabular}{@{}lll@{}}%
\toprule
\textbf{Meta-var} & \textbf{Sort} & \textbf{Ranges over / Domain}\\
\midrule
\texttt{x}  & \texttt{id}       & Identifiers (program variable names)\\
\texttt{v}  & \texttt{literal}  & Integer literals\\
\texttt{q}  & \texttt{bool}     & Boolean literals\\
\texttt{a}  & \texttt{aexp}     & Integer expressions\\
\texttt{b}  & \texttt{bexp}     & Boolean expressions\\
\texttt{s}  & \texttt{stmt}     & Statements of the language\\
\texttt{SL} & \texttt{stmt\_list} & Finite statement lists $(\texttt{SL} ::= \epsilon \mid \texttt{s :: SL'})$\\
\bottomrule
\end{tabular}
\end{wraptable}%
We formalize \IMP using a small-step structural operational semantics (\Sos).
We use two types of configurations: expression configurations 
\(
  \langle \texttt{expr},\, \sigma,\, \chi \rangle,
\)
and statement configurations:
\(
  \langle \texttt{stmt},\, \sigma,\, \chi \rangle,
\)
where $\sigma : \texttt{id} \mapsto \texttt{literal}$ is the program store mapping
identifiers to values, and $\chi$ is a last-in, first-out \emph{control stack}
of loop headers that records the dynamic nesting of currently active loops:
\(
  \chi \;::=\; \epsilon \;\mid\; \texttt{s} :: \chi'.
\)
The top of $\chi$ is the innermost executing loop.

We use standard metavariables $\texttt{x,v,q,a,b,s,SL}$ with their sorts summarized in
Table~\ref{tab:table-metavariable}. For example, $\texttt{a}$ ranges over arithmetic
expressions, so rules mentioning $\texttt{a1,a2,\ldots}$ concern arithmetic evaluation.
Auxiliary metafunctions (\texttt{push},
\texttt{pop}, \texttt{top}) for manipulating the control stack $\chi$ and concatenating 
(\,$\mathbin{++}$\,) statement lists $SL$
are given in Table~\ref{tab:table-metafunctions}.
\begin{table}[h!]
\centering%
\TableFont%
\caption{Metafunctions for control stack and statement-list concatenation.\label{tab:table-metafunctions}}
\begin{tabular}{lll}
\toprule
\textbf{Function} & \textbf{Signature} & \textbf{Definition}\\
\midrule
\texttt{push} &
$ \mathrm{stmt} \times \mathrm{Stack} \to \mathrm{Stack} $ &
$ \mathrm{push}(s,\chi) \triangleq s :: \chi $\\

\texttt{pop} &
$ \mathrm{Stack}_{\neq \epsilon} \to \mathrm{Stack} $ &
$ \mathrm{pop}(s :: \chi) \triangleq \chi $\\

\texttt{top} &
$ \mathrm{Stack} \to \mathrm{stmt} \cup \{\epsilon\} $ &
$ \mathrm{top}(\chi) \triangleq \begin{cases} \epsilon & \text{if } \chi = \epsilon, \\ s & \text{if } \chi = s :: \chi' \end{cases} $\\

$\mathbin{++}$ &
$ \mathrm{stmt\_list} \times \mathrm{stmt\_list} \to \mathrm{stmt\_list} $ &
$
SL1 \mathbin{++} SL2 \triangleq
\begin{cases}
SL2 & \text{if } SL1 = \epsilon,\\
s :: (SL1' \mathbin{++} SL2) & \text{if } SL1 = s :: SL1'.
\end{cases}
$\\
\bottomrule
\end{tabular}
\end{table}

Program execution proceeds by repeatedly applying the transition relation
$\to$ to expression configurations and $\rightrightarrows$ to statement configurations, starting from
$\langle \texttt{SL}, \sigma, \chi \rangle$, where \texttt{SL} is the program’s
statement list, until a terminal configuration is reached. We treat
$\langle \epsilon, \sigma, \chi \rangle$,
$\langle \texttt{halt}, \sigma, \chi \rangle$, and
$\langle \texttt{ERROR}, \sigma, \chi \rangle$
statement configurations as terminal configurations.

The complete set of small-step \Sos\ rules defining the semantics of \IMP\
is given in Table~\ref{tab:imp-rules}.
\begin{scriptsize}%
\begin{longtable}{|p{0.8cm}|p{11cm}|p{4cm}|}
 \caption{Small-step \Sos rules used to formalize \IMP. \label{tab:imp-rules}}\\
 \toprule%
 \textbf{Rule} & \textbf{Formalization} & \textbf{Description}\\
 \midrule%
 Rule 1 & \begin{mathpar}\inferrule{\sigma(\texttt{x}) = \texttt{v}}{\langle \texttt{x},\sigma,\chi\rangle \to \texttt{v}}\end{mathpar} & Variable lookup returns value.\\\hline
 Rule 2 & \begin{mathpar}\inferrule{\sigma(\texttt{x}) = \bot}{\langle \texttt{x},\sigma,\chi\rangle \to \langle \texttt{ERROR},\sigma,\chi\rangle }\end{mathpar} & Read of undefined variable errors.\\\hline
 Rule 3 & \begin{mathpar}\inferrule{ }{\langle \texttt{int x :: SL},\sigma,\chi\rangle \rightrightarrows \langle \texttt{SL},\sigma[\texttt{x} \mapsto 0],\chi\rangle }\end{mathpar} & Declared int variable initialized to 0.\\\hline
 Rule 4 & \begin{mathpar}\inferrule{\langle \texttt{a},\sigma,\chi\rangle \to \langle \texttt{a'},\sigma,\chi\rangle}{\langle \texttt{x := a :: SL},\sigma,\chi\rangle \rightrightarrows \langle \texttt{x := a' :: SL},\sigma,\chi\rangle }\end{mathpar} & Assignment expression steps.\\\hline
 Rule 5 & \begin{mathpar}\inferrule{\sigma(\texttt{x}) \neq \bot}{\langle \texttt{x := v :: SL},\sigma,\chi\rangle  \rightrightarrows \langle \texttt{SL},\sigma[\texttt{x} \mapsto \texttt{v}],\chi\rangle }\end{mathpar} & Writeback to existing variable.\\\hline
 Rule 6 & \begin{mathpar}\inferrule{\sigma(\texttt{x}) = \bot}{\langle \texttt{x := v :: SL},\sigma,\chi\rangle \rightrightarrows \langle \texttt{ERROR},\sigma,\chi\rangle }\end{mathpar} & Assign to undefined variable errors.\\
\end{longtable}%
\end{scriptsize}%

\clearpage
\section{\lang Program Example}
\label{sec:appendix-imp-example}

In this section, we describe the collection of \IMP programs for:
(1)~the \humanwrit, and (2)~the \fuzzgen datasets and provide
examples.

\subsection{\humanwrit Dataset}
\label{sec:appendix-imp-humanwrit-example}
\begin{figure}[h]
    \begin{subcaptionbox}{
        \UseMacro{cProgramExample}
        \label{fig:c-example}}[0.5\linewidth]{
\begin{lstlisting}[language=java-pretty]
int sumEven(int l, int r)
{
    int sum = 0;
    for (int i = l; i <= r; i++)
    {
        if (i % 2 == 0)
        {
            sum += i;
        }
    }
    return sum;
}
\end{lstlisting}
    }
    \end{subcaptionbox}
    \hfill
    \begin{subcaptionbox}{
        \UseMacro{impProgramExample}
        \label{fig:imp-example}}[0.42\linewidth]{
\begin{lstlisting}[language=java-pretty]
int sum;
int i;
int l;
int r;
l = 3;
r = 8;
i = l;
while(i <= r)
{
    if((i % 2) == 0)
    {
        sum = (sum + i);
    }
    else
    {
    
    };
    i = (i + 1);
};
\end{lstlisting}
    }
    \end{subcaptionbox}
\caption{\UseMacro{DataExample}}
\label{fig:appendix:example}
\end{figure}

In Figure~\ref{fig:appendix:example}, we show an example \CodeIn{C++}
solution to a problem from the BabelCode MBPP benchmark
(Figure~\ref{fig:c-example}) and its corresponding \IMP program
re-written by us (Figure~\ref{fig:imp-example}). To convert the
\CodeIn{C++} program into an \IMP program, we remove the function
definitions (\eg, \CodeIn{sumEven}), while keeping the body of the
function.
Unsupported syntactic constructs are either re-written (\eg, replacing
the \CodeIn{for} loop with a \CodeIn{while} loop) or removed (\eg,
removing the return statement).  One public test case is adopted as
the program input, and its output is used to verify correctness. In
this example, \CodeIn{l} is assigned to 3 and \CodeIn{r} is assigned
to 8, the test oracle 18 is used to verify the final-state of
\CodeIn{sum} after program execution.

The code-complexity profile of the \IMP program in
Figure~\ref{fig:imp-example} is: control-flow complexity (\metricCC=
3, \metricMaxnestif= 1, \metricMaxnestwhile= 1, \metricMaxtakenif= 1,
\metricMaxtakenwhile= 1), data-flow complexity (\metricDepdeg= 12,
\metricNumAssign= 12), and program-size complexity (\metricLoc= 19,
\metricVol= 294, \metricVocab= 23, \metricTracelen= 29).

\subsection{\fuzzgen Dataset}
\label{sec:appendix-imp-fuzzgen-example}

The \fuzzgen dataset is constructed using a semantic aware grammar
based fuzzer with knobs for: (1)~the generation probabilities of
different statements, (2)~the maximum nesting depth of the program
(nested loops and conditionals), (3)~the maximum and the minimum
number of statements to generate per block, (4)~the maximum number of
terms and variable terms in arithmetic expressions, (5)~the maximum
number of terms in boolean expressions (relational and logical), and
(6)~the maximum and the minimum number of variable declarations in a
program. We use the settings as shown in
Table~\ref{tab:appendix-fuzzer-settings}.

The fuzzer starts by randomly sampling an integer from the range
defined by the minimum and maximum number of variable
declarations. This integer specifies the number of variables to be
declared and used for the \IMP program being generated. The fuzzer
next samples alphabets from the set \{a-z\} and \{A-Z\} until the
required number of unique alphabets to use as variables is
obtained. Declaration statments are then generated to declare these
variables.

Following this, one assignment statement is generated per declared
variable to assign it with a randomly generated arithmetic
expression. The arithmetic expression itself is generated using the
pool of declared variables and integer constants (sampled from the set
\{0-9\}).

The fuzzer next generates statements from the set \{Assignment, While,
If, Break, Continue, Halt\} in accordance with the statement
probabilities given in Table~\ref{tab:appendix-fuzzer-settings}. No
more than three statements are generated per block. These
probabilities are used until the generation block depth reaches the
specified minimum block depth (5). Beyond this, the statement
probabilities are cosine-tapered to decrease the probabilities of
generating \CodeIn{while} and \CodeIn{if-else} statements. For
generation processes where the block depth reaches the maximum
specified block depth (10), the probabilities of further generating
\CodeIn{while} and \CodeIn{if-else} is reduced to zero.

\begin{wraptable}{r}{0.4\textwidth}%
  %\vspace{-0.4cm}%
  \centering%
  \TableFont%%
  \caption{\TableFuzzerSettingsCaption}%
  \begin{tabular}{lc}
    \toprule
    \textbf{Knob} & \textbf{Value}\\
    \midrule
    \multicolumn{2}{c}{\textbf{Structural limits}} \\
    \midrule
    Minimum number of statements per block     & 1\\
    Maximum number of statements per block     & 3\\
    Minimum block depth         & 5 \\
    Maximum block depth         & 10\\
    Minimum number of variables & 5 \\
    Maximum number of variables & 10\\
    \midrule
    \multicolumn{2}{c}{\textbf{Statement generation probabilities}} \\
    \midrule
    Assignment                   & 0.4\\
    While                        & 0.3\\
    If                           & 0.2\\
    Break                        & 0.09\\
    Continue                     & 0.005\\
    Halt                         & 0.005\\
    \midrule
    \multicolumn{2}{c}{\textbf{Expression limits}} \\
    \midrule
    Maximum number of terms in arithmetic expr & 6\\
    Maximum number of variable terms in arithmetic expr & 3\\
    Maximum number of terms in boolean expr    & 4\\
    \bottomrule
  \end{tabular}
\end{wraptable}
To ensure high probability in termination of loops, the fuzzer
generates one new variable (prefixed with \CodeIn{ble}) per loop. A
monotone update type (incrementing or decrementing) is chosen for this
variable each with a 50\% probability of being chosen. The bounds,
initial (before iteration) and expected final (after loop termination)
values are then chosen from the range [-20,20] and the size of the
update per iteration from the range [1 step, (final / 3) step]. The
variable monotone update statement is inserted towards the end of the
loop body and the bound is conjoined with the loop predicate. This
prevents infinite loops. The declaration and assignment statements for
these new generated variables is inserted right after the assignment
statements for the intially chosen variables.

The fuzzer can be used to generate extremely complex \IMP programs (as
measured by the code-complexity metrics introduced earlier) with high
probability of normal program
termination. Figure~\ref{fig:appendix:imp-fuzzgen-example} shows an
example \IMP program from the \fuzzgen
dataset that was generated using our fuzzer. Its code-complexity
metric profile is: control-flow complexity (\metricCC= 62,
\metricMaxnestif= 5, \metricMaxnestwhile= 6, \metricMaxtakenif= 3,
\metricMaxtakenwhile= 5), data-flow complexity (\metricDepdeg= 2603,
\metricNumAssign= 86), and program-size complexity (\metricLoc= 492,
\metricVol= 37140, \metricVocab= 91, \metricTracelen= 249).  This
shows that out of the maximum loop nesting depth six
(\metricMaxnestwhile) present in the program, the execution reaches a
maximum loop nesting depth of five (\metricMaxtakenwhile) implying
that the execution reached a loop contining four outer loops.

Figure~\ref{fig:appendix:imp-fuzzgen-example} shows one of the
programs from the \fuzzgen dataset that the \gemini model was
successful on in the \op task.

\section{Code-Complexity Distributions}
\label{sec:appendix-metric-dist}
\begin{figure}[h]
\centering
\begin{tikzpicture}
  \node[anchor=south west, inner sep=0] (img)
        {\includegraphics[scale=0.39]{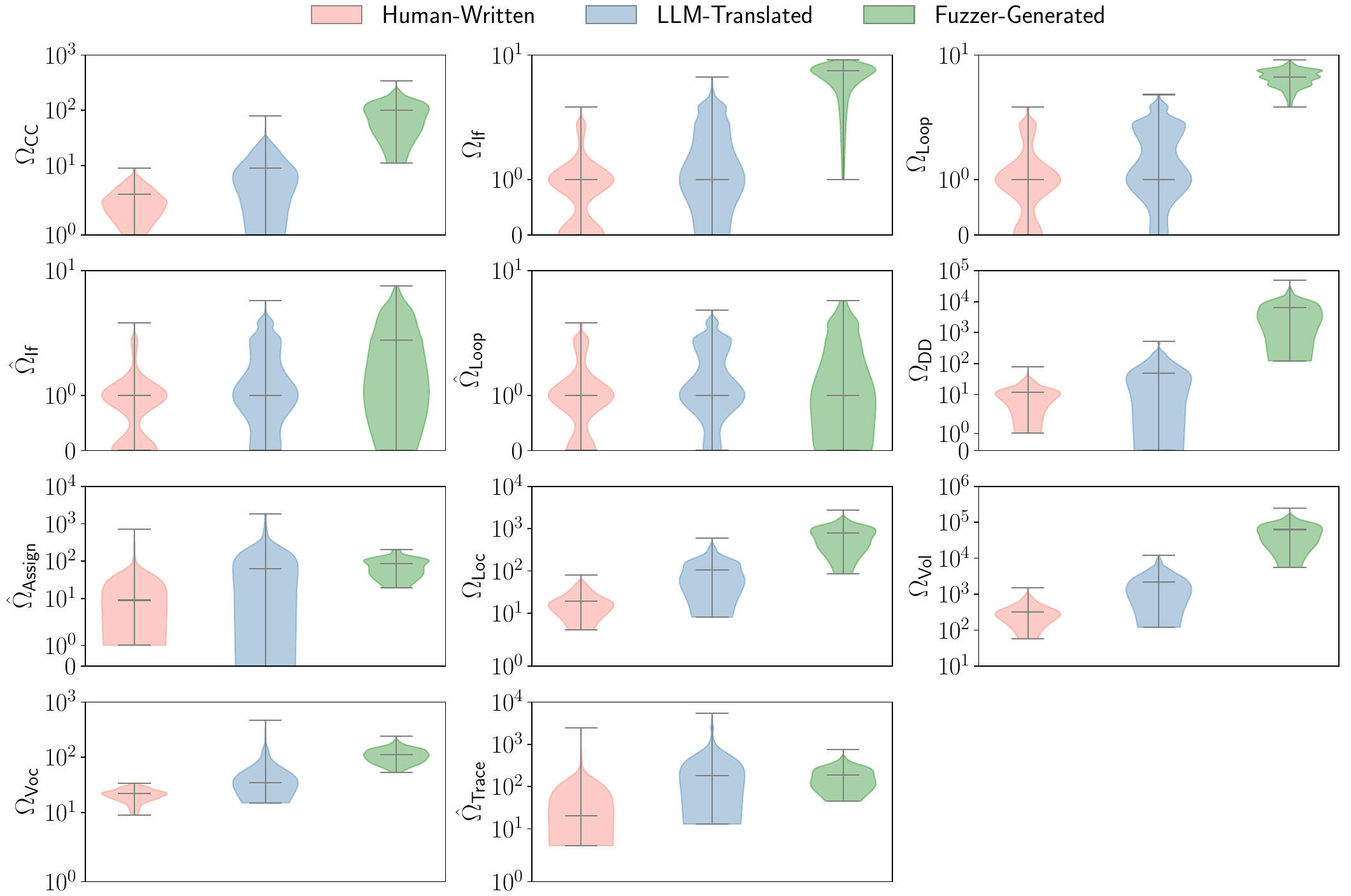}};
  \begin{scope}[x={(img.south east)}, y={(img.north west)}]
    \fill[white] (0.7,0.00) rectangle (1.00,0.23);
  \end{scope}
\end{tikzpicture}
\caption{\FigDatasetMetrics}
\end{figure}

The distributions of the code-complexity metrics used to characterize
the control-flow, data-flow, and the program size complexity are given
in Figure~\ref{figure:appendix-dataset-stats}. We mark the median and
the extremas for each distribution. We see that the
median \metricMaxnestif and
\metricMaxnestwhile is similar for the \humanwrit and the \llmtrans datasets,
whereas for every other metric, the \llmtrans has slightly higher
median values than \humanwrit and thus more complex
programs. The \fuzzgen dataset on the other hand has median values
significantly higher for every metric except \metricTracelen
and \metricNumAssign, than the other two datasets. This implies that
programs in the \fuzzgen and the \llmtrans datasets run for roughly
the same number of execution steps (measured as per the \Sos semantics)
but the programs in the former are significantly more complex than
those in the latter.

\section{Experiments Details}
\label{sec:appendix-prompts}

\subsection{Parameters}
We use the default temperature settings for reasoning models by not 
specifying a specific temperature. For other non-reasoning models, we set the temperature to
zero. All models are evaluated under one-shot setting.

\subsection{Compute Resources}
The experiments on open-weight models with fewer than 70 billion
parameters are conducted on a single compute node equipped with one
NVIDIA H200 GPU (96 GB memory), an NVIDIA Grace CPU @ 3.1 GHz with 72
cores, and 116 GB LPDDR5 memory.  For experiments involving
70B-parameter models, we use four compute nodes.

\subsection{Prompts}
\begin{prompt}
  [title={\thetcbcounter{} Prompt for \op task.}]
  % \promptsubsection{\pep}\\

\promptsubsection{\rmfamily No-semantics}\\
You are an interpreter for my language called \{language\}.\\
\\
    Here is the \{language\} program\\
    \hspace*{1cm}\{program\}\\
\\
\promptsubsection{\rmfamily \Sos}\\
You are an interpreter for a language called \{language\}. I will
describe the syntax for \{language\} in EBNF and its semantics using
small-step operational semantics. You will use this to execute a
\{language\} program. You will only use the rules described in the
semantics I provide. Assume all the rules in the semantics I give are
correct. A program has finished execution when one of the terminal
configurations $\langle\epsilon,\sigma,\chi\rangle$,$\langle$\{HALT\}$,\sigma,\chi\rangle$, $\langle$\{ERROR\}$,\sigma,\chi\rangle$ is reached.\\
\\  
    Here is the syntax of \{language\} in EBNF\\
    \hspace*{1cm}\{syntax\}\\
\\  
    Here is the small-step operational semantics of \{language\}\\
    \hspace*{1cm}\{semantics\}\\
\\  
    Here is the \{language\} program\\
    \hspace*{1cm}\{program\}\\
\\
\promptsubsection{\rmfamily \Kos-semantics}\\
You are an interpreter for a language called \{language\}. I will
describe the syntax and the semantics of the language using the
K-framework. You will use this to execute a \{language\} program. You
will only use the rules described in the semantics I provide. Assume
all the rules in the semantics I give are correct.
\\
    Here is the K-framework formalization of \{language\}\\
    \hspace*{1cm}\{semantics\}\\
\\
    Here is the \{language\} program\\
    \hspace*{1cm}\{program\}\\
\\
\\
\#\# TASK: predict the values of all the declared variables after executing the above program.\\
- If you think the program will never terminate, answer with the special word '\#\#timeout\#\#':\\
\\
    \hspace*{1cm}<answer>\#\#timeout\#\#</answer>\\
\\
- If you believe the program has an error or has undefined behavior, answer with the special word '\#\#error\#\#':\\
\\
    \hspace*{1cm}<answer>\#\#error\#\#</answer>\\
\\
- Otherwise, provide the predicted values of all the declared variables in the following format:\\   
\\
    \hspace*{1cm}<answer>[Your answer]</answer>\\
\\
Here is one example:\\
\\   
** Program **\\
int a;\\
int b;\\
int ans;\\
int c;\\
a \{ASSIGN\_OP\} 10;\\
b \{ASSIGN\_OP\} 23;\\
c \{ASSIGN\_OP\} 12;\\
ans \{ASSIGN\_OP\} a \{ADD\_OP\} b;\\
\\
\\
The final expected output is:\\
<answer>\\
\hspace*{0.3cm}<a>10</a>\\
\hspace*{0.3cm}<b>23</b>\\
\hspace*{0.3cm}<c>12</c>\\
\hspace*{0.3cm}<ans>33</ans>\\
</answer>\\
\\
\\
\promptsubsection{\rmfamily Non-\COT}Only write the answer. You **MUST** wrap your prediction with `<answer>' tags.\\
\promptsubsection{\rmfamily \COT} Explain your reasoning step-by-step **before** answering. Wrap your reasoning in `<reason>' tags.
Note that you **MUST** wrap your reasoning steps with `<reason>' tags and the prediction with `<answer>' tags.
\end{prompt}

\begin{prompt}
  [title={\thetcbcounter{} Prompt for \srp task.}]
  % \promptsubsection{\pep}\\
\promptsubsection{\rmfamily \Sos}\\
You are an interpreter for a language called \{language\}. I will
describe the syntax for \{language\} in EBNF and its semantics using
small-step operational semantics. You will use this to execute a
\{language\} program. You will only use the rules described in the
semantics I provide. Assume all the rules in the semantics I give are
correct. A program has finished execution when one of the terminal
configurations $\langle\epsilon,\sigma,\chi\rangle$,$\langle$\{HALT\}$,\sigma,\chi\rangle$, $\langle$\{ERROR\}$,\sigma,\chi\rangle$ is reached.\\
\\  
    Here is the syntax of \{language\} in EBNF\\
    \hspace*{1cm}\{syntax\}\\
\\  
    Here is the small-step operational semantics of \{language\}\\
    \hspace*{1cm}\{semantics\}\\
\\  
    Here is the \{language\} program\\
    \hspace*{1cm}\{program\}\\
\\
\#\# TASK:\\
For each question below, you'll be given:\\
1. A program\\
2. The program state ($\sigma$) (variable values) before executing the program\\
3. The control stack ($\chi$) before executing the program\\
\\
Assume that all necessary variables have been declared and have the values as
indicated in the provided program state.
\\
You must:\\
- Correctly identify and apply the small-step operational semantic rules required to evaluate the program to completion\\
- List them in the correct order of application\\
\\
A program is executed completely when its evaluation reaches one of the terminal
configurations $\langle\epsilon,\sigma,\chi\rangle$,$\langle$\{HALT\}$,\sigma,\chi\rangle$, $\langle$\{ERROR\}$,\sigma,\chi\rangle$.\\
\\
\\
Here is one example:\\
** Program:**\\
\{WHILE\} (n \{LTEQ\_OP\} 0)\\
\{\{\\
\hspace*{0.8cm}\{HALT\};\\
\}\};\\
\\
**Program state($\sigma$) before execution:**\\
\{\{'n': 100, 'sum': 0\}\}\\
\\
**Control stack($\chi$) before execution:**\\
 $\epsilon$\\
\\
\\
This is the sequence of steps:\\
1. First, we transform the \{WHILE\} into \{LOOP\} using **Rule 67**.\\
2. Reduce the loop predicate using **Rule 68**.\\
3. The loop predicate is a \{LTEQ\_OP\} operator which triggers **Rule 32** to first reduce the left-hand side 'n' to a literal using **Rule 1**.\\
4. The right-hand side is already a literal and since '100' is not less-than or equal to '0'. We use **Rule 35** to evaluate this operation to 'false'.\\
5. Since the loop predicate is 'false', we use **Rule 69** to terminate the loop.\\
6. Since there are no more statements left, we have reached the terminal configuration $\langle\epsilon,\sigma,\chi\rangle$ and the program evaluation terminates.\\
\\
Therefore, the final answer is:\\
<ans>\\
\hspace*{0.3cm}<answer id="1">\\
\hspace*{0.6cm}<rule>67</rule>\\
\hspace*{0.6cm}<rule>68</rule>\\
\hspace*{0.6cm}<rule>32</rule>\\
\hspace*{0.6cm}<rule>1</rule>\\
\hspace*{0.6cm}<rule>35</rule>\\
\hspace*{0.6cm}<rule>69</rule>\\
\hspace*{0.3cm}</answer>\\
</ans>\\
\\
\\
\#\# Questions:\\
\{questions\}\\
\\
\#\# Response Format:\\
Respond with an XML block structured as follows:\\
\\
<ans>\\
\hspace*{0.3cm}<answer id="1">\\
\hspace*{0.6cm}<rule>1</rule>\\
\hspace*{0.6cm}<rule>2</rule>\\
\hspace*{0.3cm}...\\
\hspace*{0.3cm}</answer>\\
\hspace*{0.3cm}<answer id="2">\\
\hspace*{0.6cm}<rule>1</rule>\\
\hspace*{0.6cm}<rule>2</rule>\\
\hspace*{0.3cm}...\\
\hspace*{0.3cm}</answer>\\
\hspace*{0.3cm}...\\
</ans>\\
\\
\#\#\# Notes:\\
- Each <answer id="N"> element corresponds to the N-th question.\\
- Inside each <answer> block, list each semantic rule in the correct order using <rule> tags.\\
\\
\#\# Important Notes:\\
- The **order** of rules matters and should reflect the evaluation sequence.\\
- A single rule may be needed to be applied multiple times during evaluation.\\
- You must include **all** semantic rules required for complete execution.\\
- Base your analysis solely on the provided semantics, not on general programming knowledge.\\
\\
\\    
\promptsubsection{\rmfamily \Kos-semantics}\\
You are an interpreter for a language called \{language\}. I will
describe the syntax and the semantics of the language using the
K-framework. You will use this to execute a \{language\} program. You
will only use the rules described in the semantics I provide. Assume
all the rules in the semantics I give are correct.\\
\\
    Here is the K-framework formalization of \{language\}\\
    \hspace*{1cm}\{semantics\}\\
\\
    Here is the \{language\} program\\
    \hspace*{1cm}\{program\}\\
\\
\\
\#\# TASK:\\
For each question below, you'll be given:\\
1. A program\\
2. The program state ($\sigma$) (variable values) before executing the program\\
3. The control stack ($\chi$) before executing the program\\
\\
Assume that all necessary variables have been declared and have the values as\\
indicated in the provided program state.\\
\\
You must:\\
- Correctly identify and apply the K-semantic rules required to evaluate the program to completion\\
- List them in the correct order of application\\
\\
\\
Here is one example:\\
** Program:**\\
\{WHILE\} (n \{LTEQ\_OP\} 0)\\
\{\{\\
\hspace*{0.8cm}\{HALT\};\\
\}\};\\
\\
**Program state($\sigma$) before execution:**\\
\{\{'n': 100, 'sum': 0\}\}\\
\\
**Control stack($\chi$) before execution:**\\
 $\epsilon$\\
\\
\\
This is the sequence of steps:\\
1. First, we transform the '\{WHILE\}' into '\{WHILE\}1' while also inserting a 'breakMarker' after '\{WHILE\}1' using **Rule 24**.\\
2. Next we transform the '\{WHILE\}1' into an '\{IF\}-\{ELSE\}' with the '\{WHILE\}1' as the body of the '\{IF\}' using **Rule 25**.\\
3. We then reduce the loop predicate to a boolean by first reducing left-hand-side which is a variable using **Rule 1** and then applying the '\{LTEQ\_OP\}' using **Rule 13*.\\
4. Since the loop predicate evaluates to 'false', we apply the '\{IF\}' not taken rule **Rule 23** to take the '\{ELSE\}' branch which is empty.\\
5. Finally, we evaluate the 'breakMarker' statement using **Rule 27** to conclude the program execution.\\
\\
Therefore, the final answer is:\\
<ans>\\
\hspace*{0.3cm}<answer id="1">\\
\hspace*{0.6cm}<rule>24</rule>\\
\hspace*{0.6cm}<rule>25</rule>\\
\hspace*{0.6cm}<rule>1</rule>\\
\hspace*{0.6cm}<rule>13</rule>\\
\hspace*{0.6cm}<rule>23</rule>\\
\hspace*{0.6cm}<rule>27</rule>\\
\hspace*{0.3cm}</answer>\\
</ans>\\
\\
\\
\#\# Questions:\\
\{questions\}\\
\\
\#\# Response Format:\\
Respond with an XML block structured as follows:\\
\\
<ans>\\
\hspace*{0.3cm}<answer id="1">\\
\hspace*{0.6cm}<rule>1</rule>\\
\hspace*{0.6cm}<rule>2</rule>\\
\hspace*{0.3cm}...\\
\hspace*{0.3cm}</answer>\\
\hspace*{0.3cm}<answer id="2">\\
\hspace*{0.6cm}<rule>1</rule>\\
\hspace*{0.6cm}<rule>2</rule>\\
\hspace*{0.3cm}...\\
\hspace*{0.3cm}</answer>\\
\hspace*{0.3cm}...\\
</ans>\\
\\
\#\#\# Notes:\\
- Each '<answer id="N">' element corresponds to the N-th question.\\
- Inside each '<answer>' block, list each semantic rule in the correct order using '<rule>' tags.\\
\\
\#\# Important Notes:\\
- The **order** of rules matters and should reflect the evaluation sequence.\\
- Only rules that have names indicated in '[]' adjacent to it must be reported in the answer.\\
- A single rule may be needed to be applied multiple times during evaluation.\\
- You must include **all** semantic rules required for complete execution.\\
- Base your analysis solely on the provided semantics, not on general programming knowledge.\\
\\
\\
\promptsubsection{\rmfamily Non-\COT}Only output the '<ans>' XML block. Do not include any other content.\\
\promptsubsection{\rmfamily \COT} Explain your reasoning step-by-step **before** answering. Wrap your reasoning in '<reason>' tags.\\
\end{prompt}

\begin{prompt}
  [title={\thetcbcounter{} Prompt for \etp task.}]
\promptsubsection{\rmfamily \Sos}\\
You are an interpreter for a language called \{language\}. I will
describe the syntax for \{language\} in EBNF and its semantics using
small-step operational semantics. You will use this to execute a
\{language\} program. You will only use the rules described in the
semantics I provide. Assume all the rules in the semantics I give are
correct. A program has finished execution when one of the terminal
configurations $\langle\epsilon,\sigma,\chi\rangle$,$\langle$\{HALT\}$,\sigma,\chi\rangle$, $\langle$\{ERROR\}$,\sigma,\chi\rangle$ is reached.\\
\\  
    Here is the syntax of \{language\} in EBNF\\
    \hspace*{1cm}\{syntax\}\\
\\  
    Here is the small-step operational semantics of \{language\}\\
    \hspace*{1cm}\{semantics\}\\
\\  
    Here is the \{language\} program\\
    \hspace*{1cm}\{program\}\\
\\
\#\# TASK:\\
Given a program and its semantics, predict the execution trace. Your goal is to simulate execution, step by step of executing the program using the given small-step operational semantics rules. Do not skip any rules that is needed to evaluate the program. You will output your answer in the following format.\\
\\
\#\# Response Format:\\
Respond with an XML block structured as follows:\\
\\
<answer>\\
\hspace*{0.3cm}<step>\\
\hspace*{0.6cm}<rule>1</rule>\\
\hspace*{0.6cm}<program\_state>\\
\hspace*{0.9cm}<n>0</n>\\
\hspace*{0.9cm}<sum>0</sum>\\
\hspace*{0.6cm}</program\_state>\\
\hspace*{0.3cm}</step>\\
\hspace*{0.3cm}<step>\\
\hspace*{0.6cm}<rule>2</rule>\\
\hspace*{0.6cm}<program\_state>\\
\hspace*{0.9cm}<n>100</n>\\
\hspace*{0.9cm}<sum>0</sum>\\
\hspace*{0.6cm}</program\_state>\\
\hspace*{0.3cm}</step>\\
  ...\\
</answer>\\
\\
\#\# Here is an example:\\
\\
Here is the \{language\} program:\\
int i;\\
int j;\\
i \{ASSIGN\_OP\} 0;\\
\{WHILE\} (i \{LT\_OP\} 2)\\
\{\{\\
\hspace*{0.8cm}\{HALT\};\\
\}\};\\
\\
\\
\#\# Expected output:\\
<answer>\\
\hspace*{0.3cm}<step>\\
\hspace*{0.6cm}<rule>3</rule>\\
\hspace*{0.6cm}<program\_state>\\
\hspace*{0.9cm}<i>0</i>\\
\hspace*{0.6cm}</program\_state>\\
\hspace*{0.3cm}</step\\
\hspace*{0.3cm}<step>\\
\hspace*{0.6cm}<rule>3</rule>\\
\hspace*{0.6cm}<program\_state>\\
\hspace*{0.9cm}<i>0</i>\\
\hspace*{0.9cm}<j>0</j>\\
\hspace*{0.6cm}</program\_state>\\
\hspace*{0.3cm}</step>\\
\hspace*{0.3cm}<step>\\
\hspace*{0.6cm}<rule>5</rule>\\
\hspace*{0.6cm}<program\_state>\\
\hspace*{0.9cm}<i>0</i>\\
\hspace*{0.9cm}<j>0</j>\\
\hspace*{0.6cm}</program\_state>\\
\hspace*{0.3cm}</step>\\
\hspace*{0.3cm}<step>\\
\hspace*{0.6cm}<rule>67</rule>\\
\hspace*{0.6cm}<program\_state>\\
\hspace*{0.9cm}<i>0</i>\\
\hspace*{0.9cm}<j>0</j>\\
\hspace*{0.6cm}</program\_state>\\
\hspace*{0.3cm}</step>\\
\hspace*{0.3cm}<step>\\
\hspace*{0.6cm}<rule>68</rule>\\
\hspace*{0.6cm}<program\_state>\\
\hspace*{0.9cm}<i>0</i>\\
\hspace*{0.9cm}<j>0</j>\\
\hspace*{0.6cm}</program\_state>\\
\hspace*{0.3cm}</step>\\
\hspace*{0.3cm}<step>\\
\hspace*{0.6cm}<rule>28</rule>\\
\hspace*{0.6cm}<program\_state>\\
\hspace*{0.9cm}<i>0</i>\\
\hspace*{0.9cm}<j>0</j>\\
\hspace*{0.6cm}</program\_state>\\
\hspace*{0.3cm}</step>\\
\hspace*{0.3cm}<step>\\
\hspace*{0.6cm}<rule>1</rule>\\
\hspace*{0.6cm}<program\_state>\\
\hspace*{0.9cm}<i>0</i>\\
\hspace*{0.9cm}<j>0</j>\\
\hspace*{0.6cm}</program\_state>\\
\hspace*{0.3cm}</step>\\
\hspace*{0.3cm}<step>\\
\hspace*{0.6cm}<rule>30</rule>\\
\hspace*{0.6cm}<program\_state>\\
\hspace*{0.9cm}<i>0</i>\\
\hspace*{0.9cm}<j>0</j>\\
\hspace*{0.6cm}</program\_state>\\
\hspace*{0.3cm}</step>\\
\hspace*{0.3cm}<step>\\
\hspace*{0.6cm}<rule>70</rule>\\
\hspace*{0.6cm}<program\_state>\\
\hspace*{0.9cm}<i>0</i>\\
\hspace*{0.9cm}<j>0</j>\\
\hspace*{0.6cm}</program\_state>\\
\hspace*{0.3cm}</step>\\
\hspace*{0.3cm}<step>\\
\hspace*{0.6cm}<rule>78</rule>\\
\hspace*{0.6cm}<program\_state>\\
\hspace*{0.9cm}<i>0</i>\\
\hspace*{0.9cm}<j>0</j>\\
\hspace*{0.6cm}</program\_state>\\
\hspace*{0.3cm}</step>\\
</answer>\\
\\
\\
\#\# Notes:\\
- Each '<step>' must correspond to **exactly one small-step operational semantics rule** that is needed to evaluate a statement in the given program.\\
- The '<rule>' must indicate a rule used in the evaluation of a statement.\\
- The '<program\_state>' must represent the **entire program state immediately after** the execution of that rule.\\
- The program state must list **all variables currently in scope**, using the variable names as XML tags and their current values as tag content.\\
- Include variables even if they did not change.\\
- Do not skip any step or merge multiple steps into one.\\
- Do not skip any rules (including those used to reduce expressions and variables) that are needed to evaluate the program.\\
- The program execution is complete when one of the terminal configurations $\langle\epsilon,\sigma,\chi\rangle$,$\langle$\{HALT\}$,\sigma,\chi\rangle$, $\langle$\{ERROR\}$,\sigma,\chi\rangle$ is reached\\
\\
\\
\promptsubsection{\rmfamily \Kos-semantics}\\
You are an interpreter for a language called \{language\}. I will
describe the syntax and the semantics of the language using the
K-framework. You will use this to execute a \{language\} program. You
will only use the rules described in the semantics I provide. Assume
all the rules in the semantics I give are correct.\\
\\
    Here is the K-framework formalization of \{language\}\\
    \hspace*{1cm}\{semantics\}\\
\\
    Here is the \{language\} program\\
    \hspace*{1cm}\{program\}\\
\\
\\
\#\# TASK:\\
Given a program and its semantics, predict the execution trace. Your goal is to simulate execution, step by step of executing the program using the given K-framework semantics rules. Do not skip any rules that is needed to evaluate the program. You will output your answer in the following format.\\
\\
\#\# Response Format:\\
Respond with an XML block structured as follows:\\
\\
<answer>\\
\hspace*{0.3cm}<step>\\
\hspace*{0.6cm}<rule>1</rule>\\
\hspace*{0.6cm}<program\_state>\\
\hspace*{0.9cm}<n>0</n>\\
\hspace*{0.9cm}<sum>0</sum>\\
\hspace*{0.6cm}</program\_state>\\
\hspace*{0.3cm}</step>\\
\hspace*{0.3cm}<step>\\
\hspace*{0.6cm}<rule>2</rule>\\
\hspace*{0.6cm}<program\_state>\\
\hspace*{0.9cm}<n>100</n>\\
\hspace*{0.9cm}<sum>0</sum>\\
\hspace*{0.6cm}</program\_state>\\
\hspace*{0.3cm}</step>\\
  ...\\
</answer>\\
\\
\#\# Here is an example:\\
\\
Here is the \{language\} program:\\
int i;\\
int j;\\
i \{ASSIGN\_OP\} 0;\\
\{WHILE\} (i \{LT\_OP\} 2)\\
\{\{\\
\hspace*{0.8cm}\{HALT\};\\
\}\};\\
\\
\\
\#\# Expected output:\\
\\
<answer>\\
\hspace*{0.3cm}<step>\\
\hspace*{0.6cm}<rule>36</rule>\\
\hspace*{0.6cm}<program\_state>\\
\hspace*{0.9cm}<i>0</i>\\
\hspace*{0.6cm}</program\_state>\\
\hspace*{0.3cm}</step>\\
\hspace*{0.3cm}<step>\\
\hspace*{0.6cm}<rule>36</rule>\\
\hspace*{0.6cm}<program\_state>\\
\hspace*{0.9cm}<i>0</i>\\
\hspace*{0.9cm}<j>0</j>\\
\hspace*{0.6cm}</program\_state>\\
\hspace*{0.3cm}</step>\\
\hspace*{0.3cm}<step>\\
\hspace*{0.6cm}<rule>21</rule>\\
\hspace*{0.6cm}<program\_state>\\
\hspace*{0.9cm}<i>0</i>\\
\hspace*{0.9cm}<j>0</j>\\
\hspace*{0.6cm}</program\_state>\\
\hspace*{0.3cm}</step>\\
\hspace*{0.3cm}<step>\\
\hspace*{0.6cm}<rule>24</rule>\\
\hspace*{0.6cm}<program\_state>\\
\hspace*{0.9cm}<i>0</i>\\
\hspace*{0.9cm}<j>0</j>\\
\hspace*{0.6cm}</program\_state>\\
\hspace*{0.3cm}</step>\\
\hspace*{0.3cm}<step>\\
\hspace*{0.6cm}<rule>25</rule>\\
\hspace*{0.6cm}<program\_state>\\
\hspace*{0.9cm}<i>0</i>\\
\hspace*{0.9cm}<j>0</j>\\
\hspace*{0.6cm}</program\_state>\\
\hspace*{0.3cm}</step>\\
\hspace*{0.3cm}<step>\\
\hspace*{0.6cm}<rule>1</rule>\\
\hspace*{0.6cm}<program\_state>\\
\hspace*{0.9cm}<i>0</i>\\
\hspace*{0.9cm}<j>0</j>\\
\hspace*{0.6cm}</program\_state>\\
\hspace*{0.3cm}</step>\\
\hspace*{0.3cm}<step>\\
\hspace*{0.6cm}<rule>12</rule>\\
\hspace*{0.6cm}<program\_state>\\
\hspace*{0.9cm}<i>0</i>\\
\hspace*{0.9cm}<j>0</j>\\
\hspace*{0.6cm}</program\_state>\\
\hspace*{0.3cm}</step>\\
\hspace*{0.3cm}<step>\\
\hspace*{0.6cm}<rule>22</rule>\\
\hspace*{0.6cm}<program\_state>\\
\hspace*{0.9cm}<i>0</i>\\
\hspace*{0.9cm}<j>0</j>\\
\hspace*{0.6cm}</program\_state>\\
\hspace*{0.3cm}</step>\\
\hspace*{0.3cm}<step>\\
\hspace*{0.6cm}<rule>26</rule>\\
\hspace*{0.6cm}<program\_state>\\
\hspace*{0.9cm}<i>0</i>\\
\hspace*{0.9cm}<j>0</j>\\
\hspace*{0.6cm}</program\_state>\\
\hspace*{0.3cm}</step>\\
</answer>\\
\\
\\
\#\# Notes:\\
- Each '<step>' must correspond to **exactly one K-semantics re-write rule** that is needed to evaluate a statement in the given program.\\
- Only rules that have names indicated in '[]' adjacent to it must be reported in the answer.\\
- The '<rule>' must indicate a rule used in the evaluation of a statement.\\
- The '<program\_state>' must represent the **entire program state immediately after** the execution of that rule.\\
- The program state must list **all variables currently in scope**, using the variable names as XML tags and their current values as tag content.\\
- Include variables even if they did not change.\\
- Do not skip any step or merge multiple steps into one.\\
- Do not skip any rules (including those used to reduce expressions and variables) that are needed to evaluate the program.\\
\\
\\
\promptsubsection{\rmfamily Non-\COT}Only output the `<answer>' XML block. Do not include explanations, comments, or any other text.\\
\promptsubsection{\rmfamily \COT}Explain your reasoning step-by-step **before** answering. Wrap your reasoning in '<reason>' tags.
Note that you **MUST** wrap your reasoning steps with '<reason>' tags, the prediction with '<answer>' tags.\\
\end{prompt}

\clearpage
\section{Is Keyword Obfuscation a Tokenization Artifact?}
\label{app:keyword-obf-tokenization}

\begin{wraptable}{r}{0.56\textwidth}%
\centering
\footnotesize
\vspace{-0.25cm}%
\caption{\op accuracy under \keywordObf with different symbol inventories. 
We compare the original Caucasian-Albanian symbol substitution against a
tokenization-controlled variant (\textsc{1Tok}) in which every keyword
is replaced by a symbol verified to occupy a single \gptfom token.
Accuracy remains low in both settings, indicating that failures are not
driven by token inflation.}
\label{tab:keywordobf-tokenization}
\setlength{\tabcolsep}{6pt}
\begin{tabular}{@{}c l l r@{}}
\toprule
Semantics & Variant & Model & Accuracy(\%) \\
\midrule
\multirow{4}{*}{\Sos} & \multirow{2}{*}{Caucasian-Albanian} & \gptfom  & 8.4 \\
        &           & \gptfom-\COT & 21.8 \\[4pt]
        & \multirow{2}{*}{1Tok}      & \gptfom  & 7.0 \\
        &           & \gptfom-\COT & 19.3 \\
\midrule
\multirow{4}{*}{\Kos} & \multirow{2}{*}{Caucasian-Albanian} & \gptfom  & 8.2 \\
        &           & \gptfom-\COT & 18.1 \\[4pt]
        & \multirow{2}{*}{1Tok}      & \gptfom  & 6.2 \\
        &           & \gptfom-\COT & 16.9 \\
\bottomrule
\end{tabular}
\vspace{-0.25cm}%
\end{wraptable}
A potential concern with the \keywordObf{} semantic \shift is that it
introduces rare Unicode characters (\eg, \CAADD from Caucasian-Albanian script)
that may be split into many subword tokens, artificially inflating
prompt length and degrading model performance.
If so, the observed failures would primarily reflect tokenizer
limitations rather than deficiencies in rule-conditioned reasoning.

To isolate this factor, we perform a controlled ablation that preserves
the semantic transformation of \keywordObf{} while removing tokenization
effects.

\subsection{Tokenizer-Controlled Symbol Substitution}

We construct a variant, 1Tok, in which every keyword
and operator is replaced with a symbol verified to be encoded as a
single token by \gptfomSmall tokenizer.
This \shift maintains the same distribution shift—models must map
novel surface forms to formal rules—while preventing input-length
inflation from multi-byte Unicode characters.

\subsection{Experimental Setup}

We rerun the \op task on the \humanwrit split under
\keywordObf{} semantics for both \Sos and \Kos based formalizations, comparing the
original Caucasian-Albanian symbol substitution to the new 1-token variant.
All prompts, rules, and evaluation procedures match those in
\S\ref{sec:analysis-global-rule-composition}; only the symbol inventory changes.

\subsection{Results}

Table~\ref{tab:keywordobf-tokenization} reports the results.
Replacing multi-token Unicode symbols with guaranteed single-token
alternatives yields only modest changes in accuracy.

Under \Sos, \gptfomSmall decreases slightly from 8.4\% to 7.0\%,
and \gptfomSmall-\COT from 21.8\% to 19.3\%.
A similar pattern holds for \Kos, where \gptfomSmall drops from 8.2\%
to 6.2\% and \gptfomSmall-\COT from 18.1\% to 16.9\%.

Crucially, in all cases performance remains far below that observed
under \standardSem{} semantic formalization, and the qualitative failure pattern under
\keywordObf{} is unchanged.

\subsection{Implications}

These results rule out tokenization inefficiency as the primary driver
of degraded performance under \keywordObf{}.
Even when all obfuscated symbols are atomic tokens, models still fail
to reliably apply the correct operational rules and compose them over
execution.

We therefore conclude that \keywordObf{} probes limitations in
rule-conditioned reasoning over unfamiliar formal systems, rather than
lexical or tokenizer artifacts.

\clearpage
\section{Task Extended Analysis}
\label{sec:appendix-task-extended}
\subsection{Formal Semantics Notation Comprehension}
\label{sec:appendix-notation-comprehension}

\begin{table*}[h]
\centering
\TableFont
\caption{\TableNotationComprehensionCaption}
\setlength{\tabcolsep}{7pt}
\renewcommand{\arraystretch}{1.15}
\begin{tabular}{llccl}
\toprule
\textbf{Family} & \textbf{Construct} & \textbf{Rules} & \textbf{Category} & \textbf{Semantic Role} \\
\midrule

\rowcolor{gray!70}
\multicolumn{5}{l}{\textbf{A.\ Variable \& State Access}} \\
F1 & Variable lookup & 1--2 & AEXP & Read $\sigma(x)$; distinguishes value vs unbound-variable error \\
\addlinespace[4pt]

\rowcolor{gray!70}
\multicolumn{5}{l}{\textbf{B.\ Declarations \& Assignment}} \\
F2 & Declaration & 3 & SL & Initialize state via $\sigma[x \mapsto 0]$ \\
F3 & Assignment & 4--6 & SL & Step RHS vs commit update; uninitialized-assignment error \\
\addlinespace[4pt]

\rowcolor{gray!70}
\multicolumn{5}{l}{\textbf{C.\ Arithmetic Expressions (Binary)}} \\
F4 & Addition & 7--9 & AEXP & Left-step / right-step / compute for binary $+$ \\
\bottomrule
\end{tabular}
\end{table*}
\begin{table*}[h]
  \centering
  \TableFont
  \caption{\TableNotationComprehensionKCaption}
  \setlength{\tabcolsep}{7pt}
  \renewcommand{\arraystretch}{1.15}
  \begin{tabular}{llccl}
    \toprule
    \textbf{Family} & \textbf{Construct} & \textbf{Rules} & \textbf{Category} & \textbf{Semantic Role} \\
    \midrule

    \rowcolor{gray!70}
    \multicolumn{5}{l}{\textbf{A. Variable \& State Access}} \\
    F1 & Variable lookup & 1--2 & AEXP &
      Read state binding; distinguishes value retrieval vs unbound-variable error (halts). \\

    \rowcolor{gray!70}
    \multicolumn{5}{l}{\textbf{B. Declarations \& Assignment}} \\
    F2 & Declaration & 36 & SL &
      Initialize state (introduce variable with default value $0$). \\
    F3 & Assignment & 21 & SL &
      Commit update to state once RHS is a value (update existing variable). \\

    \rowcolor{gray!70}
    \multicolumn{5}{l}{\textbf{C. Arithmetic Expressions (Binary)}} \\
    F4 & Addition & 3 & AEXP &
      Compute binary $+$ when both operands are values. \\
    F5 & Subtraction & 4 & AEXP &
      Compute binary $-$ when both operands are values. \\
    F6 & Multiplication & 5 & AEXP &
      Compute binary $\times$ when both operands are values. \\
    F7 & Division & 6--7 & AEXP &
      Case split: nonzero divisor computes division; zero divisor raises error (halts). \\
    F8 & Modulo & 8--9 & AEXP &
      Case split: nonzero divisor computes modulus; zero divisor raises error (halts). \\

    \rowcolor{gray!70}
    \multicolumn{5}{l}{\textbf{D. Arithmetic Expressions (Unary)}} \\
    F9 & Unary minus & 11 & AEXP &
      Compute unary negation once operand is a value. \\
    F10 & Unary plus & 10 & AEXP &
      Compute unary identity (operationalized as adding $0$) once operand is a value. \\

    \rowcolor{gray!70}
    \multicolumn{5}{l}{\textbf{E. Relational Comparisons}} \\
    F11 & Less-than & 12 & BEXP &
      Compute boolean result for < on value operands. \\
    F12 & Less-or-equal & 13 & BEXP &
      Compute boolean result for <= on value operands. \\
    F13 & Greater-than & 14 & BEXP &
      Compute boolean result for > on value operands. \\
    F14 & Greater-or-equal & 15 & BEXP &
      Compute boolean result for >= on value operands. \\
    F15 & Equality & 16 & BEXP &
      Compute boolean result for == on value operands. \\
    F16 & Inequality & 17 & BEXP &
      Compute boolean result for != on value operands. \\

    \rowcolor{gray!70}
    \multicolumn{5}{l}{\textbf{F. Boolean Connectives}} \\
    F17 & Boolean AND & 19 & BEXP &
      Compute conjunction on boolean operands. \\
    F18 & Boolean OR & 20 & BEXP &
      Compute disjunction on boolean operands. \\
    F19 & Boolean NOT & 18 & BEXP &
      Compute negation on boolean operand. \\

    \rowcolor{gray!70}
    \multicolumn{5}{l}{\textbf{H. Conditional Control Flow}} \\
    F21 & If--then--else & 22--23 & CTRL &
      Branch on boolean predicate. \\

    \rowcolor{gray!70}
    \multicolumn{5}{l}{\textbf{I. Loops \& Loop Context}} \\
    F22 & While entry & 24 & CTRL &
      Push loop statement onto stack; rewrite to internal form; insert markers. \\
    F23 & Loop execution & 25 & CTRL &
      Desugar internal while into if--else: iterate on true, exit on false. \\
    F24 & Loop markers & 27--28 & CTRL &
      Marker plumbing: remove loop markers when encountered to resume execution. \\

    \rowcolor{gray!70}
    \multicolumn{5}{l}{\textbf{J. Non-local Control Flow}} \\
    F25 & Break & 33--35 & CTRL &
      Handle \texttt{break} statements \\
    F26 & Continue & 29--32 & CTRL &
      Handle \texttt{continue} statements \\

    \rowcolor{gray!70}
    \multicolumn{5}{l}{\textbf{K. Termination}} \\
    F27 & Halt & 26 & CTRL &
      Terminal configuration: stops execution immediately; removes remaining computation. \\

    \bottomrule
  \end{tabular}
\end{table*}

This section provides the full details of the hierarchical distractor
sampling strategy and the rule-family coverage distributions referenced
in \S~\ref{sec:eval-framework}.

\MyPara{Hierarchical rule organization}%
Tables~\ref{tab:rule-families-full} and~\ref{tab:rule-families-full-k}
list every semantic rule used in our \nlruleSmall and \rulenlSmall
evaluation tasks for \Sos and \Kos respectively. Rules are organized
into 27 \emph{families} (F1--F27), grouped under 11 top-level
categories (A--K). Each family corresponds to a single language
construct (\eg, addition, while-entry) and may contain multiple rules
that differ in their \emph{semantic role} (\eg, left-step vs.\
right-step vs.\ compute for a binary arithmetic operator under \Sos).
\Kos rules are at a coarser granularity: because \Kos is a
big-step semantics, many of the intermediate reduction steps present
under \Sos (\eg, left-step and right-step rules for binary operators)
are collapsed into a single rule, yielding fewer rules per family.

\MyPara{Distractor sampling strategy}%
When constructing each multiple-choice sample (\numNotationCompChoices
choices), we draw distractors in a hierarchical order designed to
maximize semantic proximity to the correct answer:
\begin{enumerate}[leftmargin=*,itemsep=2pt,topsep=2pt]
  \item \textbf{Same family, different semantic role.}
        We first attempt to sample distractors from the same family as
        the correct rule. These share the same language construct but
        differ in semantic role (\eg, the compute rule vs.\ a step rule
        for addition), making them the hardest distractors.
  \item \textbf{Same category, different construct.}
        If the family does not contain enough candidate rules, we
        sample from other families within the same top-level category
        (\eg, another binary arithmetic operator from category~C).
  \item \textbf{Different category.}
        As a last resort, distractors are drawn from a different
        category entirely (\eg, a control-flow rule used as a
        distractor for an arithmetic-expression question).
\end{enumerate}
This ordering ensures that each question is discriminative: models must
distinguish among rules that govern closely related constructs rather
than exploit superficial differences in operator type or syntactic category.

\subsection{\OPTask~(\op)}
\label{sec:appendix-op-analysis}

This section analyzes (1)~the impact of code-complexity metrics on
\LLM performance in the \op task, and (2)~the average percentage
of variables per program whose final states are predicted correctly.

\subsubsection{Impact of Code-Complexity Metrics}
\label{sec:appendix-op-analysis-metric-impact}

\begin{figure}[h]
  \vspace{-0.25cm}%
  \subcaptionbox{%
    \FigPOPExtendedAnalysisDescCaption%
    \label{figure:appendix-op-task}}[\linewidth]{%
    \includegraphics[scale=0.3]{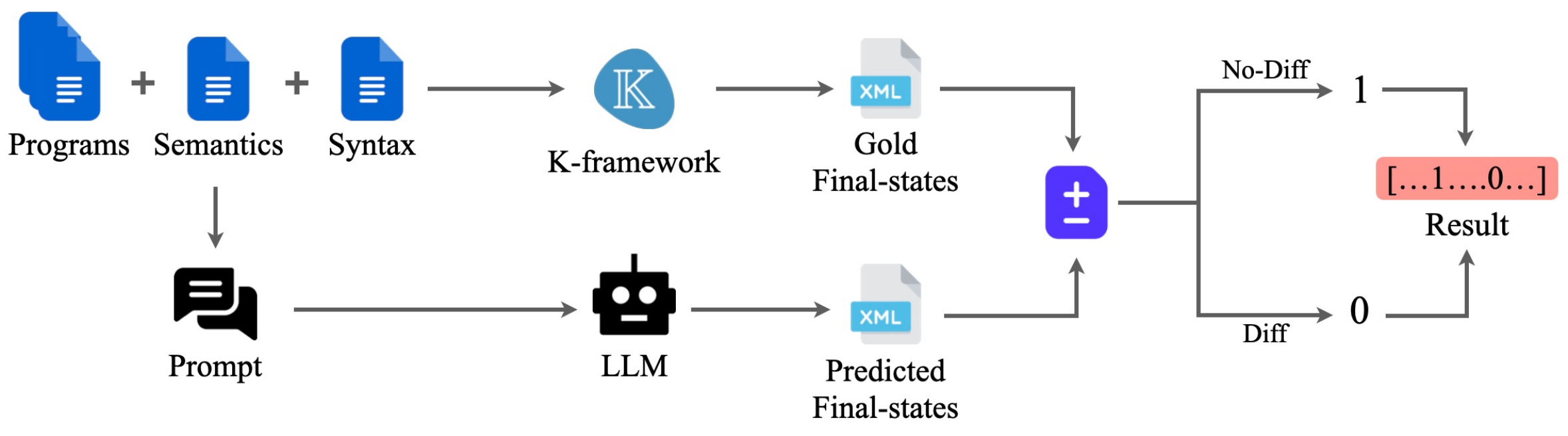}
  }%
  \vspace{0.5cm}%
  \\
  \subcaptionbox{%
    \FigPOPExtendedAnalysisModelingCaption%
    \label{figure:appendix-op-task-modeling}}[\linewidth]{%
    \includegraphics[scale=0.3]{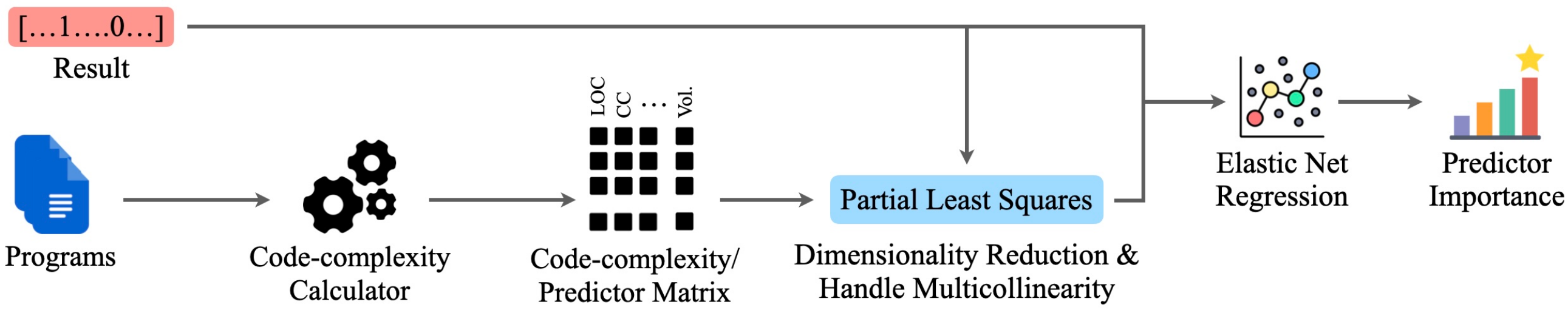}
  }%
  \caption{\FigPOPExtendedAnalysisCaption}
%  \vspace{-0.25cm}%
\end{figure}

Figure~\ref{figure:appendix-pop-ext-analysis-workflow} illustrates the
workflow of the \op task. An \IMP program, together with optional
semantics (\Kos-semantics or \Sos) and syntax, is used both to
construct prompts for the \LLMs and to obtain gold final states by
executing the program in the \KTool. The \LLM's predicted final states
are then compared with the gold states for each declared variable. A
match is recorded as \CodeIn{1} (pass), and a mismatch as \CodeIn{0}
(fail).

Different \LLMs naturally excel on different \IMP programs. To
understand why an \LLM may predict all final states correctly for one
program but fail on another, we cast this task as a classification
problem as shown in
Figure~\ref{figure:appendix-op-task-modeling}. Each \IMP program is
mapped to a predictor vector that characterizes its complexity, using
the code-complexity metrics introduced earlier. Each predictor
is then normalized using z-score normalization to ensure fair
contribution from all the variables. The resulting predictor matrix,
together with the \LLM's binary result vector of passes and fails, is
then used to train a classifier.

%% Automatically generated by: table.py make_op_odds_per_iqr_table_combined

\begin{table*}[h!]
\begin{center}
\TableFont
\caption{\UseMacro{TCap-models-op-odds-per-iqr-nk-human-written}}
\begin{tabular}{l c  c  c  c  c  c  c  c  c }
\toprule
\UseMacro{THead-op-odds-per-iqr-nk-human-written-models}
 & \multicolumn{3}{c}{\UseMacro{THead-op-odds-per-iqr-nk-human-written-Control-flow}}
 & \multicolumn{2}{c}{\UseMacro{THead-op-odds-per-iqr-nk-human-written-Data-flow}}
 & \multicolumn{4}{c}{\UseMacro{THead-op-odds-per-iqr-nk-human-written-Size}}
\\
 \cmidrule(lr){2-4}
 \cmidrule(lr){5-6}
 \cmidrule(lr){7-10}
 & \textbf{\UseMacro{THead-op-odds-per-iqr-nk-human-written-CC}}
 & \textbf{\UseMacro{THead-op-odds-per-iqr-nk-human-written-MaxTakenIf}}
 & \textbf{\UseMacro{THead-op-odds-per-iqr-nk-human-written-MaxTakenLoop}}
 & \textbf{\UseMacro{THead-op-odds-per-iqr-nk-human-written-DepDegree}}
 & \textbf{\UseMacro{THead-op-odds-per-iqr-nk-human-written-NumAssignments}}
 & \textbf{\UseMacro{THead-op-odds-per-iqr-nk-human-written-LOC}}
 & \textbf{\UseMacro{THead-op-odds-per-iqr-nk-human-written-Volume}}
 & \textbf{\UseMacro{THead-op-odds-per-iqr-nk-human-written-Vocabulary}}
 & \textbf{\UseMacro{THead-op-odds-per-iqr-nk-human-written-TraceLength}}
\\
\midrule
\multicolumn{10}{c}{\textbf{\UseMacro{THead-op-odds-per-iqr-nk-human-written-dataset}}}%
\\
\bottomrule
\end{tabular}
\end{center}
\end{table*}
%% Automatically generated by: table.py make_op_odds_per_iqr_table_combined

\begin{table*}[h!]
\begin{center}
\TableFont
\caption{\UseMacro{TCap-models-op-odds-per-iqr-uk-human-written}}
\begin{tabular}{l l c  c  c  c  c  c  c  c  c }
\toprule
 & 
\UseMacro{THead-op-odds-per-iqr-uk-human-written-models}
 & \multicolumn{3}{c}{\UseMacro{THead-op-odds-per-iqr-uk-human-written-Control-flow}}
 & \multicolumn{2}{c}{\UseMacro{THead-op-odds-per-iqr-uk-human-written-Data-flow}}
 & \multicolumn{4}{c}{\UseMacro{THead-op-odds-per-iqr-uk-human-written-Size}}
\\
 \cmidrule(lr){3-5}
 \cmidrule(lr){6-7}
 \cmidrule(lr){8-11}
& 
 & \textbf{\UseMacro{THead-op-odds-per-iqr-uk-human-written-CC}}
 & \textbf{\UseMacro{THead-op-odds-per-iqr-uk-human-written-MaxTakenIf}}
 & \textbf{\UseMacro{THead-op-odds-per-iqr-uk-human-written-MaxTakenLoop}}
 & \textbf{\UseMacro{THead-op-odds-per-iqr-uk-human-written-DepDegree}}
 & \textbf{\UseMacro{THead-op-odds-per-iqr-uk-human-written-NumAssignments}}
 & \textbf{\UseMacro{THead-op-odds-per-iqr-uk-human-written-LOC}}
 & \textbf{\UseMacro{THead-op-odds-per-iqr-uk-human-written-Volume}}
 & \textbf{\UseMacro{THead-op-odds-per-iqr-uk-human-written-Vocabulary}}
 & \textbf{\UseMacro{THead-op-odds-per-iqr-uk-human-written-TraceLength}}
\\
\midrule
\multicolumn{11}{c}{\textbf{\UseMacro{THead-op-odds-per-iqr-uk-human-written-dataset}}}%
\\
\bottomrule
\end{tabular}
\end{center}
\end{table*}

Because these complexity metrics are often highly correlated
(multicollinearity), we apply Partial Least Squares
(PLS)~\citep{wold2001plsr} for dimensionality reduction. Unlike the
unsupervised Principal Component Analysis (PCA)~\citep{wold1987pca},
which identifies linear combinations of predictors that maximize
variance, PLS is supervised: it reduces dimensionality by finding
components that maximize the covariance between predictors and the
response variables (the result vector). This makes PLS more suitable
in our setting, as it better mitigates multicollinearity while
preserving predictive power.

We next apply Elastic Net regression~\citep{zou2005elasticnet} on the
PLS-transformed predictors and the result vector to train a
classifier. In regression, each predictor is assigned a
coefficient whose magnitude reflects its relative importance and whose
sign indicates whether it contributes positively or negatively to
prediction accuracy. Elastic Net is chosen because it combines
Lasso~\citep{tibshirani1996lasso} and Ridge~\citep{hoerl1970ridge2}
regularization: the Lasso component drives irrelevant coefficients to
zero, enabling feature selection, while the Ridge component shrinks
correlated coefficients, thereby mitigating multicollinearity.

We now briefly describe the Elastic Net regression process to explain
how we use the regression coefficients to determine the impact of
different metrics. Let $n$, $p$, $\bm{y}$, and $\bm{X}$ be the total
number of samples, the total number of predictors, the response vector,
and the predictor matrix (we will use boldface font to denote vectors
and matrices) respectively. Then,

\[
\bm{y} \in \mathbb{R}^n, \quad
y_i \in \{0,1\}, \quad
\bm{x_i} \in \mathbb{R}^p, \quad
p_i(y_i = 1 | \bm{x_i}) = \frac{1}{1+e^{-(\beta_0 + \bm{x_i}^\top \bm{\beta})}}
\]

Where $p_i(y_i = 1 | \bm{x_i})$ along with
$p_i(y_i = 0 | \bm{x_i}) = (1 - p_i(y_i = 1 | \bm{x_i}))$ represent the
class-conditional probabilities and $\bm{\beta}$ is the vector of coefficients.
The Elastic Net objective function for a Negative Log-Likelihood
loss is given as~\citep{friedman2010glmnet}:

\[
\arg\min_{\beta_0,\bm{\beta}}\;\Big[
\frac{1}{n}\sum_{i=1}^n\Big[-y_i\log p_i-(1-y_i)\log(1-p_i)\Big]
+\lambda\underbrace{\!\sum_{j=1}^p\Big[\frac{1-\alpha}{2}{\beta_j}^2+\alpha\lvert\beta_{j}\rvert\Big]}_{\text{Ridge and Lasso penalties}}\Big]
\]

Let $\bm{\hat\beta}$ be the coefficient vector that minimizes this
objective function.  Then the percentage odds
ratio~\citep{agresti2013cda,cornfield1951or,harrell2015rmsbook}
$\Theta$ for the inter-quartile-range $\Delta_{j}$ of the
$j^\text{th}$ predictor can be computed as:

\[
\Theta(\Delta_{j})=100 \times \big(\exp\!\big(\hat\beta_j\,\mathrm{\Delta}_j\big) - 1\big).
\]

The percentage odds ratio per inter-quartile-range \orIqr gives the
percentage change in the odds of the classifier's positive outcome
(predicting a \CodeIn{1}) for the predictor ranging from its typical
low value ($\text{25}^\text{th}$ percentile) to its typical high value
($\text{75}^\text{th}$ percentile) in the dataset when all other
predictors are held constant. Thus if $\Theta(\Delta_j)$ for the
$j^\text{th}$ predictor is -37\%, this implies that one quartile
increase in the $j^\text{th}$ predictor lowers the odds of the
classifier's positive outcome by 37\%.

To quantify each metric's effect on accuracy, we report the odds-ratio
per interquartile range, \orIqr, in
Tables~\ref{tab:or-per-iqr-nk}-\ref{tab:or-per-iqr-uk} for all \LLMs
without and with (\Kos-semantics, \Sos) semantics. Overall patterns
are similar across settings. On the \humanwrit dataset,
\metricMaxnestwhile—the maximum executed loop-nesting depth—is the
most influential predictor: larger \metricMaxnestwhile is associated
with lower odds of a correct final-state prediction. On the \llmtrans
dataset, \metricDepdeg (data-flow complexity) and \metricVol (size)
dominate without semantics; with semantics, \metricDepdeg remains
dominant under \Sos, whereas \metricVol dominates under
\Kos-semantics. On the Fuzzer-Generated split, \metricNumAssign (total
variable assignments) is the strongest predictor both without and with
semantics, with one exception: for \gemini under \Sos, \metricTracelen
(execution-trace length) is most predictive. Collectively, these
\orIqr trends suggest that increasing control-flow depth harms models
on human code, whereas data-flow/size factors are more limiting on
translated or fuzzer generated code.

\subsubsection{Complexity-Metric Impact Patterns}
\label{sec:appendix-op-impact-patterns}
\begin{figure}[h]
  \centering%
  \includegraphics[scale=0.45]{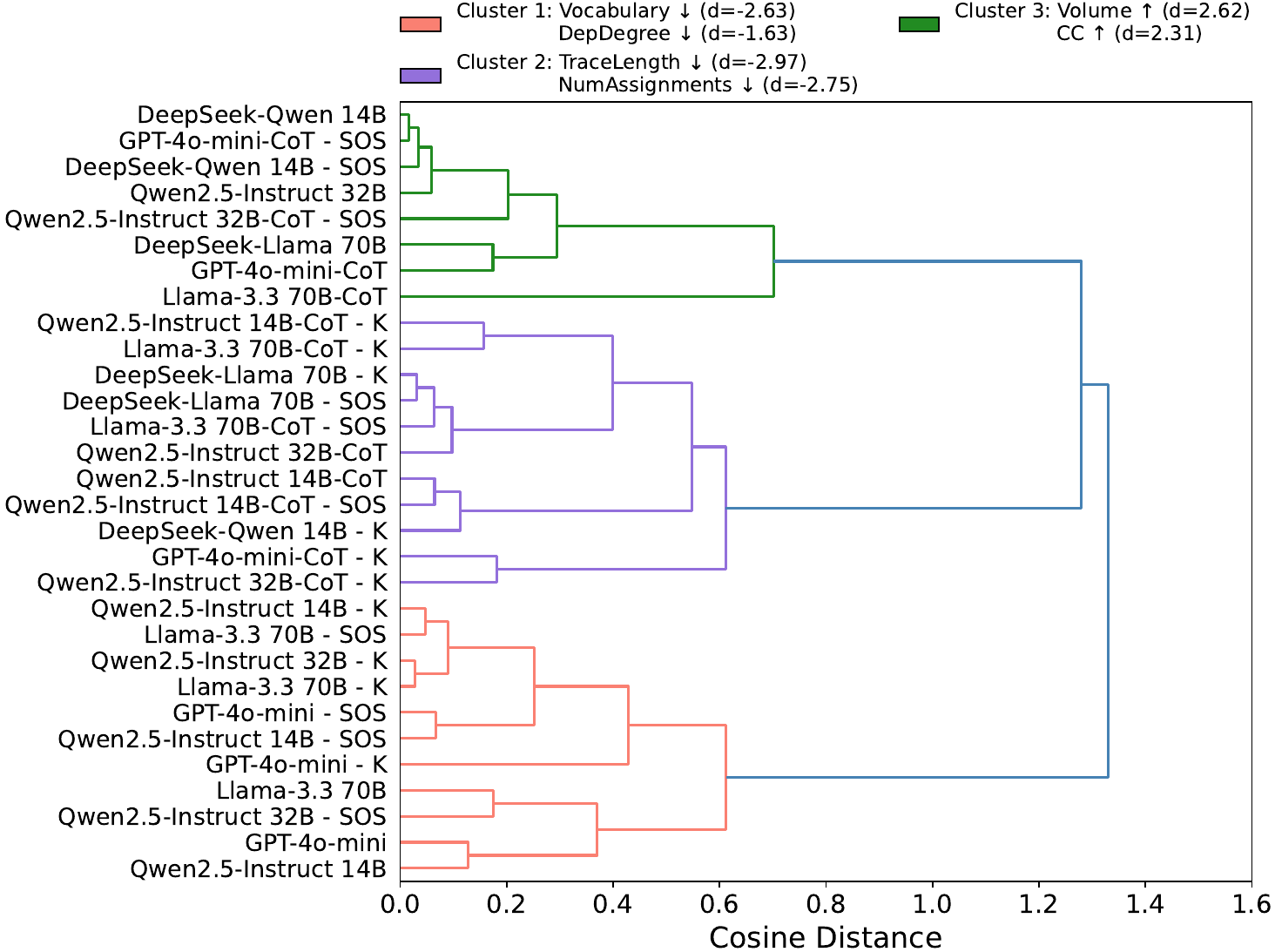}
  \caption{\FigPOPDendrogram}
\end{figure}

To identify if there is a pattern to how models perform on increasing
different code-complexity metrics, we perform hierarchical clustering
on the standardized regression coefficients ($\hat\beta_\text{SD}$) of
the metrics for the models on the \humanwrit dataset. We perform this
for the no-semantics and with \standardSem semantics (\Kos-semantics
and \Sos) cases.  We use the cosine-distance as the pair-wise distance
metric and the Cohen's d one-vs-rest test to identify the most
distinguishing metric of each cluster.
Figure~\ref{figure:appendix-op-dendrogram} shows the dendrogram of the
clustering process.

We see that there are three clusters. All the non-reasoning models
without \COT prompting are in \emph{Cluster 1} with the exception of
\qwenCoder{32} (under no-semantics case). Cluster 1 responds more
negatively to increases in the complexity metrics Vocabulary
(\metricVocab) and DepDegree (\metricDepdeg) relative to the other two
clusters.  \emph{Cluster 2} contains only the reasoning models and the
non-reasoning models with \COT prompting. It predominantly contains
models under the \Kos-semantics and responds more negatively to the
dynamically computed metrics, TraceLength (\metricTracelen) and
NumAssignments (\metricNumAssign) relative to the rest of the
clusters. The last cluster, \emph{Cluster 3} also only contains
reasoning models and non-reasoning models with \COT prompting
(\qwenCoder{32} is an exception). It predominantly contains models
under \Sos semantics and responds positively to increases in the
metrics, Volume (\metricVol) and cyclomatic-code complexity
(\metricCC) relative to the rest.

\subsubsection{Average Percentage of Variables Predicted Correctly}
\label{sec:appendix-op-analysis-metric-avg-var-prediction}

%% Automatically generated by: table.py make_two_results_table_same_task_op
\begin{table*}[h]
\begin{center}
\TableFont
\caption{\UseMacro{TCap-models-op-IMP-K-IMP-SOS-human-written-synthetic-cpp-fuzzer-generated-var-acc}\label{tab:op-IMP-K-IMP-SOS-human-written-synthetic-cpp-fuzzer-generated-var-acc}}
\begin{tabular}{l l  c  c  c  c  c  c  c }
\toprule
&
\multirow{2}{*}{\textbf{\UseMacro{THead-op-IMP-K-IMP-SOS-var-acc-models}}}
& 
& \multicolumn{3}{c}{\textbf{\Kos-Formalization}}
& \multicolumn{3}{c}{\textbf{\Sos-Formalization}}
\\
\cmidrule(lr){4-6}
\cmidrule(lr){7-9}
&
 & \textbf{$\metricAccuracyNK$}
 & \textbf{$\metricAccuracyStd$} 
 & \textbf{$\metricAccuracySwap$} 
 & \textbf{$\metricAccuracyObf$}
 & \textbf{$\metricAccuracyStd$} 
 & \textbf{$\metricAccuracySwap$} 
 & \textbf{$\metricAccuracyObf$} 
\\

\midrule
\multicolumn{9}{c}{\textbf{\UseMacro{THead-op-IMP-K-IMP-SOS-var-acc-dataset}}}%
\\
\midrule
\multirow{8}{*}{\rotatebox{90}{Non-reasoning}}
&
\UseMacro{THead-op-IMP-K-IMP-SOS-var-acc-Qwen-Qwen2.5-Coder-14B-Instruct-da}
 & \UseMacro{res-Qwen-Qwen2.5-Coder-14B-Instruct-da-op-nk-IMP-SOS-var-acc}
 & \UseMacro{res-Qwen-Qwen2.5-Coder-14B-Instruct-da-op-uk-IMP-K-var-acc}
 & \UseMacro{res-Qwen-Qwen2.5-Coder-14B-Instruct-da-op-mk-IMP-K-addSub_mulDiv_negateRelation-var-acc}
 & \UseMacro{res-Qwen-Qwen2.5-Coder-14B-Instruct-da-op-mk-IMP-K-unseen-var-acc}
 & \UseMacro{res-Qwen-Qwen2.5-Coder-14B-Instruct-da-op-uk-IMP-SOS-var-acc}
 & \UseMacro{res-Qwen-Qwen2.5-Coder-14B-Instruct-da-op-mk-IMP-SOS-addSub_mulDiv_negateRelation-var-acc}
 & \UseMacro{res-Qwen-Qwen2.5-Coder-14B-Instruct-da-op-mk-IMP-SOS-unseen-var-acc}
\\
\bottomrule
\end{tabular}
\end{center}
\vspace{-0.2cm}%
\end{table*}

We also computed on average (over the total number of declared
variables per \IMP program followed by over the total number of \IMP
programs) how many of the final-states of the declared variables per
\IMP program that are assigned to at least once are being predicted
correctly by the models. The results are shown in
Table~\ref{tab:op-IMP-K-IMP-SOS-human-written-synthetic-cpp-fuzzer-generated-var-acc}. We
see that the trend in terms of models performing better without
semantics than with semantics is similar to what is observed in the
\op task
(Table~\ref{tab:analysis-op-srp-combined}, left panel).
We also see that although models perform very poorly on the
increasingly complex datasets such as the \fuzzgen dataset on the \op
task, the average percentage of the final-states of the variables
predicted correctly per program is quite high.

\subsubsection{Standard Deviation of Task Accuracy}
\label{sec:appendix-op-analysis-metric-std-acc}

% \newcommand{\delt}[1]{%
%   % Round to integer first
%   \pgfmathtruncatemacro{\DeltaInt}{round(#1)}%
%   \pgfmathtruncatemacro{\DeltaAbsInt}{abs(\DeltaInt)}%
%   % zero-pad to 2 digits
%   \ifnum\DeltaAbsInt<10
%     \edef\DeltaFmt{0\DeltaAbsInt}%
%   \else
%     \edef\DeltaFmt{\DeltaAbsInt}%
%   \fi
%   % color + sign
%   \ifnum\DeltaInt>0
%     \textcolor{green!45!black}{\scriptsize(+\DeltaFmt)}%
%   \else\ifnum\DeltaInt<0
%     \textcolor{red!60!black}{\scriptsize(\texttt{-}\DeltaFmt)}%
%   \else
%     \textcolor{black!55}{\scriptsize(000)}%
%   \fi\fi
% }
% \newcommand{\accd}[2]{#1\,\delt{#2}}
% % needs: \usepackage{pgf}

% % acc + delta where delta is computed from (a - b), rounded to nearest int
% \newcommand{\accdfrom}[2]{%
%   \pgfmathtruncatemacro{\DeltaTmp}{round((#1) - (#2))}%
%   #1\,\delt{\DeltaTmp}%
% }
% % robust delta: Var - Std (still prints Var as the main number)
% \newcommand{\accdrobust}[2]{%
%   \pgfmathtruncatemacro{\DeltaTmp}{round((#1) - (#2))}%
%   #1\,\delt{\DeltaTmp}%
% }

\colorlet{zebra}{blue!5}

\begin{table}[h]
\begin{center}
\TableFont
\caption{Standard deviation of the accuracy metrics for the \op task. \label{tab:op-stdev}}
\setlength{\tabcolsep}{3.1pt}
\renewcommand{\arraystretch}{1}

\begin{tabular}{l S[table-format=3, table-column-width=6mm, detect-all=false]  r  r  r  c  r  r  r }
\toprule
%\\
&
& \multicolumn{3}{c}{\textbf{\KosSmall-Formalization}}
& 
& \multicolumn{3}{c}{\textbf{\SosSmall-Formalization}}
\\
\cmidrule(lr){3-5}\cmidrule(lr){7-9}
%\\
& \textbf{$\metricAccuracyNK$}
& \textbf{$\metricAccuracyStd$} & \textbf{$\metricAccuracySwap$} & \textbf{$\metricAccuracyObf$}
& 
& \textbf{$\metricAccuracyStd$} & \textbf{$\metricAccuracySwap$} & \textbf{$\metricAccuracyObf$} 
\\
\midrule

% =========================
% Reasoning
% =========================
\addlinespace[1pt]
\rowcolor{black!12}
\multicolumn{9}{l}{\textit{\humanwrit}}\\
\addlinespace[1pt]

\rowcolor{zebra}
\UseMacro{THead-op-IMP-K-IMP-SOS-acc-deepseek-ai-DeepSeek-R1-Distill-Qwen-14B-da}
& 2.3
& 2.0
& 0.3
& 1.7
&
& 1.8
& 0.0
& 2.5
\\

\UseMacro{THead-op-IMP-K-IMP-SOS-acc-deepseek-ai-DeepSeek-R1-Distill-Qwen-32B-da}
& 1.2
& 2.3
& 5.2
& 0.5
&
& 1.2
& 0.3
& 1.5
\\

\rowcolor{zebra}
\UseMacro{THead-op-IMP-K-IMP-SOS-acc-deepseek-ai-DeepSeek-R1-Distill-Llama-70B-da}
& 0.0
& 0.0
& 0.2
& 0.0
&
& 0.6
& 0.0
& 1.3
\\

\UseMacro{THead-op-IMP-K-IMP-SOS-acc-Qwen-QwQ-32B-da}
& 0.2
& 0.3
& 0.0
& 0.8
&
& 0.3
& 1.5
& 2.0
\\

\rowcolor{zebra}
\UseMacro{THead-op-IMP-K-IMP-SOS-acc-o3-mini-da}
& 1.2
& 0.0
& 1.4
& 0.6
&
& 0.0
& 1.8
& 0.3
\\
\bottomrule
\end{tabular}
\end{center}
\vspace{-0.2cm}
\end{table}
We average results over three independent runs for reasoning models
and report the standard deviation of accuracy in Table~\ref{tab:op-stdev}. 
Across all model-dataset-semantic-variant combinations, the standard 
deviation never exceeds 5.2 percentage points and is typically below 
2.0, confirming that the accuracy differences we report under semantic
shifts and across code-complexity splits are well above run-to-run
variability.

\clearpage
\subsection{\SRPTask~(\srp)}
\label{sec:appendix-srp-analysis}

In this section, we discuss: (1)~how the statements sampled from \IMP
programs are processed for the \srp task, and (2)~identify the most
mispredicted rule (first-point-of-mismatch) categories in the \srp
task.

\subsubsection{Processing \IMP Statements for \srp}
\label{sec:appendix-srp-dataset}
\begin{table}[h]%
  \centering%
  \TableFont%%
  \caption{\TablePredRuleProcessCaption}%
  \begin{tabular}{l|p{2.8cm}|c||p{5.2cm}|c}
    \toprule
    \textbf{Type} & \textbf{Statement} & \textbf{State}  & \textbf{\srp Program} & \textbf{\srp State}\\
    \midrule
    Declaration
    &
    \begin{lstlisting}[language=imp-pretty-no-lines]
int <VAR>;
    \end{lstlisting}
    &
    $\sigma$
    &
    \begin{lstlisting}[language=imp-pretty-no-lines]
int <VAR>;
    \end{lstlisting}
    &
    $\sigma$\\
    \midrule
    Assignment
    &
    \begin{lstlisting}[language=imp-pretty-no-lines]
<VAR> = <EXP>;
    \end{lstlisting}
    &
    $\sigma$
    &
    \begin{lstlisting}[language=imp-pretty-no-lines]
<VAR> = <EXP>;
    \end{lstlisting}
    &
    $\sigma$\\
    \midrule
    While
    &
    \begin{lstlisting}[language=imp-pretty-no-lines]
while(<PREDICATE>)
{
    <BODY>
};
    \end{lstlisting}
    &
    $\sigma$
    &
    \begin{lstlisting}[language=imp-pretty-no-lines]
while(<PREDICATE>)
{
    -(*@\colorbox{red!20}{<BODY>}@*)
    +(*@\colorbox{green!20}{halt;}@*)
};
    \end{lstlisting}
    &
    $\sigma$\\
    \midrule
    If-else
    &
    \begin{lstlisting}[language=imp-pretty-no-lines]
if(<PREDICATE>)
{
    <BODY>
}
else
{
    <BODY>
};
    \end{lstlisting}
    &
    $\sigma$
    &
    \begin{lstlisting}[language=imp-pretty-no-lines]
if(<PREDICATE>)
{
    -(*@\colorbox{red!20}{<BODY>}@*)
    +(*@\colorbox{green!20}{halt;}@*)
}
else
{
    -(*@\colorbox{red!20}{<BODY>}@*)
    +(*@\colorbox{green!20}{halt;}@*)
};
    \end{lstlisting}
    &
    $\sigma$\\
    \midrule
    Halt
    &
    \begin{lstlisting}[language=imp-pretty-no-lines]
halt;
    \end{lstlisting}
    &
    $\sigma$
    &
    \begin{lstlisting}[language=imp-pretty-no-lines]
halt;
    \end{lstlisting}
    &
    $\sigma$\\
\end{tabular}%
\end{table}%

The objective of the \srp task is to challenge \LLMs with predicting
the ordered sequence of semantic rules that is required to evaluate an
\IMP statement when the program state before the execution of that
statement is given. Ideally, we want to avoid requiring the \LLMs from
needing to track program state since that capability is specifically
tested for in the \etp task, and we want to avoid any
overlaps/redundancies. This is trivial for statements that are
self-contained, such as \texttt{declaration}, \texttt{assignment}, and
\texttt{halt}. However statements such as \texttt{while},
\texttt{if-else}, \texttt{break},
%and \texttt{continue}
require some
processing to make them suitable for this task.

Table~\ref{tab:appendix-srp-data} shows how each type of statement is
processed to make it suitable for the \srp task. The primary objective
behind processing is to make edits to the sampled statements such that
they can be completely evaluated by requiring the least amount of
program state updates. The first, second, and third columns lists the
type of the sampled statement, its minimal representative skeleton,
and the program state captured before its evaluation respectively. The
fourth and the fifth columns list the sampled statement after
processing and the corresponding processed program state which can now
be used in the \srp task. For the sampled \texttt{declaration},
\texttt{assignment}, and \texttt{halt} statements, the statements and
the collected program state before their executions are used as is in
the \srp task because their evaluation does not require tracking
program state nor do they require the execution of other
statements. For \texttt{while} statements, we replace the body with a
\texttt{halt} statement. This removes any possibility of needing
state updates to correctly and completely evaluate the
\texttt{while} statement. A similar approach is used for processing
the \texttt{if-else} statement. For the \texttt{break} statement,
we capture its closest enclosing loop and remove all statements
from its body up until the \texttt{break} statement.

Since the \srp task is scoped to a statement level of granularity,
it is relatively agnostic to the complexity of the program as a whole.

\subsubsection{Most Mispredicted Rules}
\label{sec:appendix-srp-mis-predicted-rules}

\begin{figure}[h]
  % \vspace{-0.6cm}%
  \hspace{2cm}\includegraphics[scale=0.28]{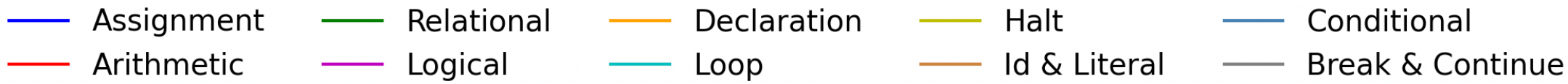}
  \vspace{2mm}
  \\
  \subcaptionbox{%
    \FigSRPUKFirstMisMatchIMPKRadarCaption%
    \label{figure:srp-uk-first-mismatch-imp-k}}[0.3\linewidth]{%
    \includegraphics[scale=0.18]{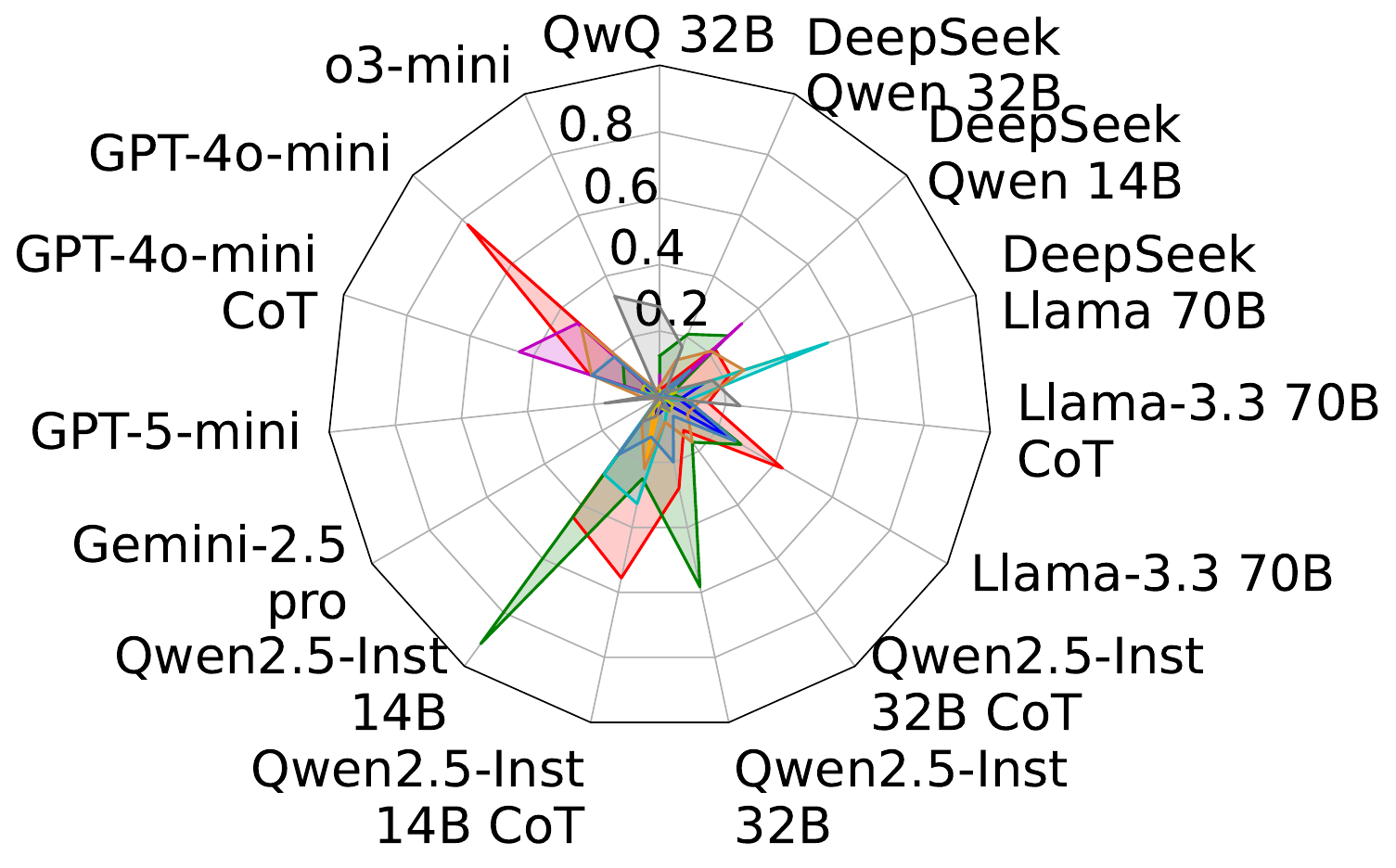}
  }%
  \hspace{5mm}%
  \subcaptionbox{%
    \FigSRPMKSwapFirstMisMatchIMPKRadarCaption%
    \label{figure:srp-mk-swap-first-mismatch-imp-k}}[0.3\linewidth]{%
    \includegraphics[scale=0.18]{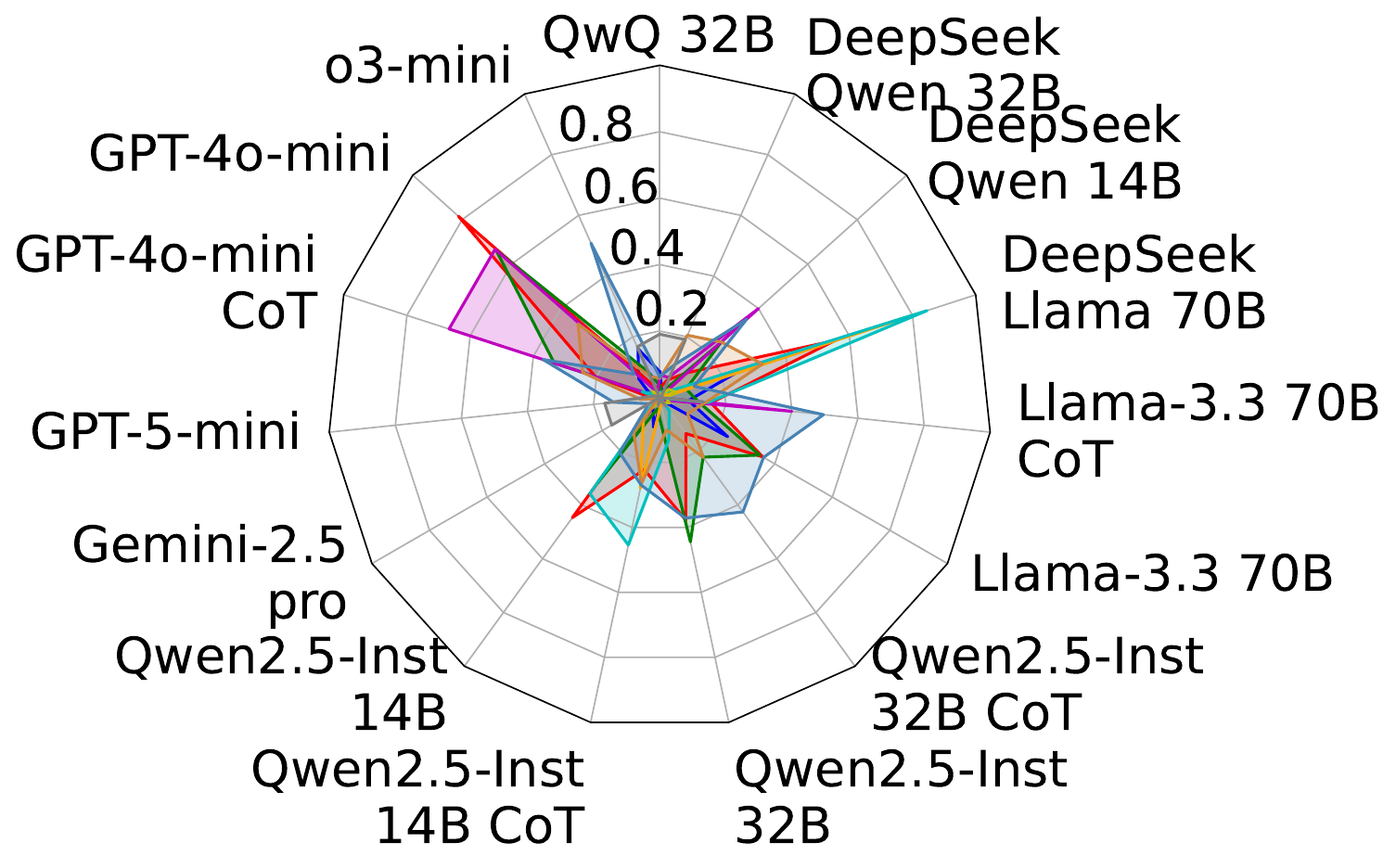}
  }%
  \hspace{5mm}%
  \subcaptionbox{%
    \FigSRPMKUnseenFirstMisMatchIMPKRadarCaption%
    \label{figure:srp-mk-unseen-first-mismatch-imp-k}}[0.3\linewidth]{%
    \includegraphics[scale=0.18]{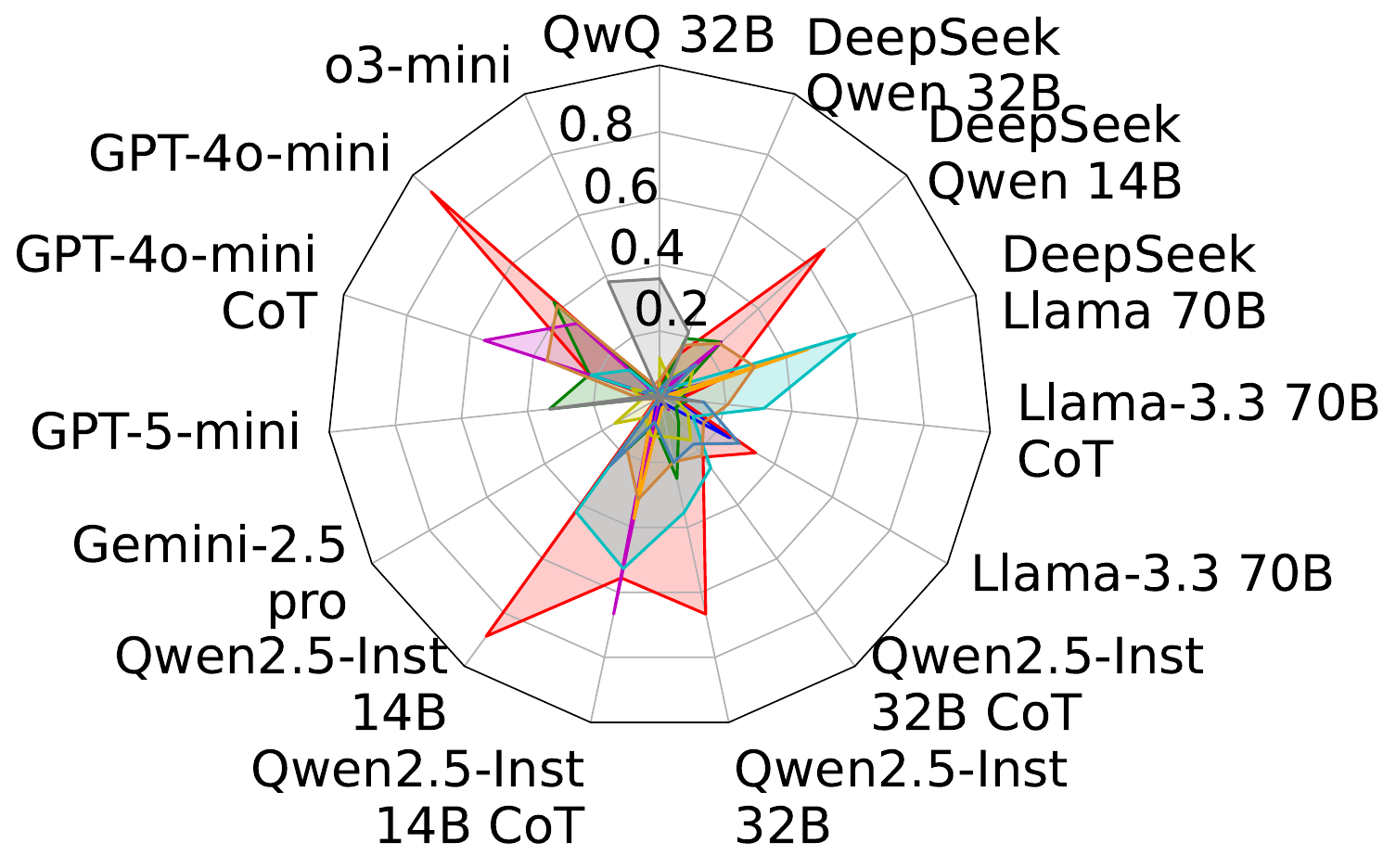}
  }%
  \\
  \subcaptionbox{%
    \FigSRPUKFirstMisMatchIMPSOSRadarCaption%
    \label{figure:srp-uk-first-mismatch-imp-sos}}[0.3\linewidth]{%
    \includegraphics[scale=0.18]{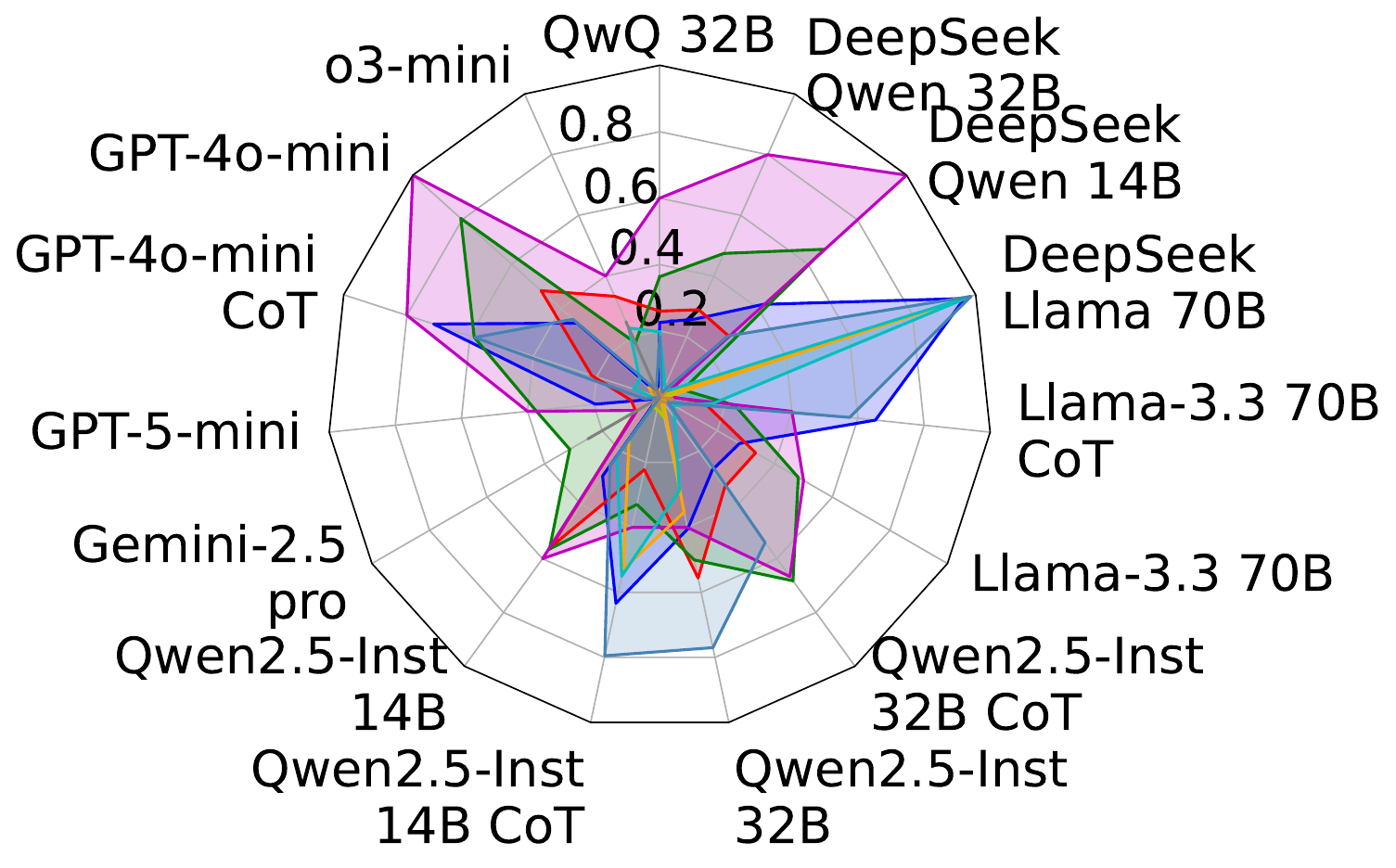}
  }%
  \hspace{5mm}%
  \subcaptionbox{%
    \FigSRPMKSwapFirstMisMatchIMPSOSRadarCaption%
    \label{figure:srp-mk-swap-first-mismatch-imp-sos}}[0.3\linewidth]{%
    \includegraphics[scale=0.18]{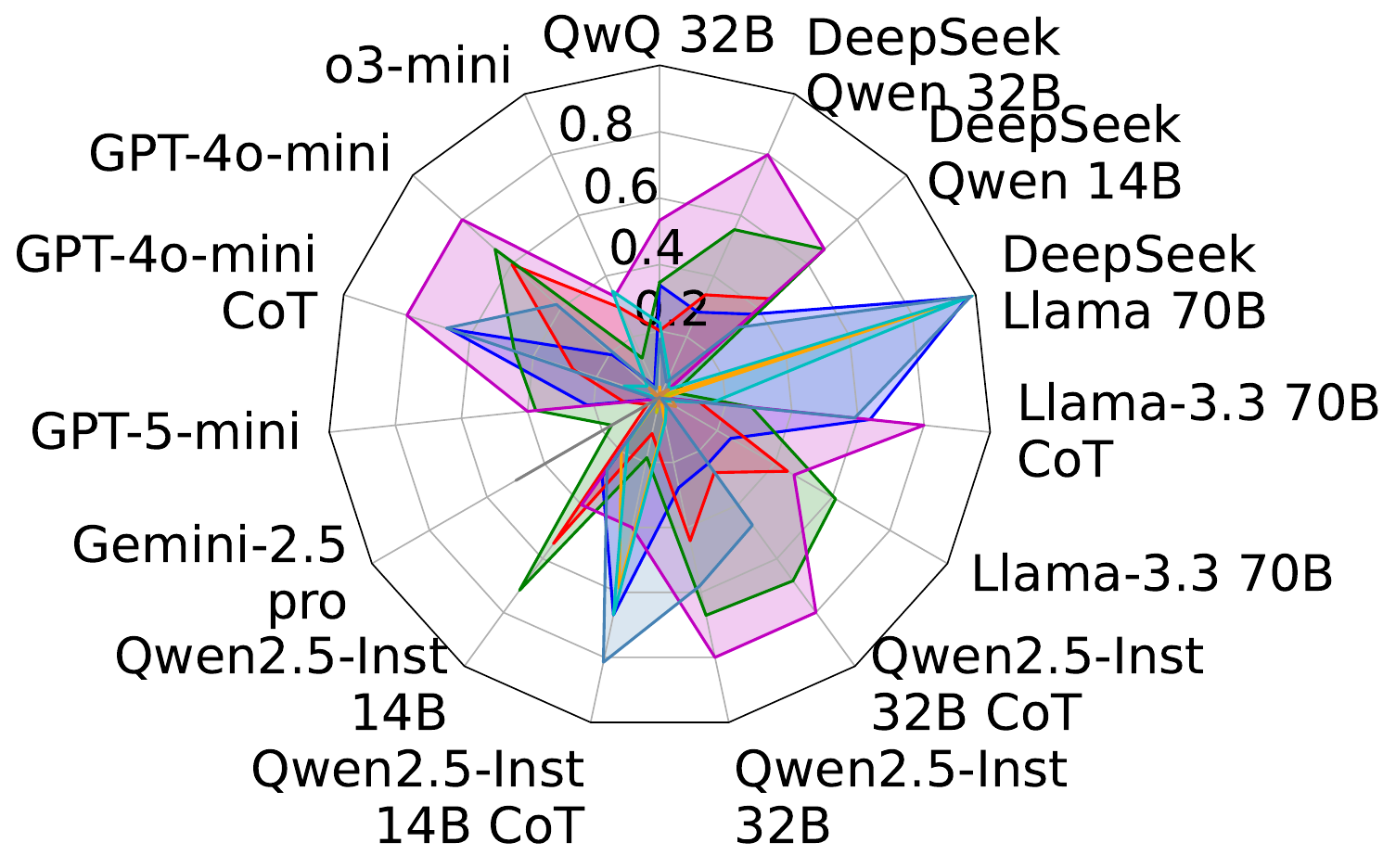}
  }%
  \hspace{5mm}%
  \subcaptionbox{%
    \FigSRPMKUnseenFirstMisMatchIMPKRadarCaption%
    \label{figure:srp-mk-unseen-first-mismatch-imp-sos}}[0.3\linewidth]{%
    \includegraphics[scale=0.18]{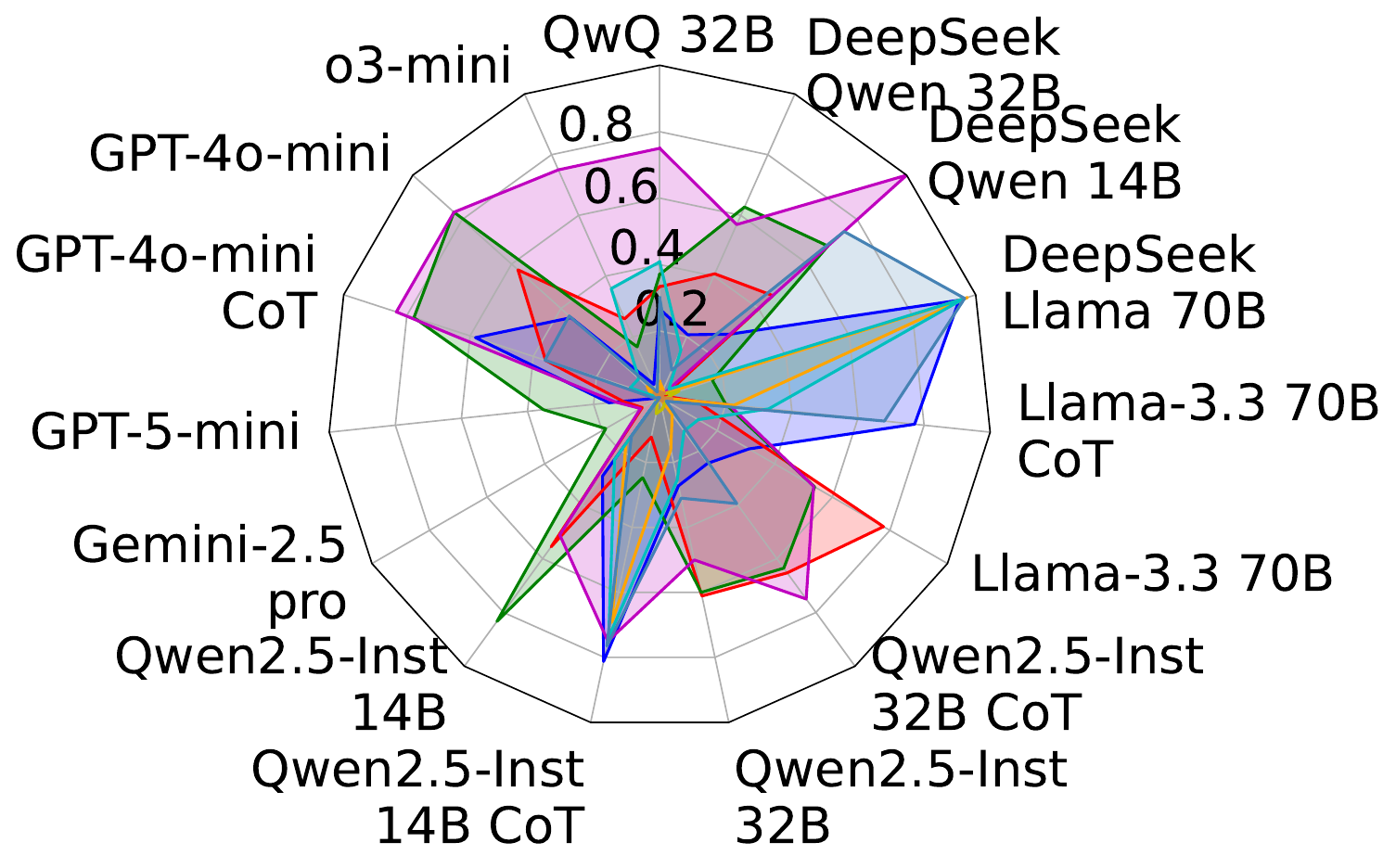}
  }%
  \caption{\FigSRPFirstMisMatchRadarCaption}
\end{figure}

%% Manually modified by Jiyang

\begin{wraptable}{r}{0.42\textwidth}
  \centering%
  \vspace{-0.25cm}%
  \TableFont%
  \caption{\TableSRPCategoryMappingCaption}
  \begin{tabular}{lll}%
    \toprule%
    \textbf{Category} & \textbf{\Kos} & \textbf{\Sos} \\
    \midrule%
    Assignment & Rule 21 & Rules 4 - 6 \\
    Arithmetic & Rules 3 - 11 & Rules 7 - 27 \\
    Relational & Rules 12 - 17 & Rules 28 - 51 \\
    Logical & Rules 18 - 20 & Rules 52 - 62 \\
    Declaration & Rule 36 & Rule 3 \\
    Loop & Rules 24 - 25 & Rules 67 - 70 \& Rule 77\\
    Break \& Continue & Rules 27 - 35 & Rules 71 - 76 \\
    Halt & Rule 26 & Rule 78\\
    Id & Rules 1 - 2 & Rules 1 - 2\\
    Conditional & Rules 22 - 23 & Rules 64 - 66\\
    \bottomrule%
  \end{tabular}
  \vspace{-0.1cm}%
\end{wraptable}%
To identify the semantic rules that models struggle with, we compute
the
\emph{first-point-of-mismatch rate} for each rule, which is the
frequency of the rule as the first mismatch between ground truth and
the model prediction, relative to its total number of occurrences in
the \srp dataset.  We group the rules into the following categories:
\emph{Assignment}, \emph{Relational}, \emph{Declaration}, \emph{Halt},
\emph{Conditional},
\emph{Arithmetic}, \emph{Logical}, \emph{Loop},
\emph{Id}, and \emph{Break \& Continue}. The mapping between the
semantic rules and these categories for the \Kos-semantics and \Sos is shown
in Table~\ref{tab:appendix-srp-mapping}.

The first-point-of-mismatch rate for a category is the maximum across
all the rules within this category.
Figure~\ref{figure:srp-first-mismatch-radar} shows the
first-point-of-mismatch rate across categories for all the models on
the \humanwrit dataset for the \standardSem and \nstandardSem
semantics, for both their \Kos-semantics (top) and \Sos (bottom)
formalizations.

Firstly, we observe that the models in general mispredict rules to a
larger extent for \Sos relative to when provided with the
\Kos-semantics. Furthermore, categories such as \texttt{Declaration},
\texttt{Id \& Literal}, and \texttt{Halt} that generally require one
or at most two rules are almost never mispredicted significantly by
any model across all the different cases. This is also observed for
the \texttt{Assignment} category under \Kos-semantics which is
formalized by just one rule and we see that its misprediction rate is
low across models. In contrast, the \texttt{Assignment}
category is heavily mispredicted under \Sos formalization for
\standardSem and \nstandardSem semantics for a large number of
models. We see a similar story with the \texttt{Logical} category
where models mispredict it more significantly under \Sos than
\Kos-semantics. The \texttt{Logical} category contains the three
logical operators (\texttt{AND}, \texttt{OR}, and \texttt{NOT}) and we
see that exactly three rules are required under \Kos-semantics thus
one rule per operator whereas \Sos requires ten rules, almost 4x more
rules per operator than \Kos-semantics. Similar trends are observed in
the \texttt{Relational} category.

\section{Use of External Assets}
\label{sec::appendix-external-assets}

In this work, we make use of several external assets, including
datasets, and pretrained models.
We acknowledge and credit the original creators of these assets as follows:

\subsection{Data}
We construct the \humanwrit dataset by rewriting the existing code solutions from the following
sources:
\let\1\relax
\begin{outline}[enumerate]
    \1 HumanEval-X
        \2 License: Apache 2.0
        \2 URL: \url{https://huggingface.co/datasets/THUDM/humaneval-x}
    \1 BabelCode MBPP
        \2 License: CC 4.0
        \2 URL: \url{https://huggingface.co/datasets/gabeorlanski/bc-mbpp}
    \1 CodeContests
        \2 License: CC 4.0
        \2 URL: \url{https://github.com/google-deepmind/code_contests}
    \1 Leetcode
        \2 We scrape only the ground-truth solutions and public test cases
        from leetcode. We use the collected problems for academic purposes only.
        \2 URL: \url{https://leetcode.com/}
\end{outline}

We construct the \llmtrans dataset by using \qwenCoder{32} to translate the C++
solutions to problems from:
\let\1\relax
\begin{outline}[enumerate]
    \1 CodeForces
        \2 License: CC 4.0 
        \2 URL: \url{https://huggingface.co/datasets/open-r1/codeforces}
\end{outline}

\subsection{Models}
We evaluate \LLMs designed for coding tasks and enhanced reasoning
ability on our \dataset :
\begin{outline}[enumerate]
    \1 \llamaBig~\citep{grattafiori2024llama},
        \2 License: llama3.3
        \2 URL:\\\url{https://huggingface.co/meta-llama/Llama-3.3-70B-Instruct}
    \1 Qwen2.5-Coder Models~\citep{hui2024qwen2},
        \2 License: Apache 2.0
        \2 URLs:\\\url{https://huggingface.co/Qwen/Qwen2.5-Coder-32B-Instruct}\\
        \url{https://huggingface.co/Qwen/Qwen2.5-Coder-14B-Instruct}
    \1 DeepSeek-R1 distilled models~\citep{guo2025deepseek}
        \2 License: MIT
        \2 URLs:\\ \url{https://huggingface.co/deepseek-ai/DeepSeek-R1-Distill-Llama-70B}\\\url{https://huggingface.co/deepseek-ai/DeepSeek-R1-Distill-Qwen-32B}\\\url{https://huggingface.co/deepseek-ai/DeepSeek-R1-Distill-Qwen-14B}
    \1 \qwq~\citep{qwq32b}
        \2 License: Apache 2.0
        \2 URL: \url{https://huggingface.co/Qwen/QwQ-32B}
    \1 \gemini. In this study, we utilized the \gemini model provided by Google AI. The use of this model is subject to the Generative AI Preview Terms and Conditions, as outlined in the Google Cloud Service Specific Terms for Pre-GA Offerings.
        \2 URL: \url{https://cloud.google.com/terms/service-terms}
    \1 OpenAI Models. In this study, the use of OpenAI's models is subject to the term of use.
        \2 URL: \url{https://openai.com/policies/row-terms-of-use/}
\end{outline}

\subsection{Icons}
We use several icons from \url{https://www.flaticon.com} which we
attribute here.

\begin{itemize}
\item Document icons created by Roman Káčerek - \url{https://www.flaticon.com/free-icons/document}
\item Robot icons created by Kiranshastry - \url{https://www.flaticon.com/free-icons/robot}
\item Xml icons created by Dimitry Miroliubov - \url{https://www.flaticon.com/free-icons/xml}
\item Diff icons created by brajaomar\_j - \url{https://www.flaticon.com/free-icons/diff}
\item Matrix icons created by meaicon - \url{https://www.flaticon.com/free-icons/matrix}
\item Logistic regression icons created by raidolicon - \url{https://www.flaticon.com/free-icons/logistic-regression}
\item Game chart icons created by Arslan Haider - \url{https://www.flaticon.com/free-icons/game-chart}
\item Gears icons created by sonnycandra - \url{https://www.flaticon.com/free-icons/gears}
\item Message icons created by Freepik - \url{https://www.flaticon.com/free-icons/message}
\end{itemize}

\begin{center}
  \vspace{0.5cm}%
\begin{lstlisting}[language=java-pretty]
int L;
int p;
int y;
int d;
int K;
int h;
int T;
int Y;
int ble0;
int ble1;
int ble2;
int ble3;
int ble4;
int ble5;
int ble6;
int ble7;
int ble8;
int ble9;
int ble10;
int ble11;
int ble12;
int ble13;
int ble14;
int ble15;
int ble16;
int ble17;
int ble18;
int ble19;
int ble20;
int ble21;
int ble22;
int ble23;
int ble24;
int ble25;
int ble26;
int ble27;
int ble28;
int ble29;
int ble30;
int ble31;
int ble32;
int ble33;
int ble34;
int ble35;
int ble36;
int ble37;
int ble38;
L = (((- y) / 4) - p);
T = ((((3 + K) + 1) - 3) + (8 / 8));
L = ((((((- p) % 1) * 7) - (- 1)) - (- 9)) - 3);
K = (((((d * (- 5)) + y) + (- 5)) - (- K)) - (- 5));
Y = (((9 + 3) - T) + 5);
K = ((7 / 7) * L);
y = ((((L + L) + 8) + 3) + 1);
p = ((((- 8) * 9) - ((- 6) % (- 8))) - T);
ble0 = (- 1);
ble1 = (- 1);
ble2 = (- 1);
ble3 = (- 1);
ble4 = (- 1);
ble5 = (- 1);
ble6 = (- 1);
ble7 = (- 1);
ble8 = (- 1);
ble9 = (- 1);
ble10 = (- 1);
ble11 = (- 1);
ble12 = (- 1);
ble13 = (- 1);
ble14 = (- 1);
ble15 = (- 1);
ble16 = (- 1);
ble17 = (- 1);
ble18 = (- 1);
ble19 = (- 1);
ble20 = (- 1);
ble21 = (- 1);
ble22 = (- 1);
ble23 = (- 1);
ble24 = (- 1);
ble25 = (- 1);
ble26 = (- 1);
ble27 = (- 1);
ble28 = (- 1);
ble29 = (- 1);
ble30 = (- 1);
ble31 = (- 1);
ble32 = (- 1);
ble33 = (- 1);
ble34 = (- 1);
ble35 = (- 1);
ble36 = (- 1);
ble37 = (- 1);
ble38 = (- 1);
if((((y - K) - 2) == ((y % 8) % 5)) || ((L + y) < ((0 / 1) * (- K))))
{
    while(((((7 % 6) % 2) > (h + (- d))) || ((T % 1) != ((T * d) * (- 3)))) && (ble0 < 0))
    {
        while((((y + (- K)) <= (((- 1) + p) + L)) || (((p - (- L)) - 9) > (T - p))) && (ble1 <= 5))
        {
            h = ((h - (4 % 2)) + (h * K));
            ble1 = (ble1 + 2);
        };
        while(((((h - p) + 2) > ((9 * h) % 3)) || (((K + 4) - L) <= ((0 - h) + Y))) && (ble2 < 20))
        {
            while(((!(((0 - d) + Y) != (h + L))) || ((d + (- d)) >= ((p - 1) - p))) && (ble3 < 15))
            {
                T = (((((1 + (- Y)) - 0) + h) - 4) - 1);
                ble3 = (ble3 + 5);
            };
            Y = (1 + (5 / 6));
            while(((!((d - L) <= ((1 * p) + L))) || ((h - K) >= ((9 - d) - (- h)))) && (ble4 < 9))
            {
                if((!(((- T) - ((- Y) % 8)) == (L + T))) || (((4 - y) + p) > (p * Y)))
                {
                    T = ((9 - Y) + (p % 1));
                    while(((((- L) + y) <= ((6 * h) - K)) || ((1 + (Y % 9)) != ((p * y) + (- 7)))) && (ble5 <= 17))
                    {
                        K = ((9 + 9) + (5 * 9));
                        ble5 = (ble5 + 3);
                    };
                    T = (((5 - 5) - 1) - (3 * 1));
                }
                else
                {
                    p = ((7 / (- 3)) + 8);
                    break;
                };
                if(((d - L) < ((5 % 9) % 1)) && (!((T * K) <= (Y - K))))
                {
                    while((((T / 4) > ((- Y) - T)) && (((- 4) - ((- y) * (- T))) < (K + (3 / 3)))) && (ble6 < 13))
                    {
                        Y = (7 + (Y / 9));
                        ble6 = (ble6 + 1);
                    };
                    if((((d + h) - 2) <= ((L - T) + 8)) && (((7 + (- h)) - h) == (L + T)))
                    {
                        while(((!((((- y) - p) + 7) <= (d + d))) && ((Y + y) > (Y + h))) && (ble7 > (- 12)))
                        {
                            break;
                            ble7 = (ble7 + (- 3));
                        };
                    }
                    else
                    {
                        h = ((8 - ((- 0) / 4)) + 2);
                    };
                    d = (((2 + (Y % 4)) - 8) - K);
                }
                else
                {
                    d = (((Y + Y) - L) - 2);
                };
                while((((T / (- 6)) < (p % 6)) || (((T + d) - 9) == ((9 + K) + h))) && (ble8 > (- 20)))
                {
                    while((((T - d) > (T + p)) && (((h - 9) + p) == ((p + 1) - p))) && (ble9 > (- 15)))
                    {
                        p = (((8 / (- 6)) + 0) + (4 / 8));
                        K = ((((d % 5) + 7) + (9 / 2)) - 7);
                        ble9 = (ble9 + (- 4));
                    };
                    ble8 = (ble8 + (- 1));
                };
                ble4 = (ble4 + 3);
            };
            ble2 = (ble2 + 6);
        };
        ble0 = (ble0 + 2);
    };
}
else
{
    p = (((L - y) - 4) + 5);
    while((((p + p) <= (d + h)) && (((L - L) - (- 1)) <= (((- h) + 7) - p))) && (ble10 <= 20))
    {
        while((((y / 7) >= ((5 - p) + (- Y))) || ((Y + (- Y)) >= ((4 * d) + (- Y)))) && (ble11 >= (- 9)))
        {
            if(((d - y) >= (h - K)) && ((T * (- T)) != ((- T) + (- Y))))
            {
                break;
                break;
                if((((- Y) % 9) > (p + y)) && (!(((3 % 6) + y) != (L + K))))
                {
                    break;
                    K = (6 - ((- 6) % 5));
                    if((((K + 6) + L) <= ((4 / 8) + Y)) || ((T * T) > (y + K)))
                    {
                        L = (d + (3 * 6));
                        while(((((T - L) + 2) > (Y + T)) || (!((K + h) <= ((d + y) - 5)))) && (ble12 < 11))
                        {
                            y = ((Y * h) - ((3 * 8) / 8));
                            break;
                            while(((((3 + d) - y) != (p / (- 6))) && (((- T) - h) >= (L / 1))) && (ble13 >= (- 17)))
                            {
                                L = (((Y / 2) * T) + (8 % (- 9)));
                                y = ((- y) + ((- p) * T));
                                p = (((- 6) + (8 * 5)) - 8);
                                ble13 = (ble13 + (- 3));
                            };
                            ble12 = (ble12 + 3);
                        };
                        p = ((((K * (- 6)) * 3) - 2) + 7);
                    }
                    else
                    {
                        Y = (((8 % 4) - (p * 4)) - 8);
                        if((((d + (- d)) + 0) == (((- d) / 4) * K)) || ((Y * y) >= ((1 % 8) + (- L))))
                        {
                            y = (((5 * p) + T) - d);
                            p = ((0 * 0) - (((0 / 3) % 1) / 9));
                        }
                        else
                        {
                            while((((K - (d * 2)) != (p / 2)) || (((L / 9) - y) < (Y - T))) && (ble14 < 20))
                            {
                                y = ((p - (0 * (- h))) + L);
                                p = (((((- 4) - 6) - y) + L) + T);
                                break;
                                ble14 = (ble14 + 1);
                            };
                            h = ((T - 4) + 9);
                            T = (((((- 5) - 3) + 2) - 1) + 8);
                        };
                        y = ((((2 + (9 / 9)) + 4) - (- 0)) + 1);
                    };
                }
                else
                {
                    break;
                };
            }
            else
            {
                while((((((- h) + 4) + K) <= (h - Y)) || (((6 * p) + d) < (T / (- 2)))) && (ble15 < 9))
                {
                    L = (Y - (((d * h) / (- 8)) % 1));
                    K = (((((- 4) * 9) + (7 * 2)) + 1) + 1);
                    d = (((6 - (L * 5)) - 0) + 8);
                    ble15 = (ble15 + 3);
                };
                while(((!((((- 1) * p) - y) != (y - h))) || ((((- 4) / (- 5)) + d) <= (y + (- h)))) && (ble16 > (- 7)))
                {
                    break;
                    K = (((d * h) + 5) - 7);
                    while((((d % (- 1)) <= ((- p) * K)) && ((h - y) == (p - h))) && (ble17 > (- 13)))
                    {
                        T = (((((0 - T) + 0) + 6) + 2) - 2);
                        break;
                        break;
                        ble17 = (ble17 + (- 3));
                    };
                    ble16 = (ble16 + (- 2));
                };
                while((((Y + (T * 3)) == (((- d) - (- 6)) + p)) || (((L % 8) - L) == (T + h))) && (ble18 >= (- 12)))
                {
                    d = (((6 % (- 6)) - (K * T)) + L);
                    if((((5 * d) + h) > ((L * 4) / 9)) && ((K + L) > (Y - y)))
                    {
                        K = ((8 + ((7 * 1) / 7)) - (- 3));
                        d = ((p - ((- 2) / 2)) - (1 * 6));
                    }
                    else
                    {
                        break;
                    };
                    if((!((0 - (d % 5)) != (L * d))) || ((d + p) >= ((8 + d) + (- K))))
                    {
                        d = (((K / 8) % 6) - (T % (- 9)));
                        if(((h / 6) >= ((- K) / 1)) || (((h - d) + (- 7)) >= ((y + 1) - h)))
                        {
                            Y = ((((T - (p * T)) - 2) - 4) + 4);
                        }
                        else
                        {
                            K = (6 + ((- 0) % 2));
                            h = (((8 + T) + 8) - 2);
                        };
                    }
                    else
                    {
                        if(((((- 3) + h) - p) < ((9 + d) + L)) && ((K + d) > (d * y)))
                        {
                            if(((h / 6) < ((y - d) - 4)) || ((h + L) != (y - (T / 3))))
                            {
                                Y = ((- 8) + ((- 5) % 7));
                            }
                            else
                            {
                                d = (((6 + 8) + 4) - (- 6));
                            };
                            while((((p + (T % 4)) <= ((6 + L) - p)) || (((0 * p) - (- K)) >= (((- 1) + (- K)) + Y))) && (ble19 < 12))
                            {
                                break;
                                T = (Y + (T / 5));
                                ble19 = (ble19 + 1);
                            };
                            Y = ((L - (5 % 8)) + ((T % 1) % 2));
                        }
                        else
                        {
                            while((((p * d) != (((- L) + T) + 9)) && (!(((8 + d) - y) < (p + y)))) && (ble20 <= 18))
                            {
                                if(((d - (3 * L)) < ((p + d) - 6)) || ((T - d) <= ((T % 1) + (- L))))
                                {
                                    break;
                                }
                                else
                                {
                                    break;
                                    h = ((((5 + 9) - 1) - (4 / 3)) - 3);
                                    L = (((p + (0 * 8)) + 0) + 8);
                                };
                                ble20 = (ble20 + 5);
                            };
                        };
                    };
                    ble18 = (ble18 + (- 3));
                };
            };
            if(((Y * (- K)) == (T - h)) || (((L % 6) - (- K)) >= ((K / (- 2)) / 6)))
            {
                h = ((((4 % 6) + (6 % 1)) + 3) + (- 1));
                while((((K + y) == ((- T) - (- L))) && ((Y / 9) != (((- p) * h) + 7))) && (ble21 >= (- 8)))
                {
                    if((!(((4 % 6) % 6) >= (p / 8))) && (((2 + y) - K) >= ((1 + y) - (- y))))
                    {
                        T = ((((- L) * 1) - (5 * (- 8))) + 6);
                    }
                    else
                    {
                        h = (((((0 - y) + L) + 5) + T) + 5);
                    };
                    if(((Y * p) <= ((4 % 2) + T)) && ((d + L) >= ((y - (- 1)) + h)))
                    {
                        h = ((((6 + 2) - (- Y)) + y) - 1);
                    }
                    else
                    {
                        d = ((((L * (- 5)) - T) - (- L)) - (2 / 8));
                        while(((((K + (- K)) + 8) != (d + h)) || ((Y * (- K)) < (T - (- K)))) && (ble22 > (- 20)))
                        {
                            while((((y % 3) >= ((6 - (- Y)) + T)) && ((y + (- K)) >= ((K + 8) + p))) && (ble23 < 18))
                            {
                                K = (((4 % (- 9)) + (- K)) + y);
                                h = ((2 * 7) - (7 % 7));
                                break;
                                ble23 = (ble23 + 1);
                            };
                            y = (((T - 8) + ((9 % 3) % 1)) + (- 8));
                            ble22 = (ble22 + (- 2));
                        };
                    };
                    y = ((T + 2) - ((((- y) / 9) % 6) / (- 9)));
                    ble21 = (ble21 + (- 3));
                };
                y = ((d - 4) - 7);
            }
            else
            {
                while((((y * K) != (L + (- Y))) || (((3 / 2) - p) > (K / 4))) && (ble24 >= (- 4)))
                {
                    while((((d + Y) < (h - L)) || (((Y + (- h)) + 7) >= (T - d))) && (ble25 <= 5))
                    {
                        while((((K - T) <= (T - (- Y))) || ((y + d) >= ((p % 9) * T))) && (ble26 > (- 15)))
                        {
                            break;
                            ble26 = (ble26 + (- 2));
                        };
                        ble25 = (ble25 + 2);
                    };
                    d = (((((- L) - h) + K) + (3 * 2)) + 7);
                    while(((!(((T - L) + (- 5)) >= ((2 * (- K)) + T))) || (((8 + (- T)) - (- y)) <= (y / 4))) && (ble27 > (- 19)))
                    {
                        while((((8 + (L * y)) >= (d / 8)) || (((y / 3) + T) < ((7 * h) + (- Y)))) && (ble28 > (- 9)))
                        {
                            break;
                            while((((Y + (Y * (- 5))) >= ((p * (- h)) - 4)) && ((p % 3) > (Y - h))) && (ble29 >= (- 10)))
                            {
                                break;
                                ble29 = (ble29 + (- 2));
                            };
                            ble28 = (ble28 + (- 1));
                        };
                        K = (((((p + d) + 0) + (- 9)) - 0) + 9);
                        if((((h * 2) + T) == (y - h)) || ((y % 7) < (L + p)))
                        {
                            while((((y - (Y % 5)) == ((L * (- L)) + 0)) || (((- 7) + (p * T)) > (L / 5))) && (ble30 >= (- 2)))
                            {
                                h = ((3 + (- d)) - ((- T) * Y));
                                T = (((((- 6) * L) - y) + 7) + 6);
                                ble30 = (ble30 + (- 2));
                            };
                            if((((h * 5) % 7) <= ((d - (- Y)) - 4)) && ((L + Y) <= ((T * 5) % 4)))
                            {
                                Y = (((3 + 7) + 3) - 9);
                                T = ((3 + L) - ((y % (- 6)) * (- h)));
                                h = ((h * T) - (7 / 4));
                            }
                            else
                            {
                                Y = ((((0 - (- 8)) + 5) + 6) - (5 * 9));
                                T = ((Y + (- 5)) + (- 6));
                            };
                        }
                        else
                        {
                            Y = (((3 - K) - Y) + (0 / 1));
                            while((((L - h) < (y + y)) && ((p * y) == (((- h) * K) % (- 2)))) && (ble31 <= 17))
                            {
                                K = (((y * y) - d) + 7);
                                while((((L / 7) != ((- p) / 6)) && (!(((4 - p) - T) == ((5 * K) / (- 2))))) && (ble32 > (- 6)))
                                {
                                    L = ((((- K) + (- 4)) - p) + (- d));
                                    d = ((y + 9) - 1);
                                    h = ((((K / 4) - 8) - (- 4)) + ((- 6) * 5));
                                    ble32 = (ble32 + (- 2));
                                };
                                K = (((6 + 6) - 6) + (p / 2));
                                ble31 = (ble31 + 6);
                            };
                            while((((8 - (h * (- p))) != (((- Y) / 2) + d)) && (((h - T) - (- 6)) > (y * d))) && (ble33 <= 0))
                            {
                                T = (((((- 2) + 7) + (- 9)) + 0) + 0);
                                break;
                                ble33 = (ble33 + 2);
                            };
                        };
                        ble27 = (ble27 + (- 5));
                    };
                    ble24 = (ble24 + (- 2));
                };
            };
            while(((((L * 2) - T) != (d + L)) || (((2 + (- T)) - d) == ((6 - h) - L))) && (ble34 > (- 7)))
            {
                d = ((K * Y) + L);
                L = ((4 % 8) % 4);
                ble34 = (ble34 + (- 1));
            };
            ble11 = (ble11 + (- 3));
        };
        T = ((((- 3) * (- K)) + 4) - (- 7));
        h = (((((7 + 4) + 4) - (- 9)) + p) - 2);
        ble10 = (ble10 + 5);
    };
    if((((Y * (- 5)) + L) > (((- 9) - Y) - T)) || (((L / 8) % 4) == (K - K)))
    {
        while(((!((h + T) != (K + ((- Y) * 7)))) || (!(((L / 9) - T) < ((Y + T) + (- 8))))) && (ble35 <= (- 3)))
        {
            Y = ((0 / 7) - 2);
            Y = ((T + 2) + 8);
            ble35 = (ble35 + 2);
        };
        while(((((3 + K) + L) != ((3 - h) + d)) && ((d + (- L)) > ((K - 8) - y))) && (ble36 < (- 1)))
        {
            if(((K + T) > (d + K)) || ((y - K) == ((L + 1) + p)))
            {
                L = (((8 * p) * p) * L);
            }
            else
            {
                if((!((((- K) * T) + 8) != (L + K))) || (((6 % 1) - p) != ((y + 1) - K)))
                {
                    y = ((((Y - 0) + 5) - h) + (- L));
                    break;
                    while((((((- Y) - h) + 9) <= (Y * p)) || ((((- 6) - K) + Y) <= (Y - y))) && (ble37 < 20))
                    {
                        K = ((p % 7) + ((d / 6) * (- 3)));
                        while((((y + (K * 7)) == ((- d) + T)) || ((h * h) >= (6 + (L * Y)))) && (ble38 > (- 18)))
                        {
                            Y = (3 - (L % 9));
                            T = ((p + Y) + L);
                            ble38 = (ble38 + (- 2));
                        };
                        K = ((p - Y) - (((- T) * 1) / 3));
                        ble37 = (ble37 + 6);
                    };
                }
                else
                {
                    h = ((9 + (- p)) + 8);
                    y = ((((K * 7) % 6) / 4) + T);
                };
            };
            h = (((L + (2 * 6)) - 1) + (- 3));
            ble36 = (ble36 + 2);
        };
    }
    else
    {
        Y = (((6 - p) - (4 * (- Y))) - L);
    };
};
\end{lstlisting}
  \captionof{figure}{\FigAppendixIMPFuzzGenExampleCaption}
\end{center}

\end{document}